\documentclass[12pt]{article}
\usepackage{amssymb,amsmath,epsfig}

\allowdisplaybreaks
\begin{document}
\title{\bf Study of Anisotropic Polytropes in $f(\mathcal{R},\mathrm{T})$ Theory}
\author{M. Sharif \thanks {msharif.math@pu.edu.pk}, Amal Majid
\thanks{amalmajid89@gmail.com} and M. Shafaqat
\thanks{shafaqatbasheer56@gmail.com}\\
Department of Mathematics, University of the Punjab,\\
Quaid-e-Azam Campus, Lahore-54590, Pakistan.}

\date{}
\maketitle

\begin{abstract}
This paper examines the general formalism and applications of
isotropic as well as anisotropic polytropic stars in
curvature-matter coupled gravity. For this purpose, we consider
static spherical and Schwarzschild spacetimes in the interior and
exterior regions, respectively. We use two polytropic equations of
state to obtain physically viable solutions of the field equations.
The hydrostatic equilibrium and Lane-Emden equations are developed
for both isotropic as well as anisotropic cases. We study the
effects of anisotropic pressure on the stellar structure. Moreover,
we graphically inspect the physical behavior of isotropic as well as
anisotropic polytropes through energy conditions and stability
criterion. Finally, we discuss Tolman mass to explore some
characteristics of the models. It is concluded that more viable and
stable polytropes are found in this theory as compared to general
relativity.
\end{abstract}

\section{Introduction}

Gravitational collapse and dark energy are considered the most
interesting and highly critical issues of astrophysics and
cosmology. A star remains in an equilibrium state when the external
pressure of the fluid and the inward gravitational pull
counter-balance each other. The stable state is disturbed when the
stellar objects do not have enough pressure to balance the
gravitational force. This results in the formation of new remnants
such as \textit{white dwarfs}, \textit{neutron stars} or
\textit{black holes} (BHs) depending on their masses. A celestial
object becomes a white dwarf if its mass is less than eight times
mass of the sun and star having a mass between eight to twenty times
solar mass transformed into a neutron star. A BH is formed when the
star has a mass more than twenty times mass of the sun. White dwarfs
and neutron stars are directly detectable while ground-based
measurements ensure the presence of BHs. The first justification for
the presence of a BH was given by astronomers in the Andromeda
galaxy and later $M$104, 4$NGC$3115, $M$106 and Milky way galaxies
confirmed the existence of BHs \cite{1}.

Over the last two decades, the expanding behavior of the universe
has been the most fascinating and dazzling result for the scientific
community. Scientists claim that a cryptic force known as dark
energy is responsible for the cosmic acceleration. This enigmatic
energy has inspired many researchers to unveil its hidden
characteristics. Einstein's theory of general relativity (GR) linked
the motion of massive bodies to gravitational fields which
completely revolutionized our understanding of the universe. In GR,
the phenomenon of expanding cosmos is explained via the cosmological
constant ($\Lambda$) in $\Lambda$CDM model \cite{9a}-\cite{101}.
However, the $\Lambda$CDM model does not provide a suitable
explanation for the difference between the inferred value of
$\Lambda$ (120 orders of magnitude lower) and the predicted value of
the vacuum energy density. Moreover, the $\Lambda$CDM model fails to
explain why the present value is comparable to the matter density.
In cosmology, the presence of the flatness, monopole and horizon
problems together with the big-bang singularity indicate that the
standard cosmological model of GR cannot adequately describe the
cosmos at extreme regimes. On the other hand, a classical theory
like GR does not provide a full quantum description of spacetime and
gravity.

For these reasons, various alternative theories of gravity are
proposed which attempt to formulate a semiclassical scheme that
could replicate GR and its successes. These theories can be
established by adding curvature invariants and their associated
functions in the geometric part of the Einstein-Hilbert action. The
simplest approach to modify GR is $f(\mathcal{R})$ gravity \cite{2}.
The extended theories of gravity naturally overcome the shortcomings
of the standard big-bang model in GR by admitting an inflationary
behavior. The related inflationary scenarios seem capable of
matching the current observations of the cosmic microwave background
(CMB) \cite{6a, 102}. Finally, these extended schemes could
naturally solve the graceful exit problem, avoiding the shortcomings
of previous inflationary models \cite{6b, 8a}. However,
$f(\mathcal{R})$ theory is not in agreement with the solar system
tests \cite{13a, 103} and fails to justify the existence of a stable
stellar configuration \cite{15a}-\cite{105}. Moreover,
$f(\mathcal{R})$ gravity is inconsistent with the CMB radiation
tests as well as the strong lensing regime \cite{14a}-\cite{107}.
These limitations of $f(\mathcal{R})$ gravity led to its
generalizations which include coupling between the scalar curvature
and matter distribution.

Harko et al. \cite{3} developed such couplings which led to
$f(\mathcal{R},\mathrm{T})$ gravity. This theory is consistent with
the solar system observations as well as effectively describes the
late time accelerated expansion of the universe \cite{6}.
Furthermore, a specific form of $f(\mathcal{R}, \mathrm{T})$ gravity
emerged in a quantum gravitating system considered by Dzhunushaliev
et al. \cite{19a, 120}. The validity of $f(\mathcal{R}, \mathrm{T})$
theory is further strengthened by its compliance with the the dark
matter galactic effects and gravitational lensing test \cite{23a}.
In this framework, an additional force appears due to the existence
of non-conserved energy-momentum tensor which leads to non-geodesic
motion of particles. It also indicates that the flow of energy from
the gravitational field to the newly created matter is irreversible.
According to some studies, the non-conservation of the
energy-momentum tensor is supported by the accelerated expansion of
the universe \cite{22a, 108}.

Different cosmic scenarios have been explored in the context of
$f(\mathcal{R}, \mathrm{T})$ gravity. Sharif and Zubair \cite{5,
109} analyzed thermodynamics laws, energy conditions and anisotropic
universe models in this theory. Sharif and Yousaf \cite{7} explored
the stability of collapsing objects with isotropic matter
configuration in curvature-matter coupled gravity. Singh and Singh
\cite{8} found that if the ordinary matter is not included in the
system then the exotic matter acts as a cosmological constant due to
this coupling. Moraes et al. \cite{9} investigated the presence of
physically viable and stable compact stars in this framework. Yousaf
and Bamba \cite{10} analyzed the evolutionary behavior of compact
objects in this theory. The effect of curvature-matter coupling with
isotropic and anisotropic matter configurations on compact objects
has also been examined in \cite{11}-\cite{111}. Recently, Maurya and
Tello-Ortiz \cite{111a} focused on developing anisotropic spherical
structures that can successfully represent stellar structures.

Polytropes belong to a class of equation of state (EoS) which
relates pressure and density in a power-law form. These
self-gravitating spheres are the solutions of Lane-Emden equation
(LEE). The polytropic equation motivated many researchers to explore
physical characteristics of the polytropic models. Tooper \cite{12}
was the pioneer in studying relativistic polytropes who constructed
hydrostatic equilibrium and mass equations. He solved these
equations numerically and obtained physical quantities for
polytropes. Nilsson and Uggla \cite{12a} studied spherically
symmetric perfect fluid distribution with polytropic EoS and
concluded that the developed models have finite/infinite radii for
suitable values of the polytropic index. Ferrari et al. \cite{12b}
found numerical solution of both equations for two values of
polytropic index and explored the effects of increasing relativistic
parameter. Herrera and Barreto \cite{13} considered a relation
between radial and tangential pressures to construct relativistic
spherical polytropic models from known solution for isotropic
matter. They concluded that dimensionless density parameter attains
larger values for large anisotropy parameter. Azam et al. \cite{14}
discussed physical properties of anisotropic polytropes with
conformally flat conditions and investigated the stability of these
polytropes through Tolman mass. Nasim and Azam \cite{15} studied
charged anisotropic polytropes with generalized EoS. Polytropic EoS
has also been used in Einstein-Gauss-Bonnet gravity to model
spherically symmetric stars \cite{111b}.

There are two important sources of anisotropy given as follows. The
strong magnetic field visible in the dense objects \cite{17, 112}
and viscosity in neutron stars and extremely dense matter \cite{18,
113}. Maurya et al. \cite{19} studied spherically symmetric
anisotropic fluid distribution and obtained physically realistic
stellar models. Chowdhury and Sarkar \cite{20} discussed small
anisotropy due to rotation and inclusion of a magnetic field in
stellar objects. Abellan et al. \cite{21} developed the procedure to
study the anisotropic stars such that the radial and tangential
pressures satisfy the polytropic EoS.

Modified theories of gravity have great importance in the study of
self-gravitating objects. Henttunen et al. \cite{22} investigated
stellar configurations in $f(\mathcal{R})$ theory through different
polytropic EoS and explored the analytic solutions near star's core.
Sharif and Waseem \cite{23} analyzed spherically symmetric
anisotropic polytropes in $f(\mathcal{R},\mathrm{T},Q)$ theory.
Sharif and Siddiqa \cite{24} used the polytropic EoS to explain the
viability and stability of anisotropic dense objects in
$f(\mathcal{R},\mathrm{T})$ gravity. Bhatti and Tariq \cite{25}
described the detailed description of conformally flat spacetime
governed by a polytropic EoS in $f(\mathcal{R})$ gravity. Wojnar
\cite{26} examined the polytropic stars in the Palatini
$f(\mathcal{R})$ gravity.

In this paper, we study the general formalism and applications of
polytropic stars with isotropic/anisotropic matter configurations in
the framework of $f(\mathcal{R},\mathrm{T})$ gravity. The paper is
planned as follows. In section \textbf{2}, we derive the field
equations and Tolman-Oppenheimer-Volkoff (TOV) equation of this
theory. We examine isotropic and anisotropic polytropes with
polytropic EoS in sections \textbf{3} and \textbf{4}, respectively.
Finally, we discuss viability and stability of anisotropic
polytropes in section \textbf{5}. We summarize our results in the
last section.

\section{Curvature-Matter Coupled Gravity}

In this section, we establish the field equations with isotropic
matter configuration in the context of $f(\mathcal{R},\mathrm{T})$
gravity. The action of this modified theory is given as
\begin{equation}\label{1}
S=\int \left(\frac{f\left(\mathcal{R},\mathrm{T}\right)}
{16\pi}+\mathcal{L}_{m}\right){\sqrt{-g}}d^4x,
\end{equation}
where matter Lagrangian density and determinant of the metric tensor
are represented by $\mathcal{L}_{m}$ and $g$, respectively. By
varying the action with respect to $g_{\xi\eta}$, the field
equations become
\begin{equation}\label{2}
f_{\mathcal{R}}\mathcal{R}_{\xi\eta}+(g_{\xi\eta}\Box
-\nabla_{\xi}\nabla_{\eta})f_{\mathcal{R}}-\frac{1}
{2}g_{\xi\eta}f=8\pi\mathrm{T}_{\xi\eta}-f_{\mathrm{T}}
(\Theta_{\xi\eta}+\mathrm{T}_{\xi\eta}),
\end{equation}
where $\Box= \nabla_{\xi}\nabla^{\xi}$, $f\equiv
f(\mathcal{R},\mathrm{T})$, $f_{\mathrm{T}}= \frac{\partial
f}{\partial \mathrm{T}}$, $f_{\mathcal{R}}= \frac{\partial
f}{\partial \mathcal{R}}$ and
\begin{equation}\label{3}
\Theta_{\xi\eta}=g^{\alpha\beta}\frac{\delta{\mathrm{T}
_{\alpha\beta}}}{\delta{g^{\xi\eta}}}, \quad
\mathrm{T}_{\xi\eta}=g_{\xi\eta}\mathcal{L}_{m}
-\frac{\partial\mathcal{L}_{m}}{\partial{g^{\xi\eta}}}.
\end{equation}
To examine the geometry of a star, we consider the interior
spacetime as
\begin{equation}\label{4}
ds^{2}_{-}=e^{\vartheta(r)}dt^{2}-e^{\lambda(r)}dr^{2}-r^{2}
(d\theta^{2}+\sin^{2}\theta d\phi^{2}).
\end{equation}
The matter distribution in the interior region is
\begin{equation}\label{5}
\mathrm{T}_{\xi\eta}=(\rho+P)u_{\xi}u_{\eta}-Pg_{\xi\eta},
\end{equation}
where $\rho$ and $P$ define the energy density and isotropic
pressure, respectively. Manipulating Eq.(\ref{3}), we have
\begin{equation}\label{8}
\Theta_{\xi\eta}=-2\mathrm{T}_{\xi\eta}-Pg_{\xi\eta}.
\end{equation}

In order to solve the field equations, we take a specific model of
this theory given as $f(\mathcal{R},\mathrm{T})=\mathcal{R}+2\mu
\mathrm{T}$ and analyze the impact of curvature-matter coupling on
anisotropic polytropes, where $\mu$ is a coupling parameter
\cite{27, 114}. Houndjo and Piattella \cite{91} proposed that this
model, together with a pressureless matter component, could
introduce the effects of holographic dark energy. Moroever, it is
also consistent with the standard conservation of the matter
distribution \cite{93}. Chakraborty \cite{94} showed that minimal
matter-curvature coupling produces structures composed entirely of
two non-interacting matter components (the test particles follow the
geodesic path) where the second fluid is produced from the
interaction between matter and geometry. Further, a linear model in
$\mathcal{R}$ does not involve extra degrees of freedom. This
particular model has also been used extensively to study the
features of different astrophysical objects \cite{24a}-\cite{118}.

The corresponding field equations for this model turn out to be
\begin{eqnarray}\label{9}
&&\frac{1}{r^{2}}+e^{-\lambda}\left(\frac{\lambda'}{r}
-\frac{1}{r^{2}}\right)=8\pi\rho+\mu\left(3\rho-P\right),
\\\label{10}&&
\frac{-1}{r^{2}}+e^{-\lambda}\left(\frac{\vartheta'}{r}
+\frac{1}{r^{2}}\right)=8\pi P+\mu\left(3P-\rho\right),
\\\label{11}&&
\frac{e^{-\lambda}}{4}\left(2\vartheta''+\vartheta'^{2}
-\lambda'\vartheta'+2\frac{(\vartheta'-\lambda')}{r} \right)=8\pi
P+\mu\left(3P-\rho\right).
\end{eqnarray}
Here prime is the derivative corresponding to $r$. As we have
considered a linear model in $f(\mathcal{R},\mathrm{T})$ gravity
therefore, only the matter sector of Einstein field equations is
modified. Thus, it might be possible to obtain similar results in GR
by applying a modified EoS on the effective terms related to the
matter sector. However, it is not an easy task to find the exact EoS
that reproduces the effects of a linear $f(\mathcal{R},\mathrm{T})$
model in GR. This theory is non-conserved that yields the existence
of an additional force given by
\begin{equation}\nonumber
\nabla^{\xi}\mathrm{T}_{\xi\eta}=\frac{f_{\mathrm{T}}}
{8\pi-f_{\mathrm{T}}}\left((\mathrm{T}_{\xi\eta}
+\Theta_{\xi\eta})\nabla^{\xi}\ln
f_{\mathrm{T}}+\nabla^{\xi}\Theta_{\xi\eta}-\frac{1}{2}
g_{\xi\eta}\nabla^{\xi}\mathrm{T}\right).
\end{equation}
For the considered model, the above equation takes the form
\begin{equation}\label{12}
\nabla^{\xi}\mathrm{T}_{\xi\eta}=\frac{2\mu}{8\pi-2\mu}
\left(\nabla^{\xi}\Theta_{\xi\eta}-\frac{1}{2}g_{\xi\eta}
\nabla^{\xi}\mathrm{T}\right),
\end{equation}
which provides the TOV equation given by
\begin{eqnarray}\label{13}
P'+\frac{\vartheta'}{2}\left(\rho+P\right)
+\frac{\mu}{8\pi+2\mu}\left(P'-\rho'\right)=0.
\end{eqnarray}
Using Misner and Sharp \cite{51} mass function for spherically
symmetric object, we obtain
\begin{equation}\label{14}
e^{-\lambda}=1-\frac{2m(r)}{r}.
\end{equation}
Substituting this value in Eqs.(\ref{9}), it follows that
\begin{equation}\label{15}
m'=4\pi\rho r^{2}+\frac{\mu}{2}(3\rho-P)r^{2}
\end{equation}
Using Eq.(\ref{14}) in Eqs.(\ref{10}) and (\ref{13}), we have
\begin{equation}\label{16}
P'=-\left(\rho+P\right)\frac{\left[4\pi P
r+\frac{m}{r^2}-\frac{\mu}{2}(\rho-3P)r\right]}{\left
(1-\frac{2m}{r}\right)\left[1+\frac{\mu}{2(4\pi+\mu)}
(1-\frac{\rho'}{P'})\right]}.
\end{equation}
It is mentioned here that both the above equations reduce to GR when
$\mu=0$.

The smooth joining of interior and exterior geometries at the
boundary of the stellar structure ensures a finite matter
distribution within a confined radius. In GR, the exterior vacuum of
an uncharged static sphere is described by the well-known
Schwarzschild spacetime. However, the description of the outer
manifold is not clear in the context of $f(\mathcal{R},\mathrm{T})$
gravity. It is possible that the non-minimal coupling between matter
and geometry modifies the exterior spacetime. Consequently, the
usual junction conditions in GR must be appropriately redefined in
this scenario \cite{122, 123}. In the background of
$f(\mathcal{R},\mathrm{T})=\mathcal{R}+2\mu\mathrm{T}$ model, the
general field equation (\ref{2}) reduces to
\begin{equation*}
G_{\xi\eta}=8\pi\mathrm{T}_{\xi\eta}+\mu\mathrm{T}
g_{\xi\eta}+2\mu(\mathrm{T}_{\xi\eta}+Pg_{\xi\eta}),
\end{equation*}
whose trace is given as
\begin{equation*}
\mathcal{R}=-(8\pi+6\mu)\mathrm{T}-8\pi P.
\end{equation*}
In the absence of matter
($\mathrm{T}_{\xi\eta}=0\Rightarrow\mathrm{T}=0$), the above
equation yields $\mathcal{R}=0$. Thus, for the considered form of
$f(\mathcal{R},\mathrm{T})$, the boundary conditions of GR can be
applied if the exterior of the sphere matches the Schwarzschild
solution given as
\begin{equation}\label{6}
ds^{2}_{+}=\left(1-\frac{2M}{r}\right)dt^{2}-\left(1-\frac{2M}
{r}\right)^{-1}dr^{2}-r^{2}\left(d\theta^{2}+\sin^{2}\theta
d\phi^{2}\right).
\end{equation}
For the smooth matching of two geometries at the surface boundary
$(r=r_{\Sigma})$, we use Darmois matching conditions that yield
\begin{eqnarray}\label{7}
e^{\vartheta_{\Sigma}}=\left(1-\frac{2M}{r_{\Sigma}}\right), \quad
e^{-\lambda_{\Sigma}}=\left(1-\frac{2M}{r_{\Sigma}}\right), \quad
P_{\Sigma}^{eff}=0.
\end{eqnarray}

In subsequent sections, we employ polytropic EoS,
$P=K\rho_{o}^\gamma=K \rho^{1+\frac{1}{n}}$ ($K$, $\gamma$ and $n$
represent the polytropic constant, polytropic exponent and
polytropic index, respectively), to generate isotropic as well as
anisotropic stellar models. Polytropes were the first models used to
represent stars. They yield useful information regarding the
structure and mechanism of stars and provide relation between total
mass (M) and total radius (R). In order to explore the mass-radius
relation of isotropic spheres, we have used the polytropic EoS to
numerically solve Eqs.(\ref{15}) and (\ref{16}) with $n=1$ and the
initial conditions $m(0)=0$ and $\rho(0)=\rho_c$. Figure \textbf{1}
shows that total mass (in terms of solar mass $M_{\bigodot}$) of the
spherical structure is less than the mass of GR counterpart when
$\mu>0$ and vice-versa. In physically relevant stellar models,
pressure is dependent on density as well as temperature. However, a
pre-defined relation between density and temperature simplifies
complicated scenarios. In these special scenarios, the polytropic
EoS provides a suitable relation between pressure and density.
\begin{figure}\center
\epsfig{file=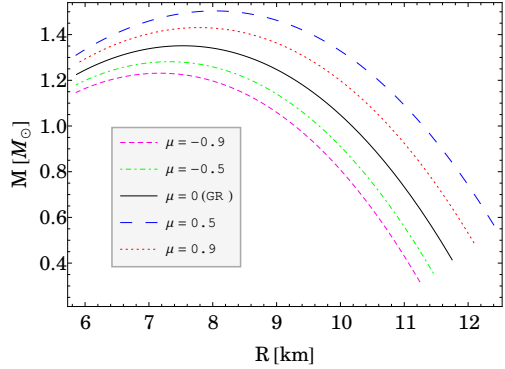,width=0.48\linewidth} \caption{Mass-radius
relation for isotropic sphere with $n=1$.}
\end{figure}

\section{Isotropic Polytropes}

Here, we study isotropic polytropes through two polytropic EoS
\cite{28} and develop LEE for both mass (baryonic) density
$(\rho_{o})$ and energy density $(\rho)$ cases.

\subsection*{Case I}

In this case, we take the polytropic EoS as
\begin{equation}\label{17}
P=K\rho_{o}^\gamma=K \rho_{o}^{1+\frac{1}{n}}.
\end{equation}
The correlation between mass (baryonic) density and energy density
is expressed as
\begin{equation}\nonumber
\rho=\rho_{o}+nP.
\end{equation}
To develop the LEE, we define dimensionless variables as
\begin{eqnarray}\nonumber
&&\alpha=P_{c}/\rho_{c},\quad A^2=4\pi\rho_{c}/\alpha(n+1), \quad
\psi_{o}^n=\rho_{o}/\rho_{oc}, \\\label{18} &&r=\zeta/A,\quad
\upsilon(\zeta)=m(r)A^3/(4\pi\rho_{c}),
\end{eqnarray}
where $c$ means that the corresponding term is evaluated at the
center. Inserting these values in TOV equation, it follows that
\begin{eqnarray}\nonumber
&&\frac{a\zeta^2}{\alpha(n+1)}\left[\frac{L(n(1-n\alpha)
+(n^2-1)\alpha\psi_{o})\psi_{o}^{-1}-\alpha(n+1)}{(1-n\alpha)+
\left( n+1 \right) \alpha\,\psi_{o} }\right]
\frac{d\psi_{o}}{d\zeta}\\\label{19}&&-\zeta^{3}\psi_{o}^{n}
\frac{\mu}{8\pi}((3-n)\alpha\psi_{o}-(1-n\alpha))
-\upsilon-\alpha\zeta^{3}\psi_{o}^{n+1}=0,
\end{eqnarray}
where $L=\frac{\mu}{8\pi+2\mu}$ and
$a=1-\frac{2(n+1)\alpha\upsilon}{\zeta}$. Substituting the
dimensionless variables in Eq.(\ref{15}), we obtain
\begin{equation}\label{21}
\frac{d\upsilon}{d\zeta}=\zeta^{2}\psi_{o}^{n}((1-n\alpha
+n\alpha\,\psi_{o})+\frac{1}{8\pi}\, {\mu\,
\left(3(1-n\alpha)+\alpha\,\psi_{o} \left( 3\,n-1 \right) \right)}).
\end{equation}
Now we combine Eqs.(\ref{19}) and (\ref{21}), it follows that
\begin{eqnarray}\nonumber
&&a\{L(n(1-n\alpha)+(n^{2}-1)\alpha\psi_{o})\psi_{o}^{-1}-\alpha(n+1)\}
\frac{d^{2}\psi_{o}}{d\zeta^{2}}\\\nonumber&&
+\frac{2}{\zeta}\frac{d\psi_{o}}{d\zeta}(a\{L(n(1-n\alpha)+(n^{2}-1)\alpha\psi_{o})\psi_{o}^{-1}
-\alpha(n+1)\}\\\nonumber&&
+\frac{1}{2}a(L(n^{2}-1)\alpha\psi_{o}^{-1}
-L(n(1-n\alpha)+(n^{2}-1)\alpha\psi_{o})\psi_{o}^{-2})\zeta\frac{d\psi_{o}}
{d\zeta}\\\nonumber&&
-\zeta\alpha(n+1)(\zeta(\{L(n(1-n\alpha)+(n^{2}-1)
\alpha\psi_{o})\psi_{o}^{-1}-\alpha(n+1)\})\\\nonumber&&
\times(1-n\alpha+n\alpha\psi_{o}
+\frac{1}{8\pi}\mu(3-3n\alpha+\alpha\psi_{o}(3n-1))\psi^{n}\\\nonumber&&
-\frac{\upsilon}{\zeta^{2}}\{L(n(1-n\alpha)
+(n^{2}-1)\alpha\psi_{o})\psi_{o}^{-1}-\alpha(n+1)\}+\frac{1}{2}\alpha\zeta(n+1)
b\psi_{o}^{n}\\\nonumber&&
+\frac{1}{2b}a\{L(n(1-n\alpha)+(n^{2}-1)\alpha\psi_{o})\psi_{o}^{-1}-\alpha(n+1)\}\frac{d\psi_{o}}{d\zeta}
\\\nonumber&&
+\frac{1}{16\pi}nb\mu\zeta(n((3-n)\alpha\psi_{o}+n\alpha-1)
+(3-n)\alpha\psi_o)\psi_{o}^{n-1})))\\\nonumber&&
-(3\alpha\psi_{o}+1-n\alpha+n\alpha\psi_{o}+\frac{1}
{8\pi}\mu(3-3n\alpha
+\alpha\psi_{o}(3n-1))b\alpha(n+1)\psi_{o}^{n}\\\label{22}&&
-\frac{3}{8\pi}b\mu\alpha(n+1)((3-n)\alpha\psi_{o}+n\alpha-1)\psi_{o}^{n}=0,
\end{eqnarray}
where $b=1-n\alpha+(n+1)\alpha\psi_{o}$. This is the LEE that
describes the polytropic stars in the hydrostatic equilibrium. It is
obvious that the above equation reduces to Eq.(34) in {\cite{13}
when $\mu=0$.
\begin{figure}\center
\epsfig{file=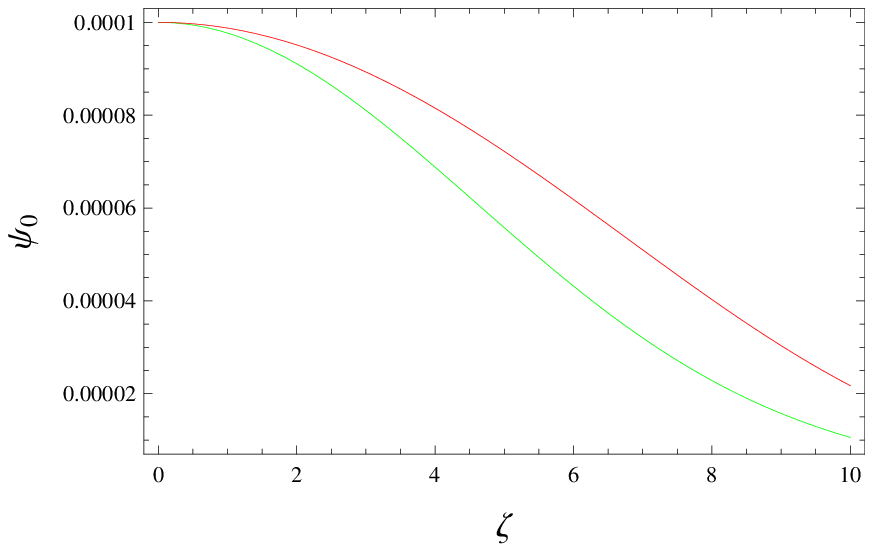,width=0.48\linewidth}\epsfig{file=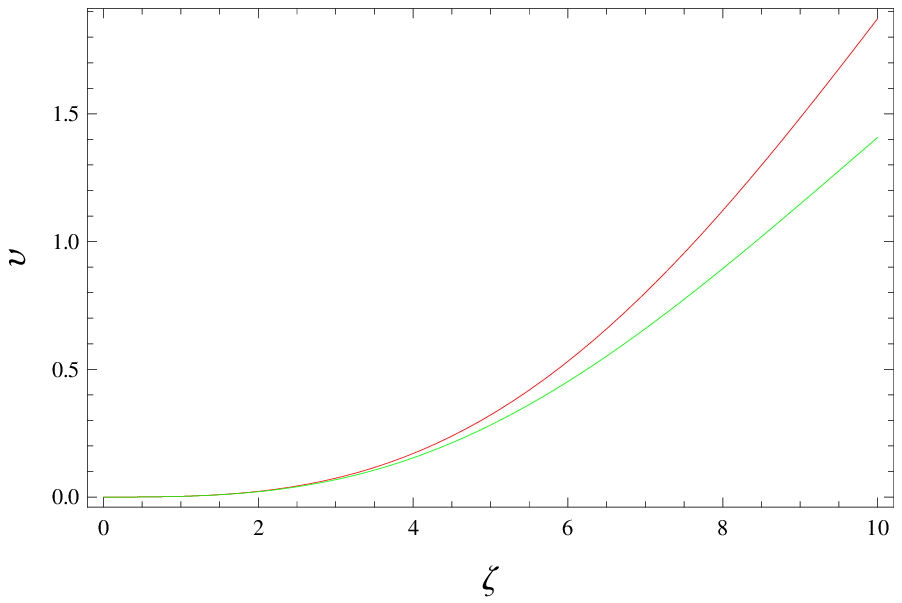,width=0.48\linewidth}
\caption{Plots of $\psi_{o}$ and $\upsilon$ corresponding to case
\textbf{I} for $\mu=-0.9$, $n=0.5$ and $\alpha=0.1$ (red), $0.2$
(green).}
\end{figure}

Since Eqs.(\ref{19}) and (\ref{21}) are complicated, so, it is
difficult to find exact solution of these equations. Thus, we find
numerical solution of these equations by considering different
values of the involved parameters. Figure \textbf{2} describes that
the value of $\psi_{o}$ is maximum at the center and decreases as
the radius increases. This provides a viable behavior of $\psi_{o}$
as it is positive inside the star and decreases as moves away from
the center of the star. The graphical behavior of total mass is
represented in Figure \textbf{2}. This shows that the total mass has
an increasing trend and is minimum at the center which represents
compactness of the polytropic stars.

\subsection*{Case II}

Here, we take the second polytropic EoS as
\begin{equation}\label{23}
P=\emph{K}\rho^\gamma=\emph{K}\rho^{1+\frac{1}{n}},
\end{equation}
where $\rho=\frac{\rho_{o}}{(1-\emph{K}\rho_{o}^\frac{1}{n})^{n}}$
and define another dimensionless variable as $\psi^n=\rho/\rho_{c}$
for this case. Kaisavelu et al. \cite{121} generated superdense
stellar structures obeying a polytropic EoS and also discussed their
stability. The TOV equation and Eq.(\ref{15}) turn out to be
\begin{eqnarray}\nonumber
&&a\zeta^2\frac{d\psi}{d\zeta}
\left[\frac{L(n-\alpha(n+1)\psi)\psi^{-1}-\alpha(n+1)}{\alpha(n+1)
(1+\alpha\psi)}\right]-\zeta^{3}\psi^{n}\frac{\mu}{8\pi}(3\alpha\psi-1)
\\\label{24}&&-\upsilon-\alpha\zeta^{3}\psi^{n+1}=0,
\\\label{25}&&
\frac{d\upsilon}{d\zeta}=\zeta^{2}\psi^{n}(1+\frac{1}{8\pi}\mu
(3-\alpha\psi)).
\end{eqnarray}
The corresponding LEE becomes
\begin{eqnarray}\nonumber
&&a(L(n-\alpha(n+1)\psi)\psi^{-1}-\alpha(n+1))\frac{d^{2}\psi}
{d\zeta^{2}}+\frac{2}{\zeta}\frac{d\psi}{d\zeta}(a\{(L(n\\\nonumber&&-\alpha
(n+1)\psi)\psi^{-1}-\alpha(n+1)\})-\frac{1}{2}a\zeta
L\{(n-\alpha(n+1)\psi)\psi^{-2}\\\nonumber&&+\alpha(n+1)\psi^{-1}\}\frac{d\psi}
{d\zeta}-\frac{1}{2c}a\zeta\alpha\{L(n-\alpha(n+1)
\psi)\psi^{-1}+\alpha(n+1)\}\frac{d\psi}{d\zeta}\\\nonumber&&+\frac{1}{\zeta}
\alpha(n+1)\upsilon\{L(n-\alpha(n+1)\psi)
\psi^{-1}-\alpha(n+1)\}-\frac{1}{2}\alpha^{2}(n+1)^{2}c\zeta^{2}
\psi^{n}\\\nonumber&&-\zeta^{2}\alpha(n+1)\{L(n-\alpha(n+1)\psi)
\psi^{-1}-\alpha(n+1)\}\{1+\frac{1}{8\pi}\mu(3-\alpha
\psi)\}\psi^{n}\\\nonumber&&-\frac{1}{16\pi}\zeta
c\mu\alpha(n+1)(n(3\alpha\psi-1)
+3\zeta\alpha\psi)\psi^{n-1})-3c(n+1)\alpha^{2}\psi^{n+1}
\\\label{26}&&-c\alpha(n+1)\{1+\frac{1}{8\pi}
\mu(3-\alpha\psi)\}\psi^{n}-\frac{3}{8\pi}
c\mu\alpha(n+1)(3\alpha\psi-1)=0,
\end{eqnarray}
where $c=1+\alpha\psi$. This equation reduces to Eq.(42) in
{\cite{13} when $\mu=0$. The numerical solution of Eqs.(\ref{24})
and (\ref{25}) is given in Figure \textbf{3}. Figure \textbf{3}
shows that the behavior of $\psi$ is physically viable as it is
maximum at the center and decreases towards the surface boundary. It
also represents that the total mass has increasing behavior and is
minimum at the center. Equations (\ref{22}) and (\ref{26}) are two
LEE that correspond to the mass (baryonic) density and total energy
density, respectively. These are the nonlinear ordinary differential
equations which describe the internal configuration of
self-gravitating polytrophic stars and help to understand different
astrophysical phenomena.
\begin{figure}\center
\epsfig{file=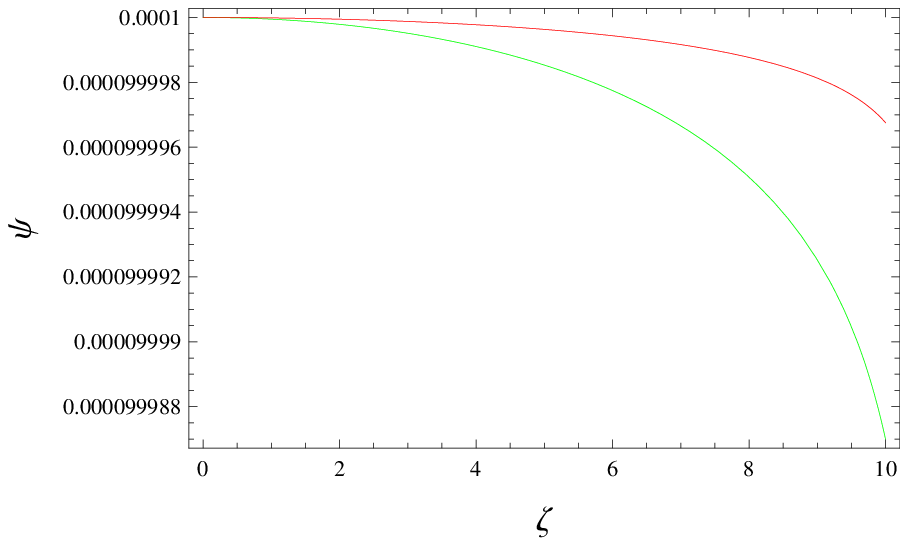,width=0.5\linewidth}\epsfig{file=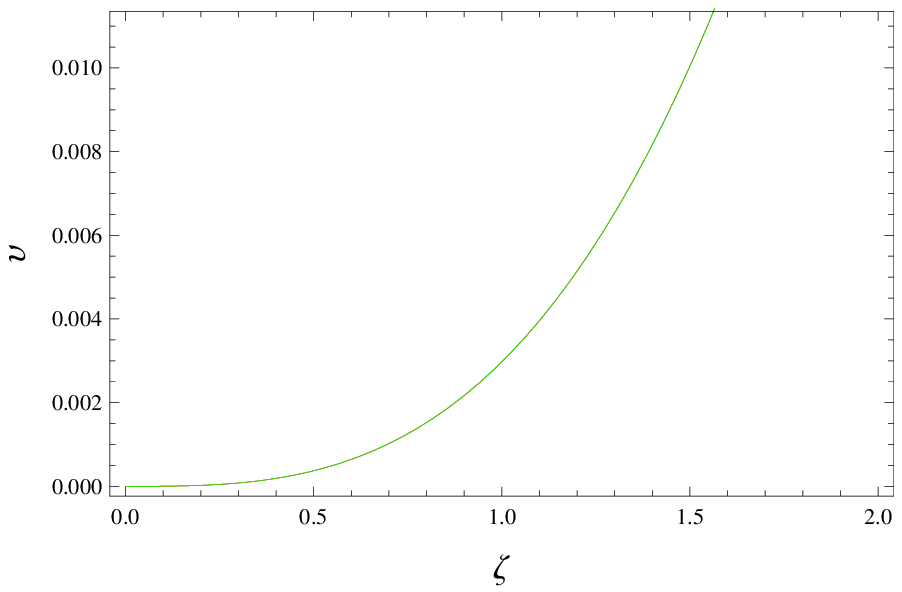,width=0.46\linewidth}
\caption{Plots of $\psi$ and $\upsilon$ versus $\zeta$ corresponding
to case \textbf{II} for $\mu=-0.9$, $n=0.5$ and $\alpha=0.1$ (red),
0.2 (green).}
\end{figure}

\subsection{Physical Features of Isotropic Polytropes}

A physically acceptable stellar object must have monotonically
decreasing density and pressure away from the center. It is shown in
Figure \textbf{4} that the density as well as pressure of both
isotropic models are positive and finite with a decreasing trend
towards the boundary. Moreover, it is necessary that the
energy-momentum tensor (representing normal matter distribution
inside the compact structure) is consistent with the following
energy conditions
\begin{eqnarray*}
&&\text{null energy condition:}\quad\rho+P\geq0,\quad\rho+P\geq0,\\
&&\text{weak energy condition:}\quad\rho\geq0,\quad\rho+P\geq0,\\
&&\text{strong energy condition:}\quad\rho+3P\geq0,\\
&&\text{dominant energy condition:}\quad\rho-P\geq0,\\
&&\text{trace energy condition:}\quad\rho-3P\geq0
\end{eqnarray*}
Figure \textbf{4} indicates that first three bounds are fulfilled
for both isotropic models. The models are viable as the dominant and
trace energy conditions are also satisfied as shown in Figure
\textbf{5}.
\begin{figure}\center
\epsfig{file=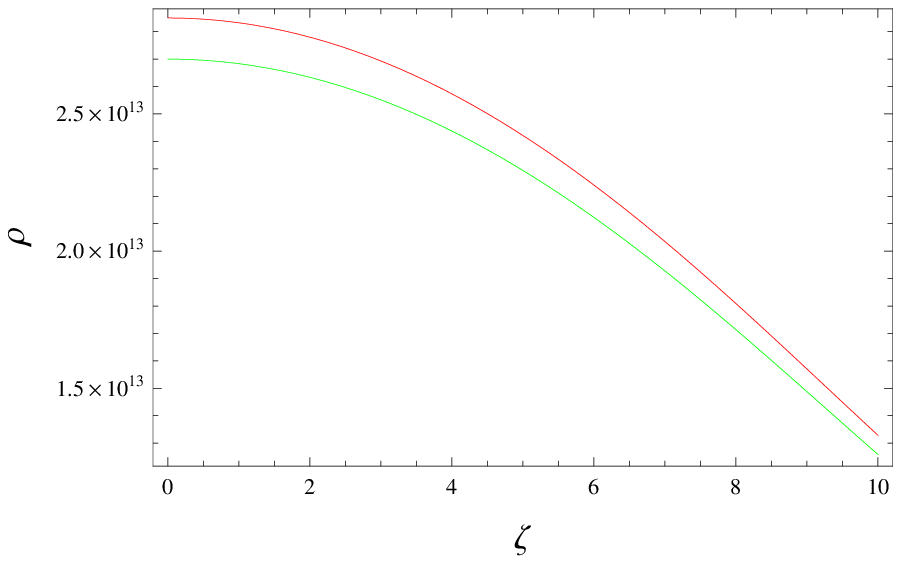,width=0.45\linewidth}\epsfig{file=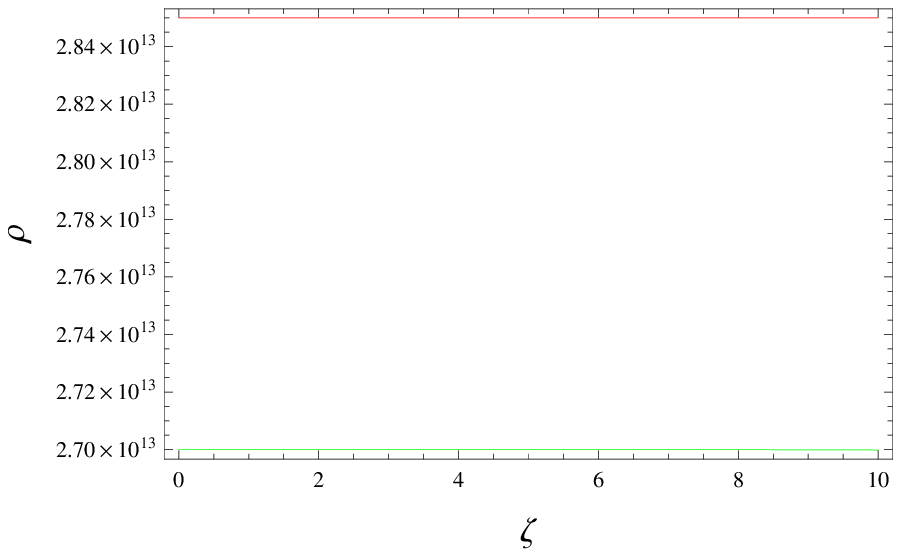,width=0.45\linewidth}
\epsfig{file=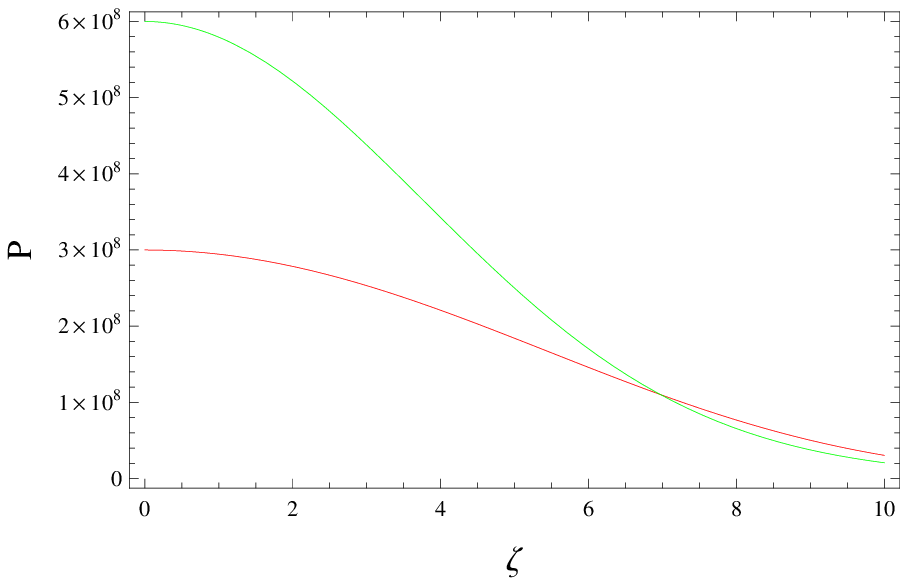,width=0.45\linewidth}\epsfig{file=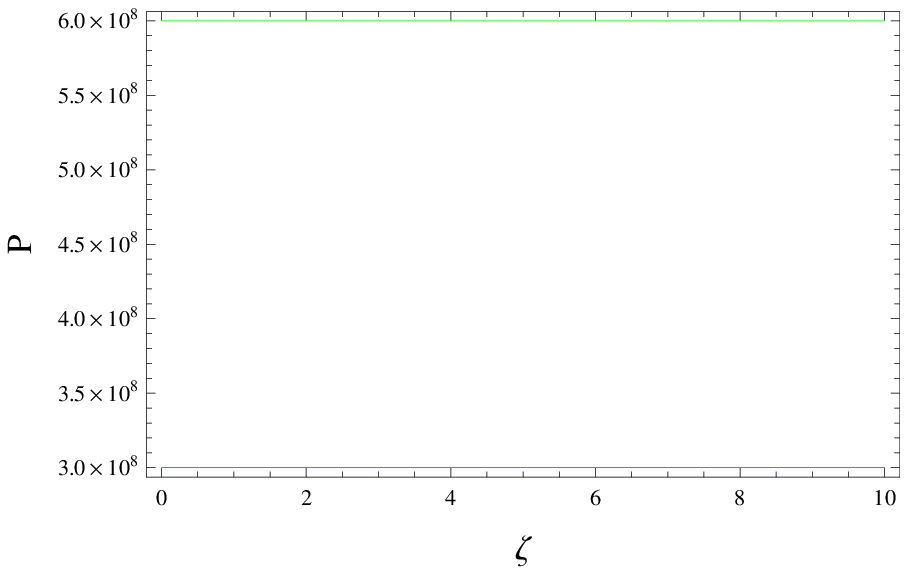,width=0.45\linewidth}
\caption{Plots of $\rho$ and $P$ versus $\zeta$ for $\mu=-0.9$,
$n=0.5$ and $\alpha=0.1$ (red), 0.2 (green) for cases \textbf{I}
(left) and \textbf{II} (right).}
\end{figure}
\begin{figure}\center
\epsfig{file=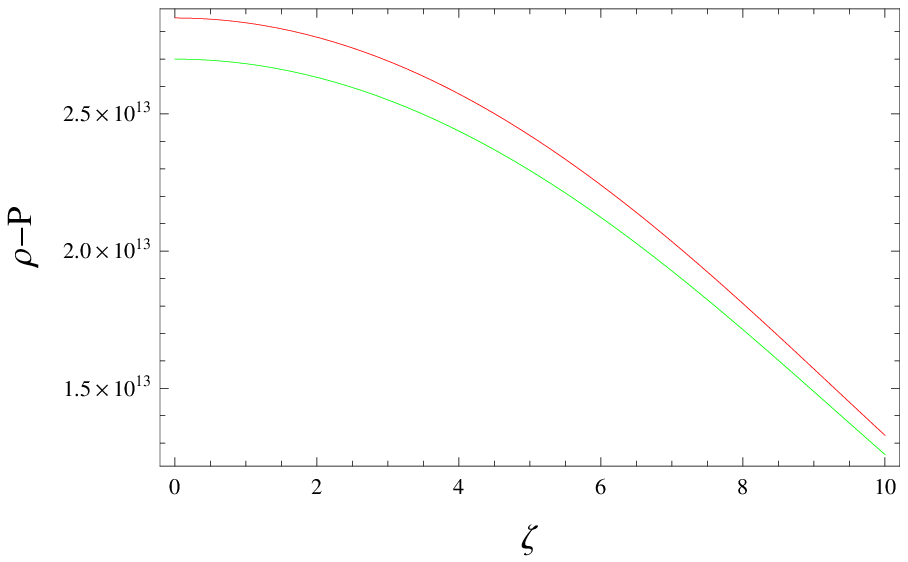,width=0.45\linewidth}\epsfig{file=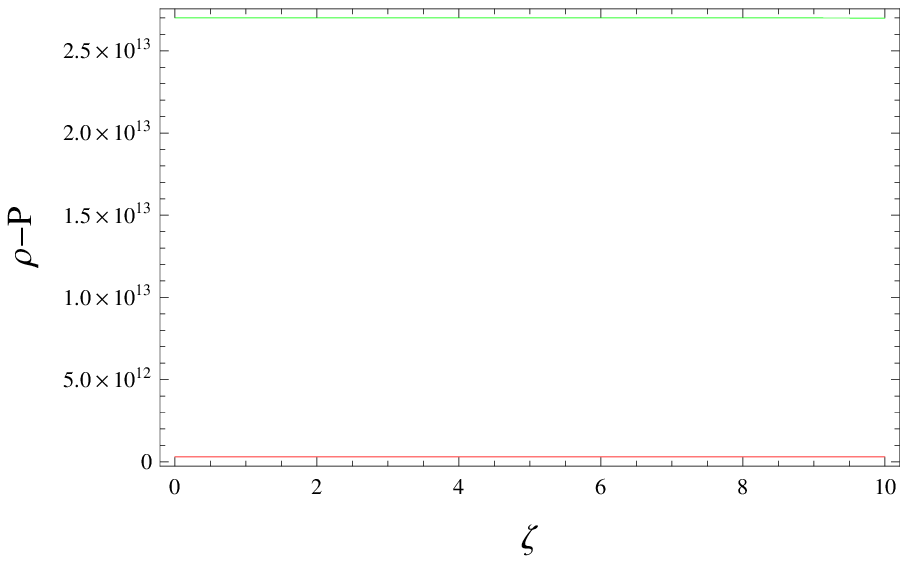,width=0.45\linewidth}
\epsfig{file=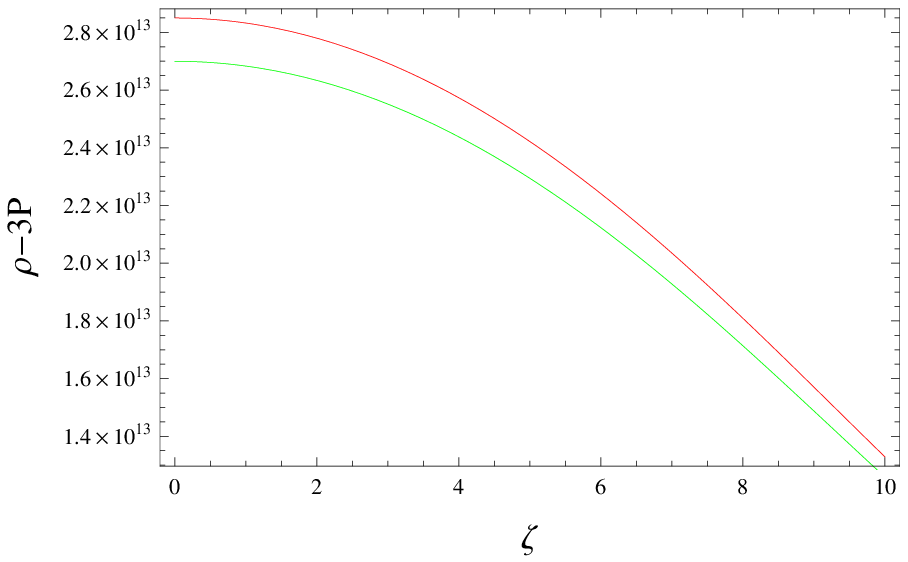,width=0.45\linewidth}\epsfig{file=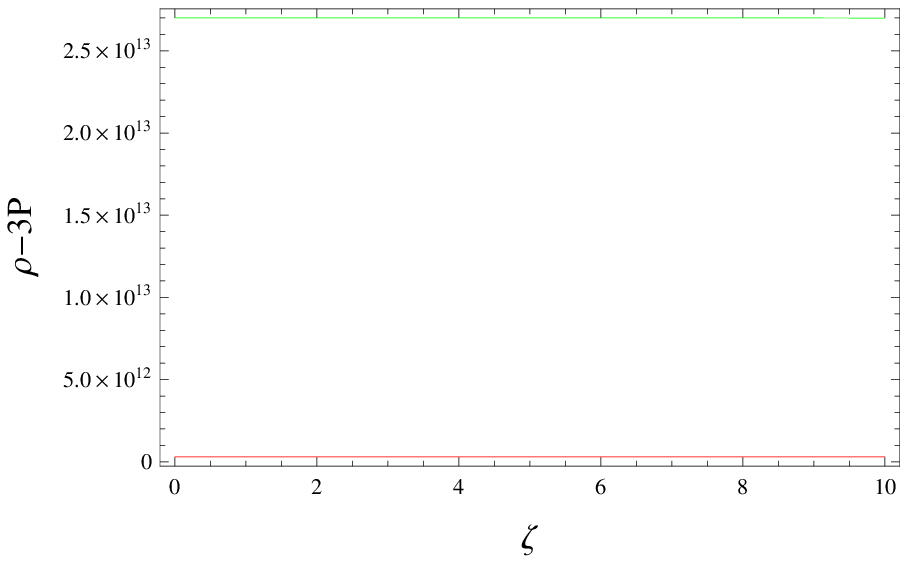,width=0.45\linewidth}
\caption{Plots of $\rho-P$ and $\rho-3P$ versus $\zeta$ for
$\mu=-0.9$, $n=0.5$ and $\alpha=0.1$ (red), 0.2 (green) for cases
\textbf{I} (left) and \textbf{II} (right).}
\end{figure}
\begin{figure}\center
\epsfig{file=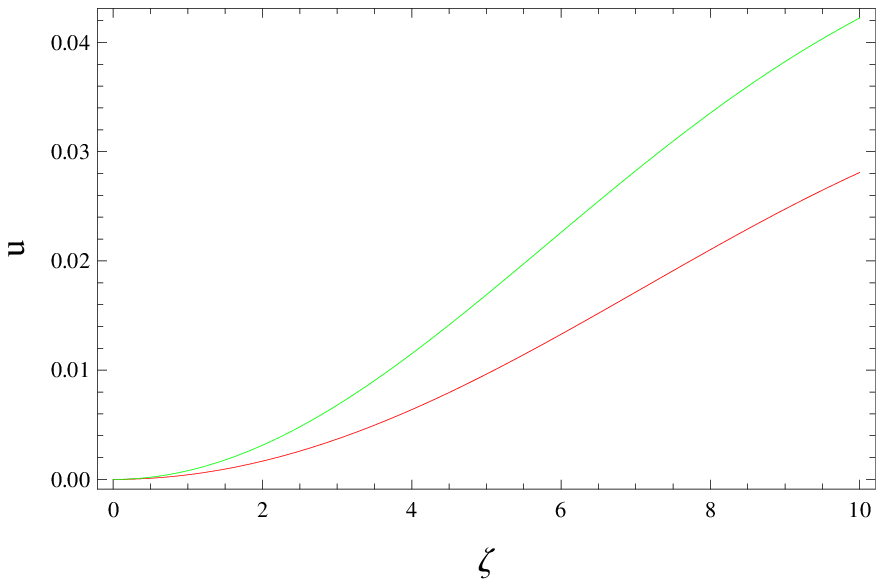,width=0.45\linewidth}\epsfig{file=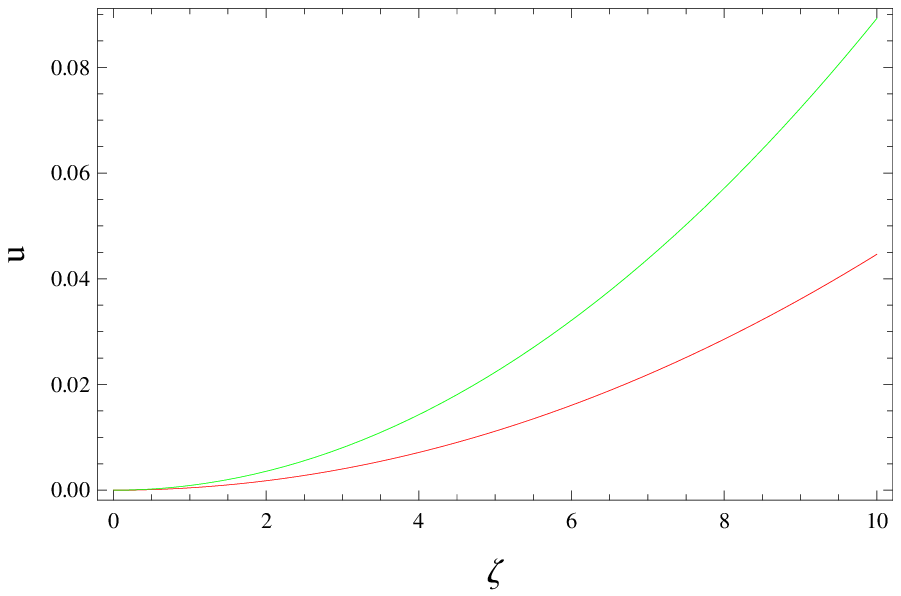,width=0.45\linewidth}
\epsfig{file=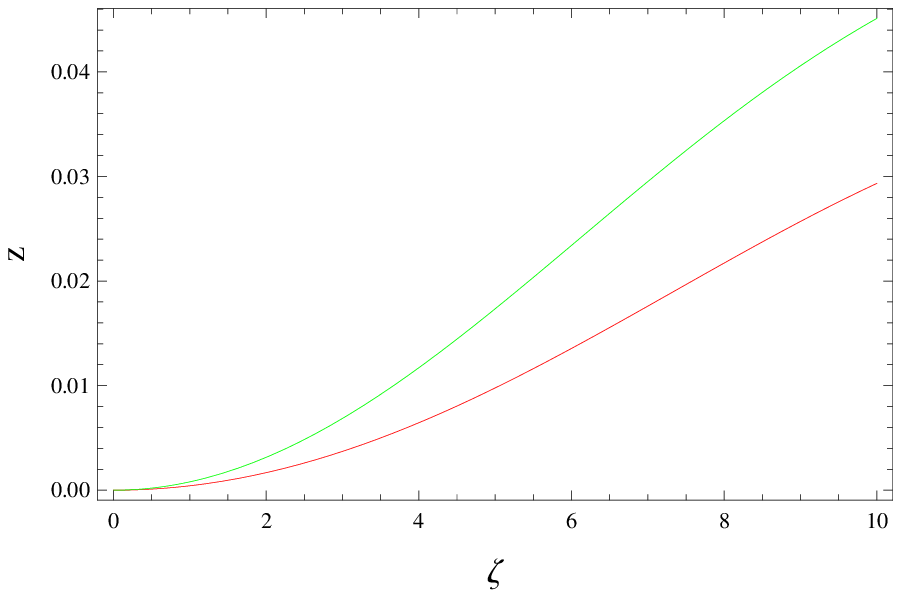,width=0.45\linewidth}\epsfig{file=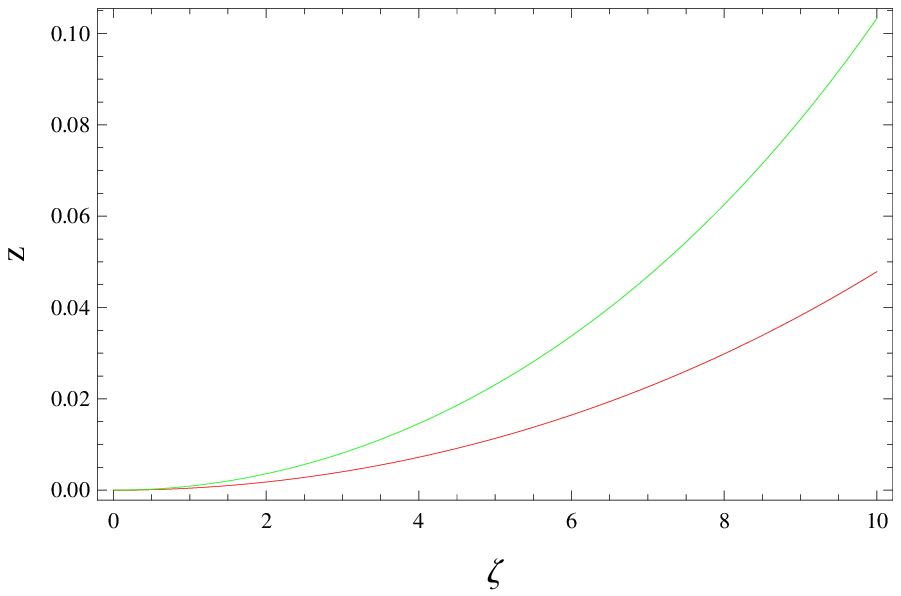,width=0.45\linewidth}
\caption{Plots of $u$ and $Z$ versus $\zeta$ for $\mu=-0.9$, $n=0.5$
and $\alpha=0.1$ (red), 0.2 (green) for cases \textbf{I} (left) and
\textbf{II} (right).}
\end{figure}

The ratio of mass to radius provides a metric for the compactness of
a self-gravitating system. This ratio, known as compactness factor
$(u)$, must be less than $4/9$ throughout the interior of the sphere
\cite{28a}. The compactness factor is further employed to calculate
the possible redshift produced in an electromagnetic wave due to the
cosmic object's gravitational field. The redshift measured as
\begin{equation*}
Z(r)=\frac{1-\sqrt{1-2u}}{\sqrt{1-2u}},
\end{equation*}
lies below 2 for isotropic structures \cite{28a}. Moreover,
compactness of a structure increases if more matter is packed within
the same radius. On the other hand, gravitational redshift measures
the effect of the spherical object's gravitational field on
electromagnetic waves. Consequently, an increase in the compactness
of the spherical stellar structure will lead to a stronger
gravitational field, i.e., the redshift of electromagnetic waves
will be higher. The compactness as well as redshift parameters
related to both models (plotted in Figure \textbf{6}) obey the
required limits.

A fluid distribution is stable if a sound wave travels at a speed
($\nu^2=\frac{dP}{d\rho}$) less than that of light, i.e., the
condition of causality does not fail at any point within the medium.
Plots of sound speed in Figure \textbf{7} imply that the constructed
models are stable. Further, the adiabatic index
($\Gamma=\frac{P+\rho}{P}\frac{dP}{d\rho}$) gauges the stiffness of
the configuration. A stiff system (corresponds to
$\Gamma>\frac{4}{3}$ \cite{29}) is difficult to compress as a slight
increase in density significantly enhances the outward pressure. The
scenarios under consideration are stiff as they correspond to
$\Gamma=3$ (refer to Figure \textbf{7}).
\begin{figure}\center
\epsfig{file=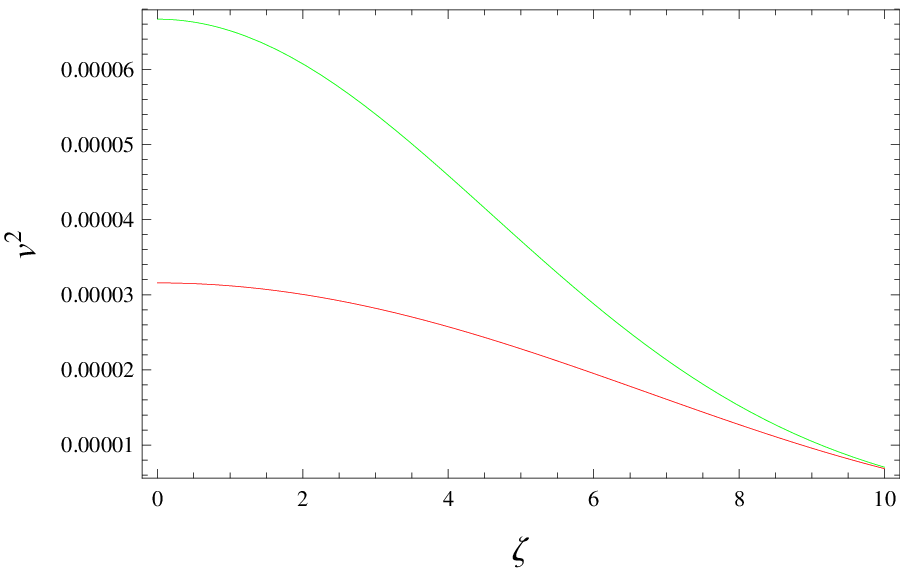,width=0.45\linewidth}\epsfig{file=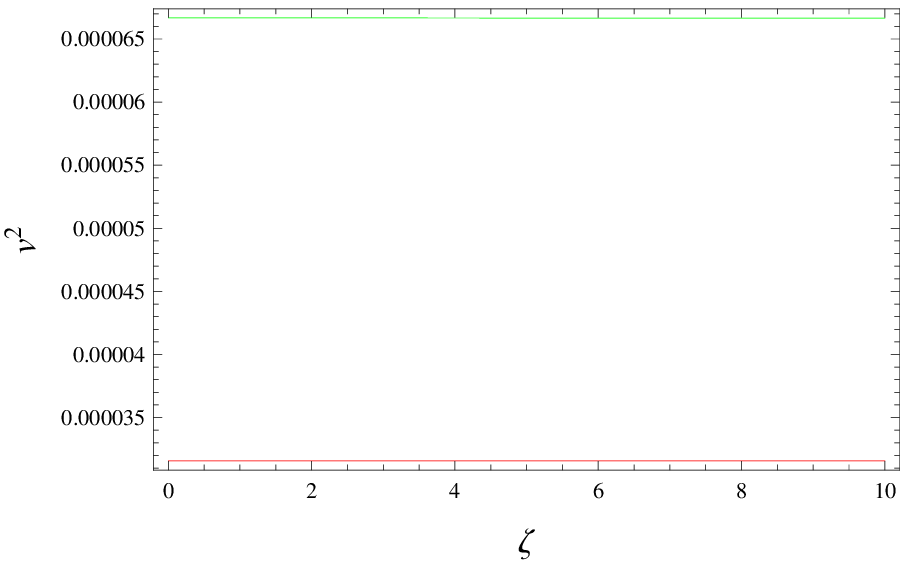,width=0.45\linewidth}
\epsfig{file=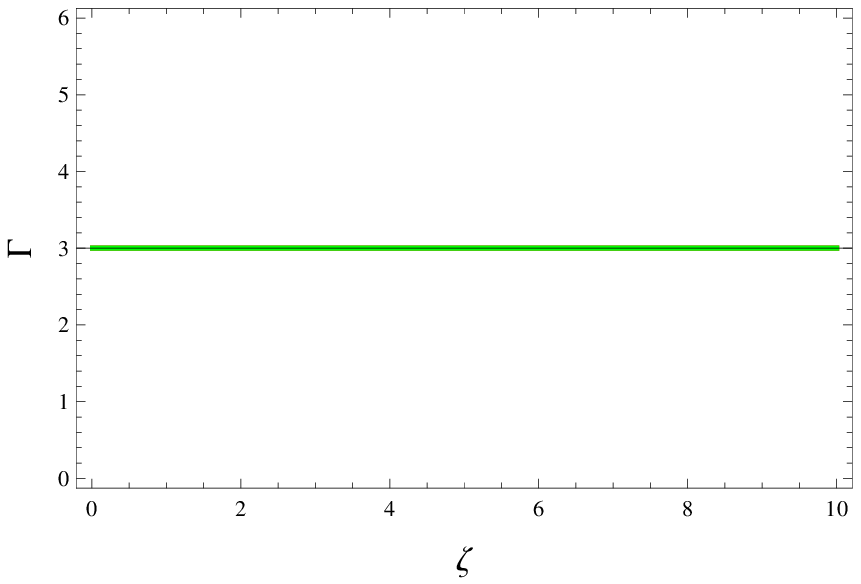,width=0.45\linewidth}\epsfig{file=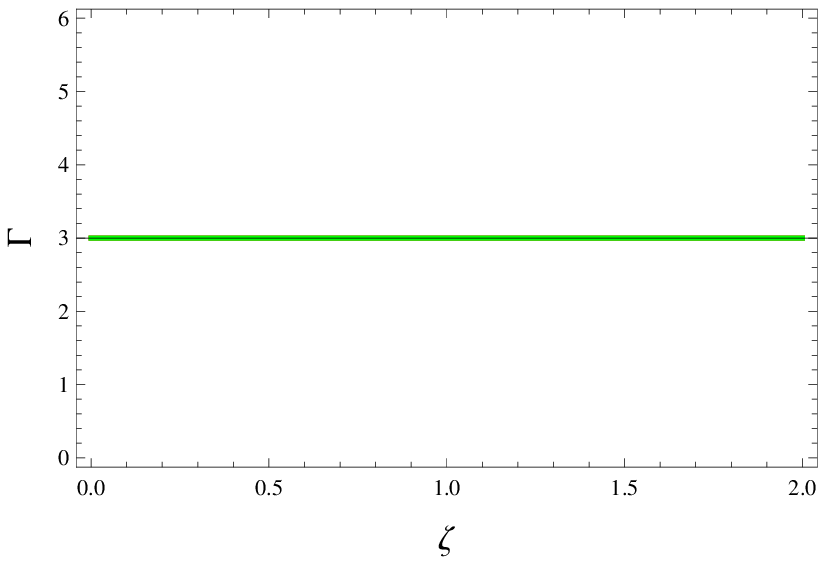,width=0.45\linewidth}
\caption{Plots of $\nu^2$ and $\Gamma$ versus $\zeta$ for
$\mu=-0.9$, $n=0.5$ and $\alpha=0.1$ (red), 0.2 (green) for cases
\textbf{I} (left) and \textbf{II} (right).}
\end{figure}

\section{Anisotropic Polytropes}

Anisotropy plays an important role in various dynamical phases of
stellar evolution. Mostly stellar models are anisotropic in nature.
The effect of anisotropy appears when the radial stress differs from
the tangential stress. The anisotropic matter distribution is
\begin{equation}\label{27}
\mathrm{T}_{\xi\eta}=(\rho+P_{\perp})u_{\xi}u_{\eta}
-P_{\perp}g_{\xi\eta}+(P_{r}-P_{\perp})s_{\xi}s_{\eta},
\end{equation}
where $s_{\xi}$  is the four-vector, $P_r$ and $P_{\perp}$ are
radial and tangential pressures, respectively. The corresponding
field equations and TOV equation are
\begin{eqnarray}\label{28}
&&\frac{1}{r^{2}}+e^{-\lambda}(\frac{\lambda'}{r}-\frac{1}
{r^{2}})=8\pi\rho+\mu(\rho-P_{r}-2P_{\perp}), \\\label{29}&&
\frac{-1}{r^{2}}+e^{-\lambda}(\frac{\vartheta'}{r}+\frac{1}
{r^{2}})=8\pi P_{r}+\mu(\rho+3P_{r}+2P_{\perp}),
\\\label{30}&&
\frac{e^{-\lambda}}{4}(2\vartheta''+\vartheta'^{2}-
\lambda'\vartheta'+2\frac{(\vartheta'-\lambda')}{r})=8\pi
P_{\perp}+\mu(\rho+P_{r}+4P_{\perp}), \\\label{31}
&&P'_{r}+\frac{\vartheta'}{2}(\rho+P_{r})-
\frac{2\Delta}{r}+\frac{\mu}{8\pi+2\mu}(\rho'+P_{r}'
+2P_{\perp}')=0,
\end{eqnarray}
where $\Delta=-\Pi=P_{\perp}-P_{r}$. Heintzmann and Hillebrandt
\cite{29} established a relation between radial and tangential
pressures as
\begin{equation}\label{32}
P_{\perp}=(1+q)P_{r}, \quad q=\frac{3\epsilon}{2(\epsilon-1)}, \quad
\epsilon<1.
\end{equation}
This relation is used to examine the effects of anisotropic stress
on self-gravitating objects. Using this relation in the above
equations, we have
\begin{eqnarray}\label{33}
&&\frac{1}{r^{2}}+e^{-\lambda}(\frac{\lambda'}{r}-\frac{1}{r^{2}})=8\pi
\rho+\mu(\rho-c_1P_{r}), \\\label{34}&&
\frac{-1}{r^{2}}+e^{-\lambda}(\frac{\vartheta'}{r}+\frac{1}{r^{2}})=8\pi
P_{r}+\mu(\rho+c_2P_{r}), \\\label{35}&&
\frac{e^{-\lambda}}{4}(2\vartheta''+\vartheta'^{2}-\lambda'\vartheta'
+2\frac{(\vartheta'-\lambda')}{r})=8\pi
P_{\perp}+\mu(\rho+c_3P_{r}),
\\\label{36}
&&P'_{r}+\frac{\vartheta'}{2}(\rho+P_{r})-\frac{2\Delta}{r}
+\frac{\mu}{8\pi+2\mu}(\rho'+c_1P_{r}')=0,
\end{eqnarray}
where $c_1=3+2q$, $c_2=5+2q$ and $c_3=5+4q$. Using Eqs.(\ref{14})
and (\ref{33}), we have
\begin{equation}\label{37}
m'=4\pi\rho r^{2}+\frac{\mu}{2}(\rho-c_1P_{r})r^{2}.
\end{equation}

In the following, we use two polytropic EoS in terms of radial
pressure defined in Eqs.(\ref{17}) and (\ref{23}) as
\begin{eqnarray}\label{38}
P_{r}=\emph{K}\rho_{o}^\gamma=\emph{K}\rho_{o}^{1+\frac{1}{n}},
\quad P_{r}=\emph{K}\rho^\gamma=\emph{K}\rho^{1+\frac{1}{n}}.
\end{eqnarray}
We also plot the mass-radius curves for different values of $\mu$ by
employing the relation $P_{r}=\emph{K}\rho^{2}$ along side
Eqs.(\ref{37}) and (\ref{38}) (refer to Figure \textbf{8}). We
observe that mass of the stellar structure approaches to that of GR
model as $\mu\rightarrow0$.
\begin{figure}\center
\epsfig{file=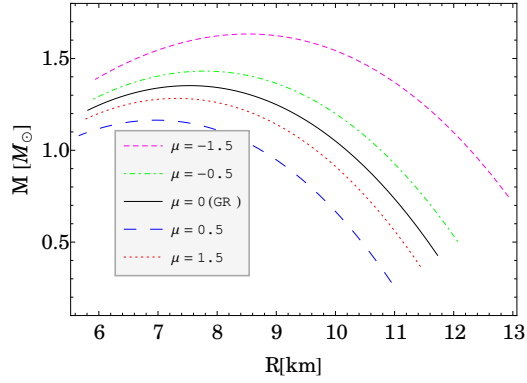,width=0.5\linewidth} \caption{Mass-radius
relation for anisotropic configuration with $n=2$.}
\end{figure}

\subsection*{Case I}

We formulate LEE using the same procedure as in the isotropic case.
The TOV equation in terms of dimensionless parameters becomes
\begin{eqnarray}\nonumber
&&a\zeta^{2}\frac{d\psi_{0}}{d\zeta}\bigg[P_{rc}(n+1)+\frac{P_{rc}}{\alpha}
\psi_{0}^{-1}L(n(\alpha\psi_{0}(n+1)-(1-n\alpha))+(n+1)
\alpha\psi_{0}c_1)\bigg]
\\\nonumber&&\times
[P_{rc}(n+1)((n+1)\alpha\psi_{0}+(1-n\alpha)) ]^{-1}
-\frac{2\Delta\psi_{0}^{-n}\zeta}{P_{rc}(n+1)}\frac{a}{b}
+\upsilon+\alpha\zeta^{3}\psi_{0}^{n+1}
\\\label{39}
&&+\frac{\mu}{8\pi}\zeta^{3}
\psi_{0}^{n}((1-n\alpha)+(n+c_2)\alpha\psi_{0})=0,
\end{eqnarray}
where $\alpha=P_{rc}/\rho_c$. The conservation of mass in terms of
dimensionless variables takes the form
\begin{eqnarray}\label{40}
\frac{d\upsilon}{d\zeta}=\zeta^{2}\psi_{o}^n\left(((1-n\alpha)
+n\alpha\psi_{o})+\frac{\mu}{8\pi}((1-n\alpha)+
(n-c_1)\alpha\psi_{o})\right).
\end{eqnarray}
Manipulating Eqs.(\ref{39}) and (\ref{40}), it follows that
\begin{eqnarray}\nonumber
&&a\left[1+\frac{L(n(1-n\alpha+(n+1)\alpha\psi_{o})+\alpha\psi_{o}
c_1(n+1))}{\psi_{o}\alpha(n+1)}\right]\frac{d^{2}\psi_{o}}{d\zeta^{2}}
+\frac{1}{P_{rc}(n+1)\zeta}
\\\nonumber
&&\times\frac{d\psi_{o}}{d\zeta}\bigg[a\left(P_{rc}(n+1)+\frac{P_{rc}L(n(1-n\alpha
+(n+1)\alpha\psi_{o})+\alpha\psi_{o}
c_1(n+1))}{\psi_{o}\alpha}\right)
\\\nonumber&&
+aP_{rc}(n+1)+2\bigg[P_{rc}(n+1)+P_{rc}L(n(1-n\alpha+(n+1)\alpha\psi_{o})
\\\nonumber&& +\alpha\psi_{o}c_1(n+1))(\psi_{o}\alpha)^{-1}
\bigg]\alpha\upsilon(n+1)\zeta^{-1}+\frac{aP_{rc}}{\alpha\psi_{o}}L
\bigg[n(1-n\alpha \\\nonumber&& +(n+1)\alpha\psi_{o})+\alpha\psi_{o}
c_1(n+1)\bigg]-a\zeta P_{rc}L\bigg[n(1-n\alpha
\\\nonumber&&+(n+1)\alpha\psi_{o})+\alpha\psi_{o}
c_1(n+1)\bigg](\alpha\psi_{o}^{2})^{-1}\frac{d\psi_{o}}{d\zeta}-2\zeta^{2}(n+1)\alpha
\bigg[1-n\alpha
\\\nonumber&&+n\alpha\psi_{o}+\frac{1}{8\pi}\mu(1-n\alpha+n\alpha
\psi_{o}-c_1\alpha\psi_{o})\bigg]\bigg(P_{rc}(n+1)
\\\nonumber&&+\frac{P_{rc}L(n(1-n\alpha
+(n+1)\alpha\psi_{o})+\alpha\psi_{o}
c_1(n+1))}{\psi_{o}\alpha}\bigg)\psi_{o}^{-n}-a\alpha\zeta(n+1)\frac{1}{b}
\\\nonumber&&\times\left(P_{rc}(n+1)+\frac{P_{rc}L(n(1-n\alpha+(n+1)\alpha\psi_{o})
+\alpha\psi_{o}c_1(n+1))}{\psi_{o}\alpha}\right)\frac{d\psi_{o}}{d\zeta}
\\\nonumber&&+b\zeta\alpha(n+1)^{2}\alpha P_{rc}\psi_{o}^{n}
+\frac{LP_{rc}(n+1)(n+c_1)}{\psi_{o}}
\frac{d\psi_{o}}{d\zeta}\bigg]-\frac{2\psi_{o}^{-n}}{\alpha
P_{rc}\zeta(n+1)}
\\\nonumber&&\times\left(a\frac{d\Delta}{d\zeta}
-\frac{na\Delta} {\psi_{o}}\frac{d\psi_{o}}{d\zeta}
+\frac{2\alpha\upsilon(n+1)\Delta}{\zeta^{2}}
+\frac{a\Delta}{\zeta}-\frac{a\alpha(n+1)\Delta}
{b}\frac{d\psi_{o}}{d\zeta} \right.\\\nonumber&&\left.
-2\zeta\alpha\psi_{o}^{n}\Delta\left[1-n\alpha+n\alpha\psi_{o} +
\frac{1}{8\pi}\mu(1-n\alpha+n\alpha\psi_{o}-c_1\alpha\psi_{o})\right]
(n+1)\right)
\\\nonumber&&+b\psi_{o}^{n}\left(3\alpha\psi_{o}
+\left[1-n\alpha+n\alpha\psi_{o}
+\frac{1}{8\pi}\mu(1-n\alpha+n\alpha\psi_{o}-c_1\alpha\psi_{o})\right]
\right)
\\\nonumber&&+\frac{1}{8\pi}\frac{d\psi_{o}}{d\zeta}b\mu\zeta\psi_{o}^{n-1}
\bigg(n(1-n\alpha+(n+c_2)\alpha\psi_{o})+\alpha\psi_{o}(n+c_2)\bigg)
\\\label{41}&&+\frac{3}{8\pi}
b\mu\bigg(1-n\alpha+(n+c_2)\alpha\psi_{o}\bigg)\psi_{o}^{n}=0.
\end{eqnarray}
This is the LEE equation which helps to comprehend the internal
structure of the anisotropic polytropic stars. It is interesting to
mention here that the above equation reduces to GR equation (58) in
\cite{13} when the model parameter ($\mu$) vanishes.

\subsection*{Case II}

In this case, the TOV equation reduces to
\begin{eqnarray}\nonumber
&&a\zeta^{2}\frac{d\psi}{d\zeta}\left[\frac{P_{rc}(n+1)+
\frac{P_{rc}}{\alpha}\psi^{-1}L(n+(n+1)\alpha\psi c_1)}
{P_{rc}(n+1)(1+\alpha\psi)}\right]+\alpha\zeta^{3}\psi^{n+1}
\\\label{42}&&-\frac{2a\Delta\psi^{-n}\zeta}{P_{rc}(n+1)(1+\alpha\psi)}
+\upsilon+\frac{\mu}{8\pi}(1+\alpha c_2\psi)\zeta^{3}\psi^{n}=0.
\end{eqnarray}
Using dimensionless variables in Eq.(\ref{37}), we have
\begin{eqnarray}\label{43}
\frac{d\upsilon}{d\zeta}=\zeta^{2}\psi^n
\left[1+\frac{\mu}{8\pi}(1-\alpha c_1\psi)\right].
\end{eqnarray}
The resulting LEE turns out to be
\begin{eqnarray}\nonumber
&&a\left[1+\frac{L(n+\alpha\psi
c_1(n+1))}{\psi\alpha(n+1)}\right]\frac{d^{2}\psi}
{d\zeta^{2}}-\frac{2\psi^{-n}}{\zeta
P_{rc}(n+1)}\left[a\frac{d\Delta}{d\zeta}
-\frac{na\Delta}{\psi}\frac{d\psi}{d\zeta}+\frac{a\Delta}{\zeta}
\right.\\\nonumber&&\left.+
\frac{2(n+1)\alpha\upsilon\Delta}{\zeta^{2}}
-\frac{a\alpha\Delta}{c} \frac{d\psi}{d\zeta}
-2\Delta\zeta\alpha(n+1)\left(1+\frac{1}{8\pi}
{\mu(1-c_1\alpha\psi)}\right)\psi^{n}\right]
\\\nonumber&&
+\frac{1}{\zeta P_{rc}(n+1)}\frac{d\psi}{d\zeta}\bigg[a (P_{rc}(n+1)
+\frac{P_{rc}L(n+\alpha\psi c_1(n+1))}{\alpha\psi})
\\\nonumber&&+\frac{2}{\zeta}(n+1)\alpha\upsilon\left(P_{rc}(n+1)
+\frac{P_{rc}L(n+\alpha\psi c_1(n+1))}{\alpha\psi}\right)+\alpha
P_{rc}(n+1)
\\\nonumber&&+\frac{a P_{rc}L(n+\alpha\psi
c_1(n+1))}{\alpha\psi}+\frac{\zeta\alpha P_{rc}L
c_1(n+1)}{\psi}\frac{d\psi}{d\zeta}-\frac{\zeta
a\alpha}{c}\bigg[P_{rc}(n+1)
\\\nonumber&&+\frac{P_{rc}L(n+\alpha\psi
c_1(n+1))}{\alpha\psi}\bigg]\frac{d\psi}{d\zeta}-2\alpha\zeta^{2}\psi^{n}(n+1)
\bigg[P_{rc}(n+1)
\\\nonumber&&+\frac{P_{rc}L(n+\alpha\psi
c_1(n+1))}{\alpha\psi}\bigg]\left(1+\frac{1}{8}
\frac{\mu(1-c_1\alpha\psi)}{\pi}\right)+\frac{a
P_{rc}}{\alpha\psi^{2}}\frac{d\psi}{d\zeta}
\\\nonumber&&-c\zeta\alpha
P_{rc}(n+1)^{2}\psi^{n}\bigg]
+c\psi^{n}\left(3\alpha\psi+1+\frac{1}{8}
\frac{\mu(1-c_1\alpha\psi)}{\pi}\right)
\\\label{44}&&+\frac{1}{8\pi}c\mu\zeta\psi^{n-1}
(n+(n+1)\alpha\psi c_2)\frac{d\psi} {d\zeta}+\frac{3}{8\pi}\mu
c\psi^{n}(1+\alpha c_2\psi)=0.
\end{eqnarray}
In the limiting case ($\mu=0$), we can recover Eq.(60) of GR in
\cite{13}.

In order to proceed further with the modeling of a compact object,
we need some additional information based on the particular physical
problem under consideration. For this purpose, we define specific
anisotropy $(\Delta)$ for the modeling of relativistic anisotropic
stars \cite{31,30} which helps to study the effect of anisotropy of
the structure of the compact object.

\section{Modeling of Anisotropic Polytropes}

In this section, we use the method used in \cite{31} to obtain
specific models. This will help to find solutions for anisotropic
matter. This procedure is briefly mentioned as follows. We assume
anisotropic factor as
\begin{equation}\label{45}
\bigtriangleup=Cf(P_{r},r)(\rho+P_{r})r^N,
\end{equation}
where $C$ is the anisotropic parameter. It is mentioned here that
the function $f$ and the number $N$ are specified for each model. We
further assume that
\begin{equation}\label{46}
f(P_{r},r)r^{N-1}=\frac{\vartheta'}{2}.
\end{equation}
Using Eqs.(\ref{45}) and (\ref{46}) in (\ref{37}), we have
\begin{equation}\label{47}
P_{r}'=-\frac{\vartheta'}{2}h(\rho+P_{r})-L(\rho'+P_{r}'c_1),
\end{equation}
where $h=1-2C$. For the sake of convenience, we assume that $h$ is
constant which doest not mean that either pressure is constant. Thus
we obtain the following two equations for the case \textbf{I} as
\begin{eqnarray}\nonumber
&&a\zeta^{2}\frac{d\psi_{0}}{d\zeta}
\left[\frac{\alpha(n+1)+L(n(\alpha\psi_{o}(n+1)+(1-n\alpha))
+\alpha(n+1)c_1\psi_{o})\psi_{o}^{-1}}{\alpha(n+1)((1-n\alpha)+(n+1)\alpha\psi_{o})}\right]
\\\label{48}&&
+h\left[\upsilon+\zeta^3\alpha\psi_{o}^{n+1}+\frac{\mu}{8\pi}
(\alpha\psi_{o}(n+c_2)+(1-n\alpha))\zeta^{3}\psi_{o}^{n}\right]=0,
\\\label{49}
&&\frac{d\upsilon}{d\zeta}=\zeta^{2}\psi_{o}^n\left(((1-n\alpha)
+n\alpha\psi_{o})+\frac{\mu}{8\pi}((1-n\alpha)+
(n-c_1)\alpha\psi_{o})\right).
\end{eqnarray}
The analytic solution of these equations is too difficult to find
due to their complicated nature. Therefore, we solve these equations
numerically by taking different values of the involved parameters.
\begin{figure}\center
\epsfig{file=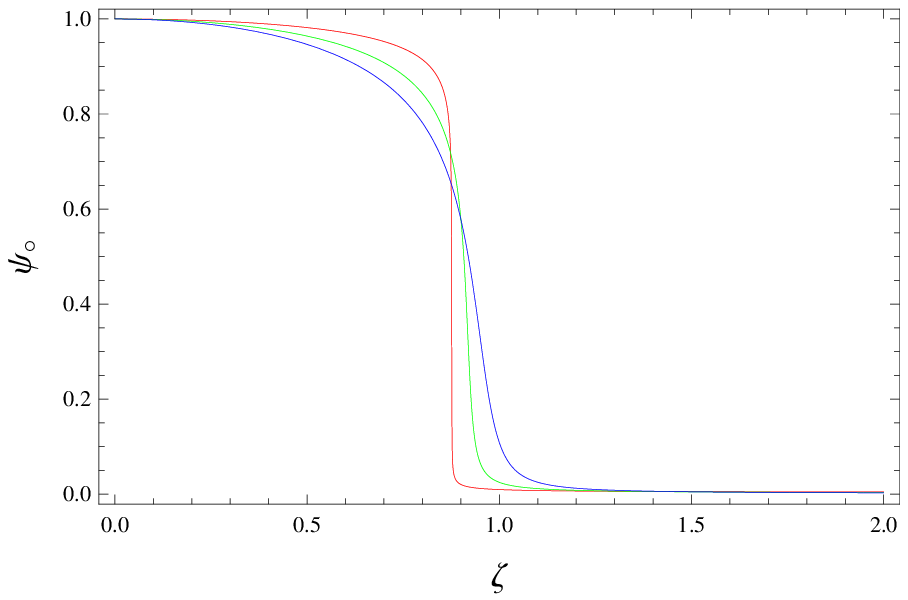,width=0.4\linewidth}\epsfig{file=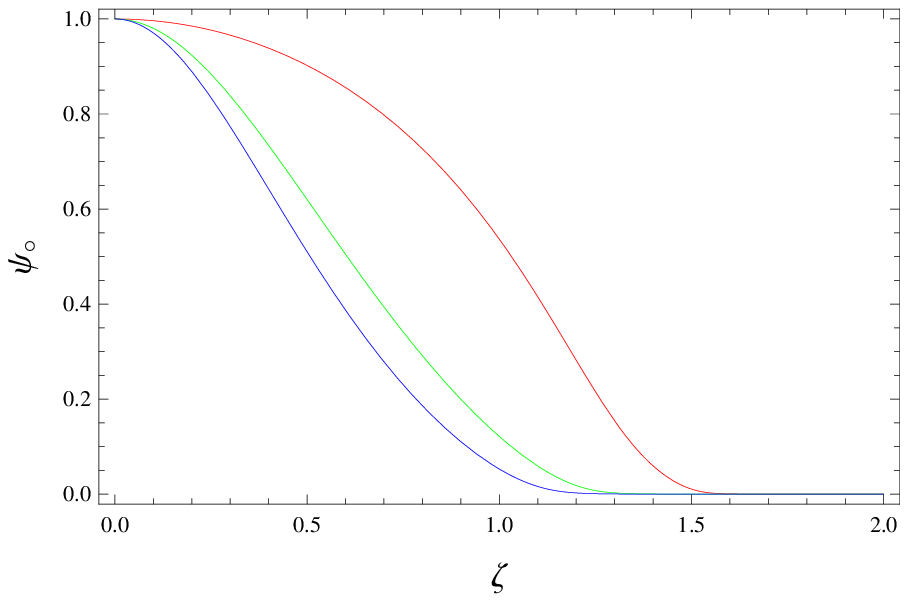,width=0.4\linewidth}
\epsfig{file=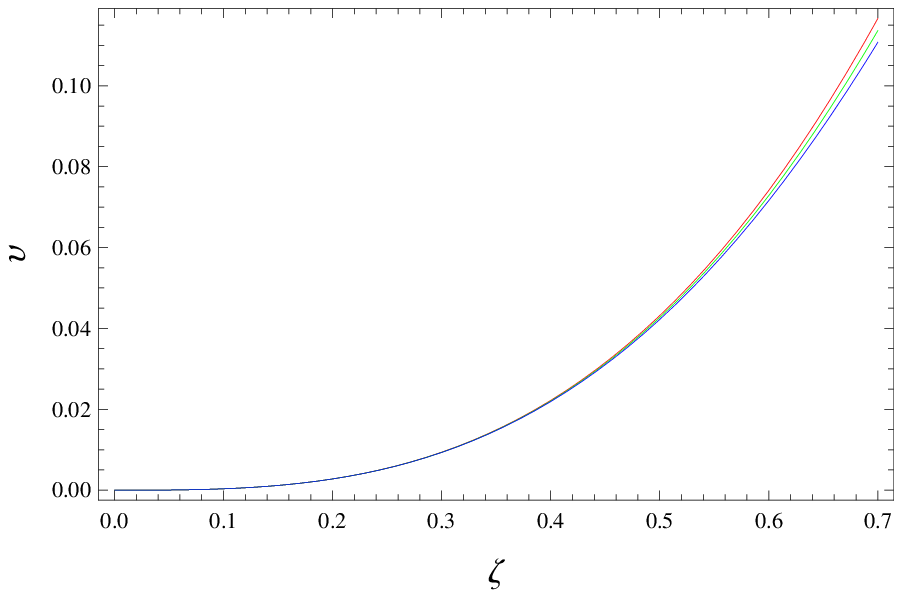,width=0.4\linewidth}\epsfig{file=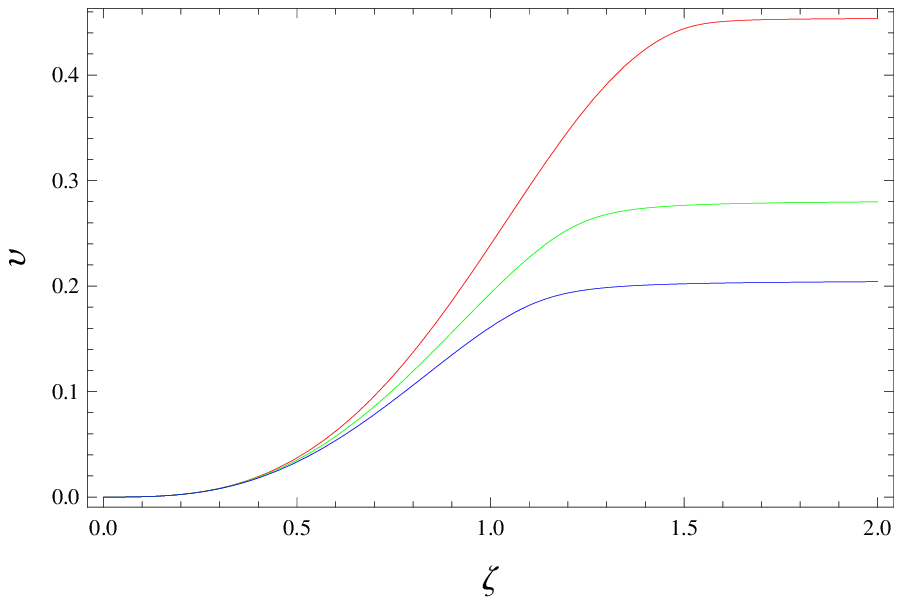,width=0.4\linewidth}
\caption{Plots of $\psi_{o}$ and $\upsilon$ corresponding to
anisotropic case \textbf{I }for $\mu=1.5$, $c_1=2$, $h=0.5$ (red), 1
(green), 1.5 (blue) with $\alpha=0.1$ (left), $n=1$ (left) and
$\alpha=1$ (right), $n=0.5$ (right).}
\end{figure}

Figure \textbf{9} shows that the value of $\psi_{o}$ at the center
is maximum and decreases as the radius increases. Thus there is a
decreasing behavior of the curve for the involved parameters. Also,
we obtain large values of $\psi_{o}$ corresponding to a small value
of $h$. Hence, there is a viable behavior of $\psi_{o}$, i.e., it
must be non-negative inside the star and decreases as we move away
from the center of the star. The graph of total mass for different
values of $h$ is represented in Figure \textbf{9}. We have a larger
value of $\upsilon$ for a smaller value of $h$ and there is no
irregular pattern in both plots. This shows that at the center, the
value of $\upsilon$ is minimum and gradually increases as we move
away from the center. Thus, for small values of the anisotropic
parameter, we have a more compact model. Similarly, for case
\textbf{II}, we have
\begin{eqnarray}\nonumber
&&a\zeta^{2}\frac{d\psi}{d\zeta} \left[\frac{\alpha(n+1)+L(n+\psi
c_1\alpha(n+1))\psi^{-1}}{\alpha(n+1)(1+\alpha\psi)}\right]\\\label{50}&&
+h\left[\upsilon+\zeta^3\alpha\psi^{n+1}+\frac{\mu}{8\pi} (1+\alpha
c_2\psi)\zeta^{3}\psi^{n}\right]=0,\\\label{51}&&
\frac{d\upsilon}{d\zeta}=\zeta^{2}\psi^n
\left[1+\frac{\mu}{8\pi}(1-\alpha c_1\psi)\right].
\end{eqnarray}
The numerical solution of these equations is represented in Figure
\textbf{10}. Figure \textbf{10} shows that the value of $\psi$ is
maximum at the center and decreases towards the surface boundary.
Figure \textbf{10} indicates that we have a more compact model for
small values of anisotropic parameter.
\begin{figure}\center
\epsfig{file=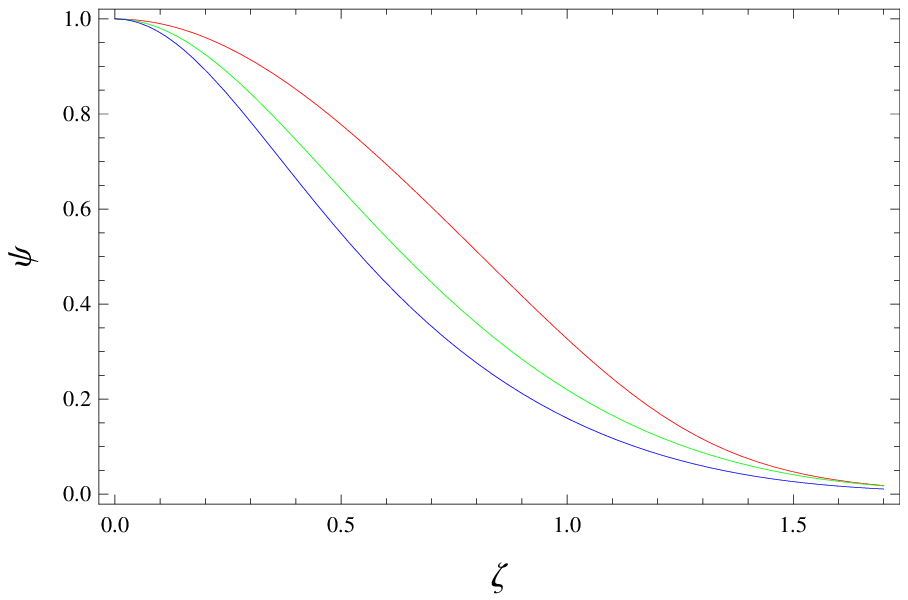,width=0.4\linewidth}\epsfig{file=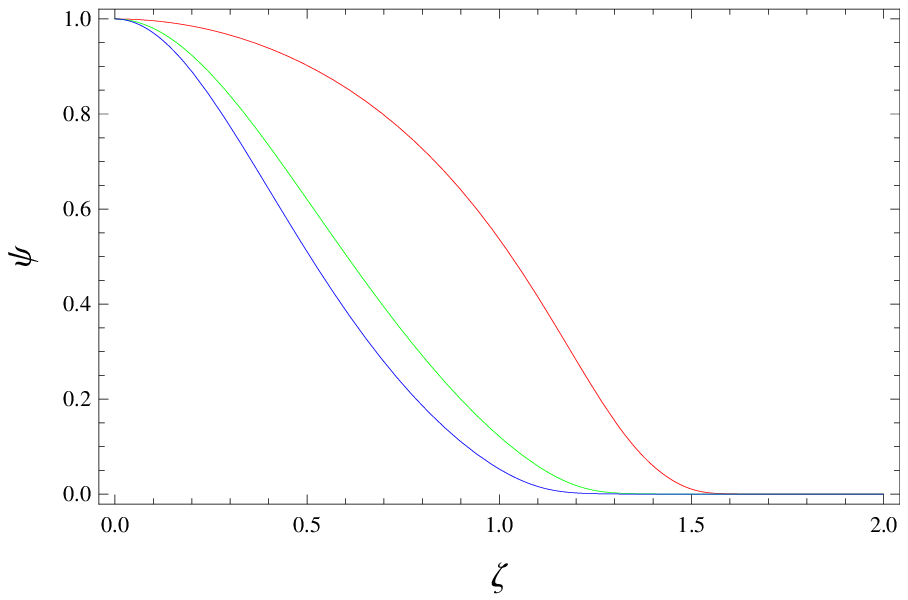,width=0.4\linewidth}
\epsfig{file=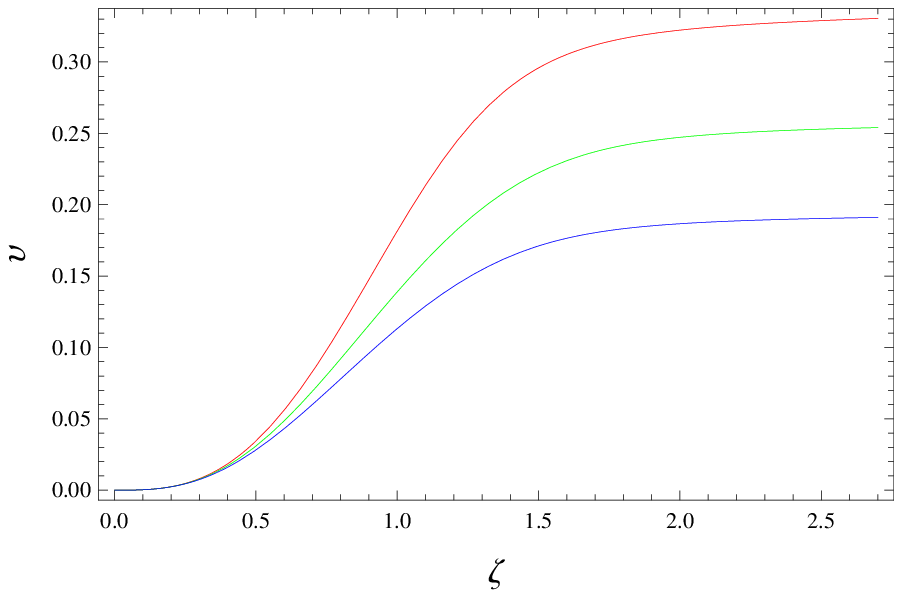,width=0.4\linewidth}\epsfig{file=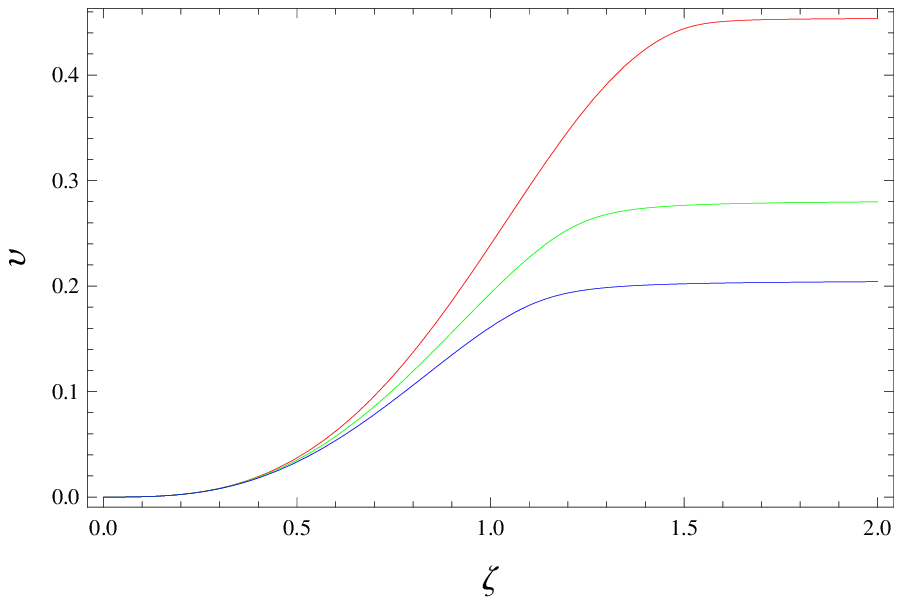,width=0.4\linewidth}
\caption{Plots of $\psi$ and  $\upsilon$ corresponding to
anisotropic case \textbf{II} for $\mu=1.5$, $c_1=2$ and $h=0.5$
(red), 1 (green), 1.5 (blue) with $\alpha=1$ (left), $n=1$ (left)
and $\alpha=0.1$ (right) and $n=2$ (right).}
\end{figure}

\subsection{Physical Features of Anisotropic Polytropes}
\begin{figure}\center
\epsfig{file=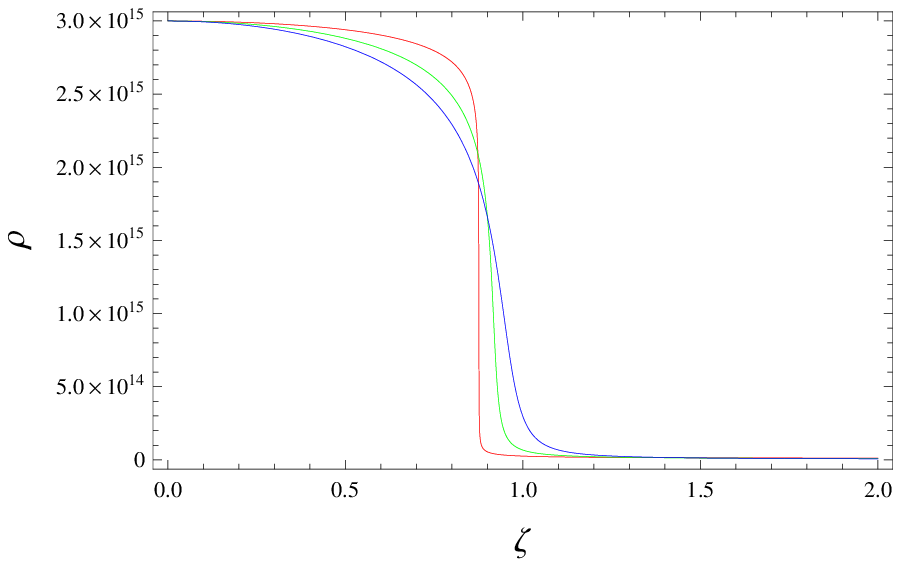,width=0.5\linewidth}\epsfig{file=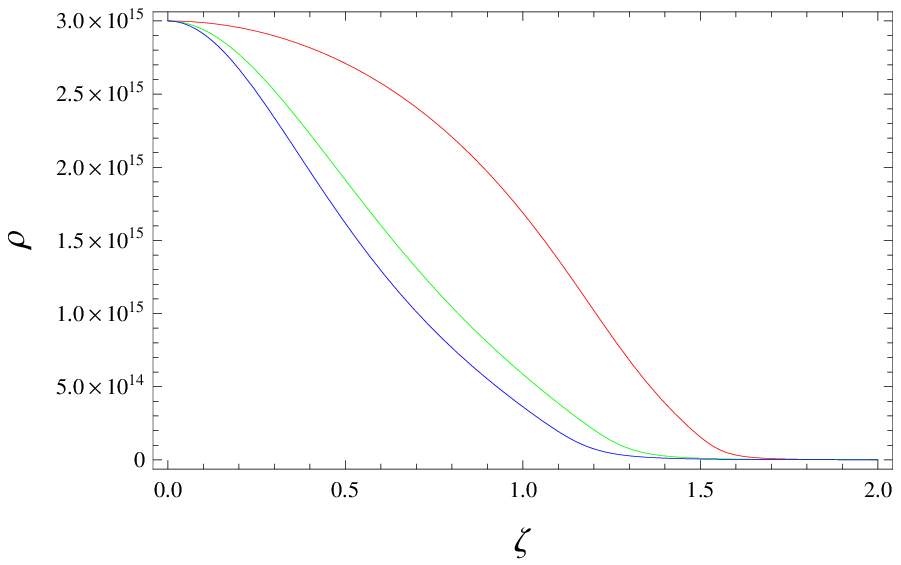,width=0.5\linewidth}
\epsfig{file=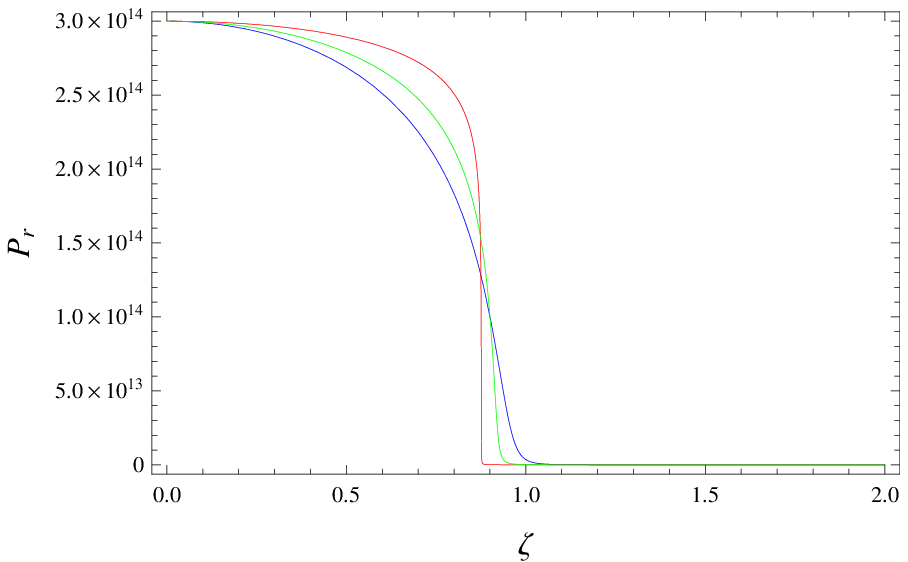,width=0.5\linewidth}\epsfig{file=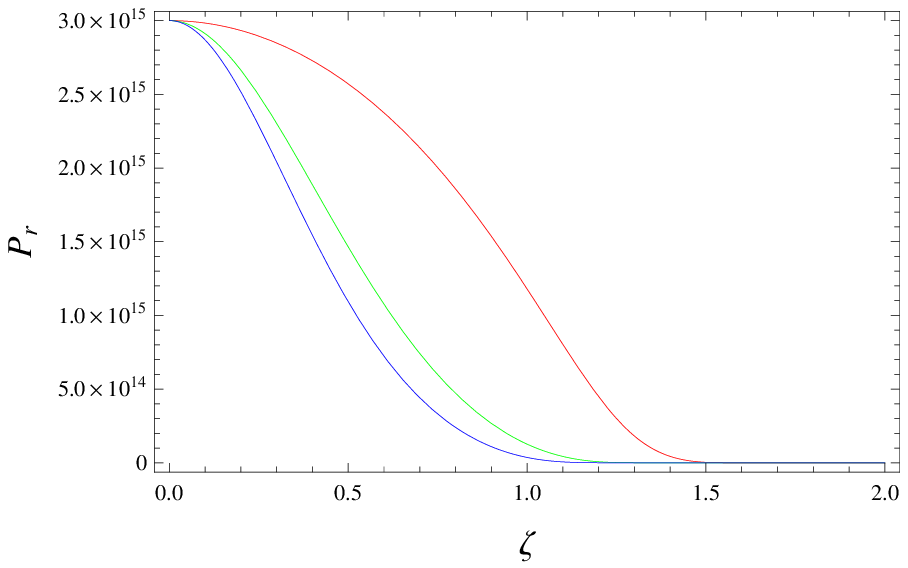,width=0.5\linewidth}
\epsfig{file=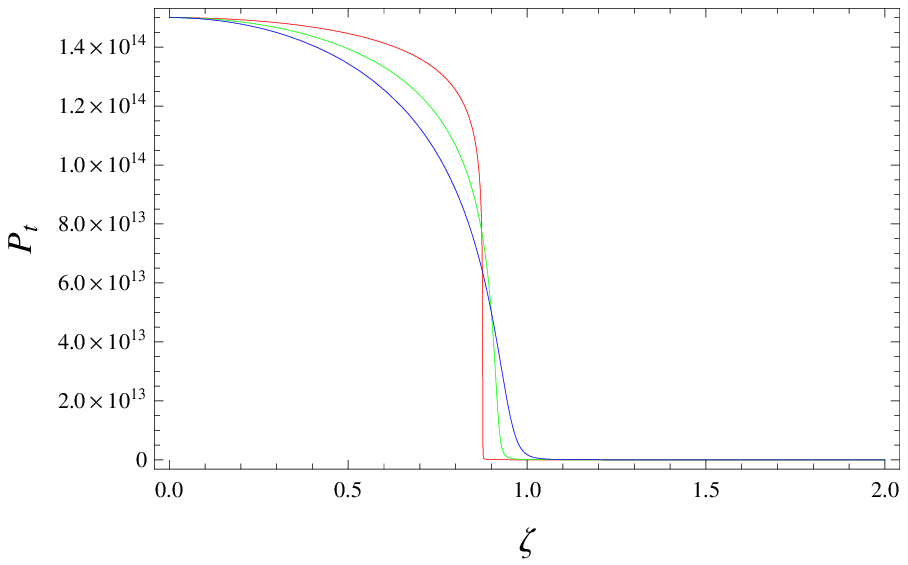,width=0.5\linewidth}\epsfig{file=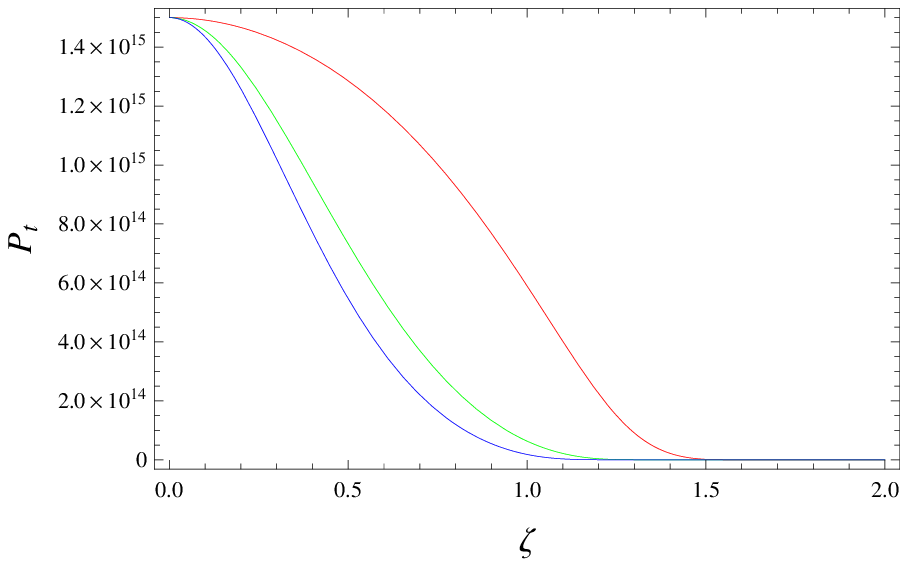,width=0.5\linewidth}
\epsfig{file=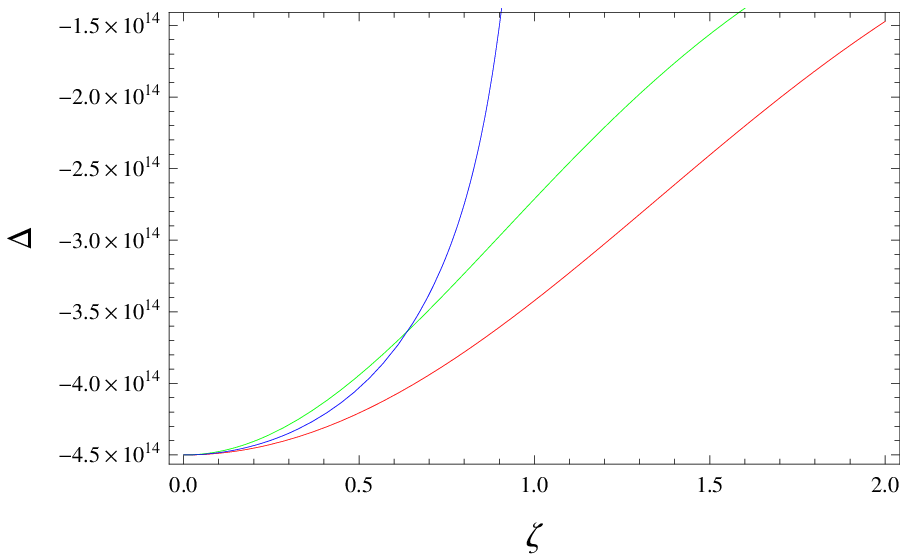,width=0.5\linewidth}\epsfig{file=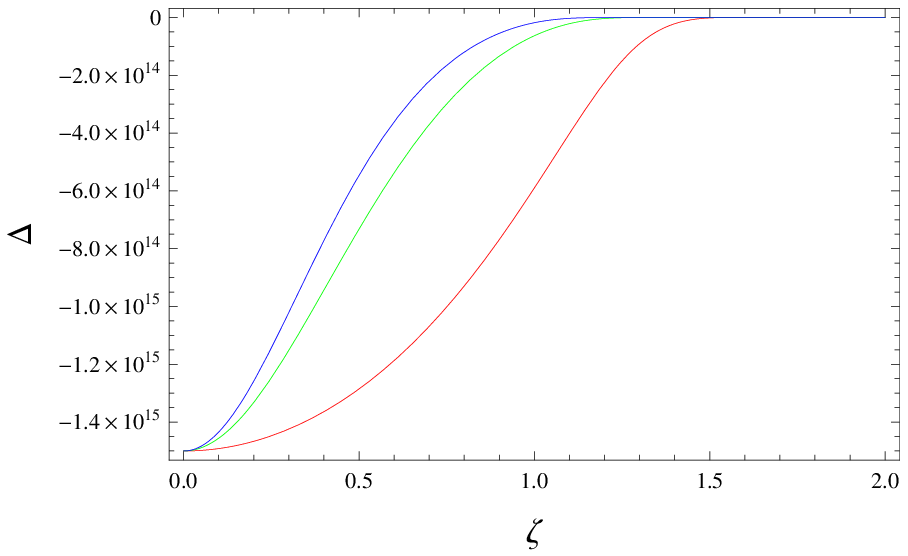,width=0.5\linewidth}
\caption{Plots of $\rho,~P_r,~P_\perp$ and $\Delta$ corresponding to
anisotropic case \textbf{I} for $\mu=1.5$, $c_1=2$, $h=0.5$ (red), 1
(green), 1.5 (blue) with $\alpha=0.1$ (left), $n=1$ (left) and
$\alpha=1$ (right), $n=0.5$ (right).}
\end{figure}
\begin{figure}\center
\epsfig{file=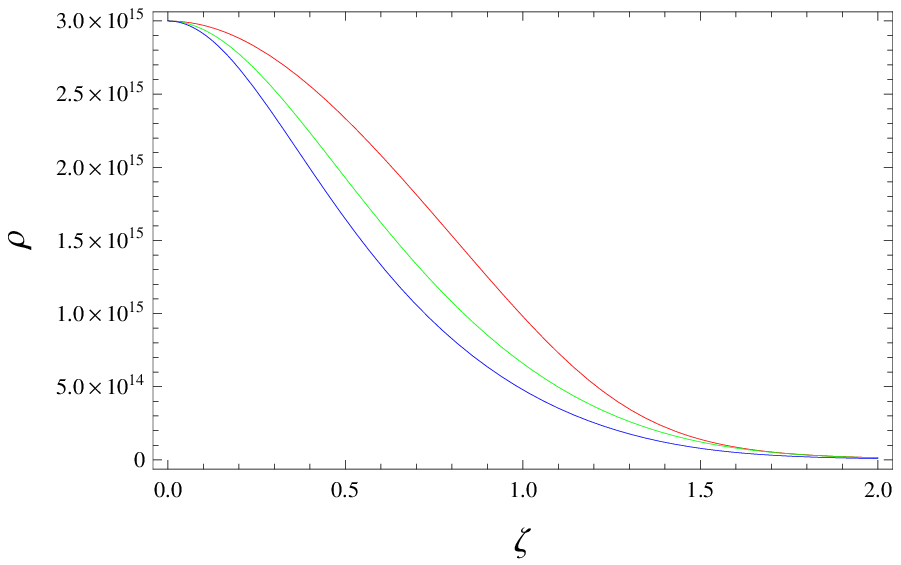,width=0.5\linewidth}\epsfig{file=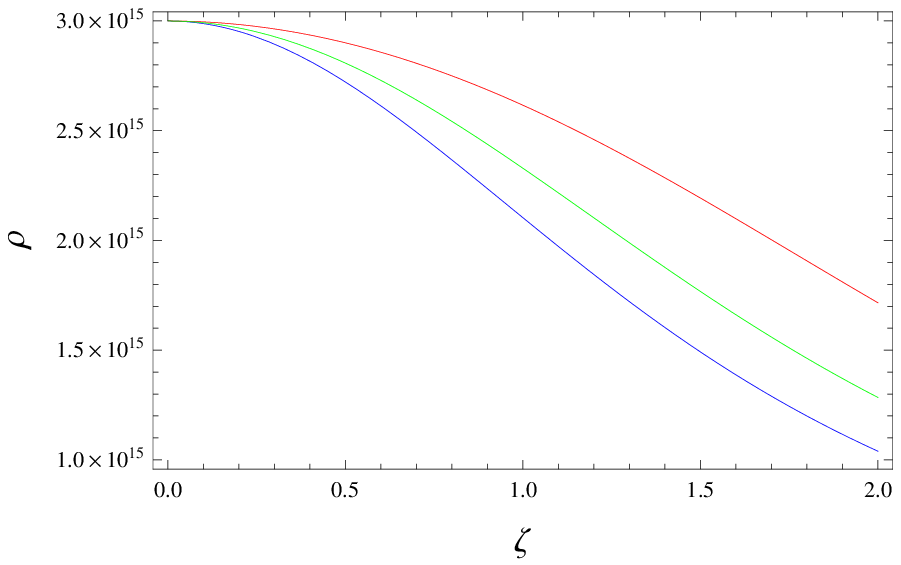,width=0.5\linewidth}
\epsfig{file=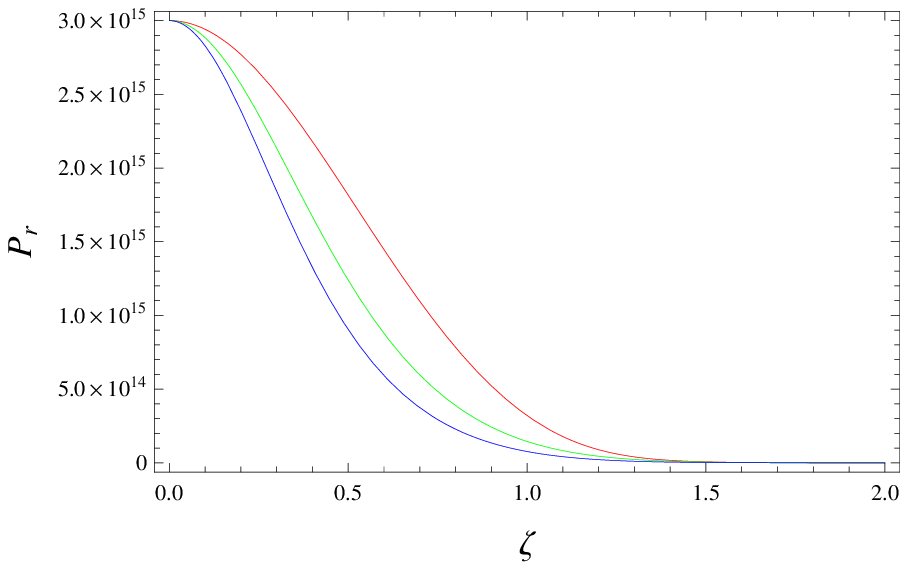,width=0.5\linewidth}\epsfig{file=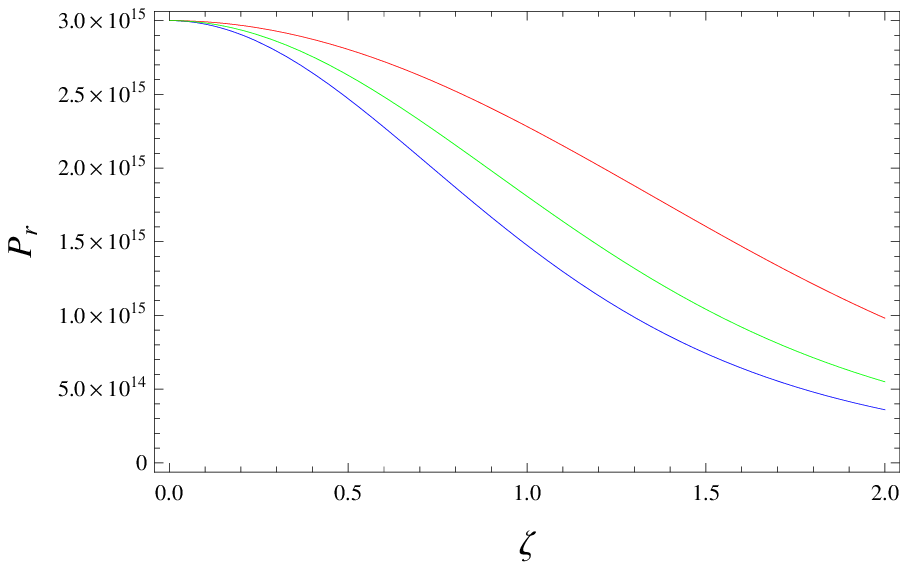,width=0.5\linewidth}
\epsfig{file=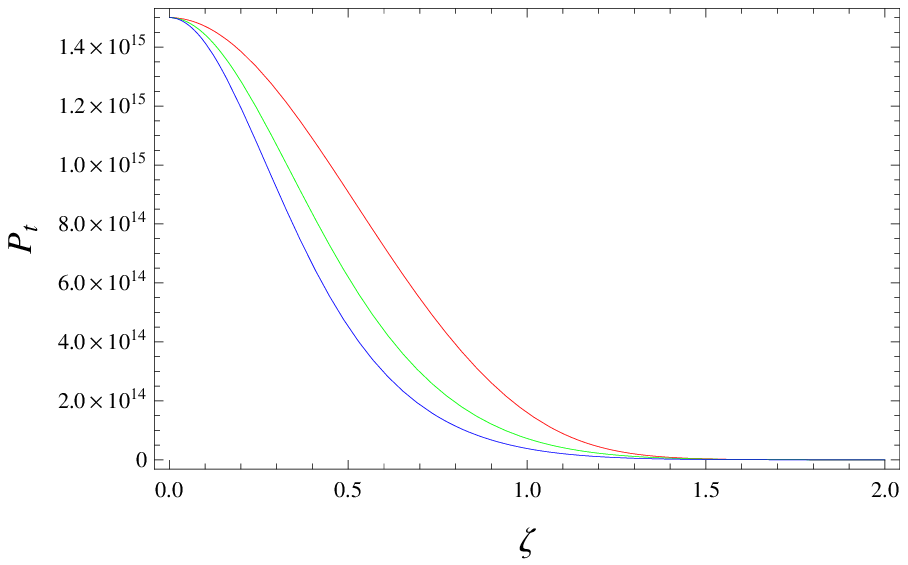,width=0.5\linewidth}\epsfig{file=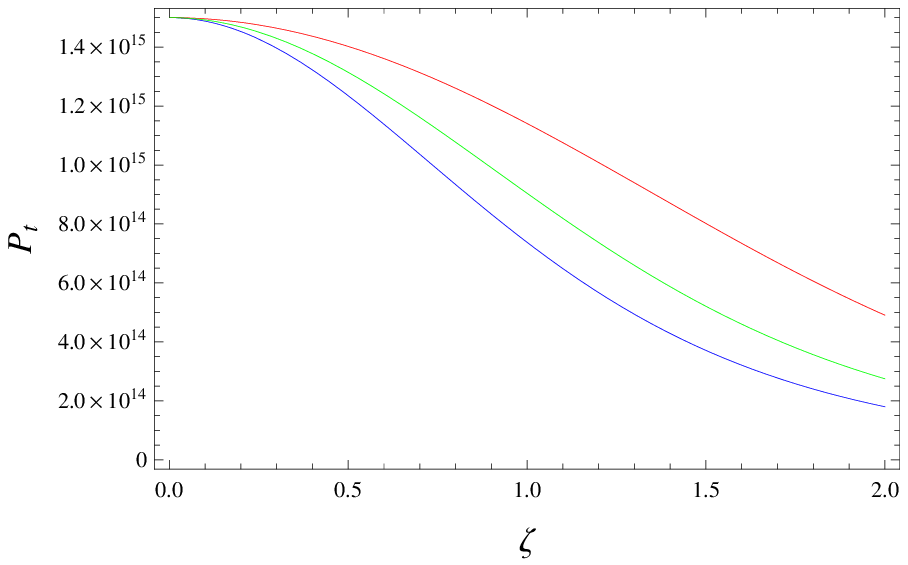,width=0.5\linewidth}
\epsfig{file=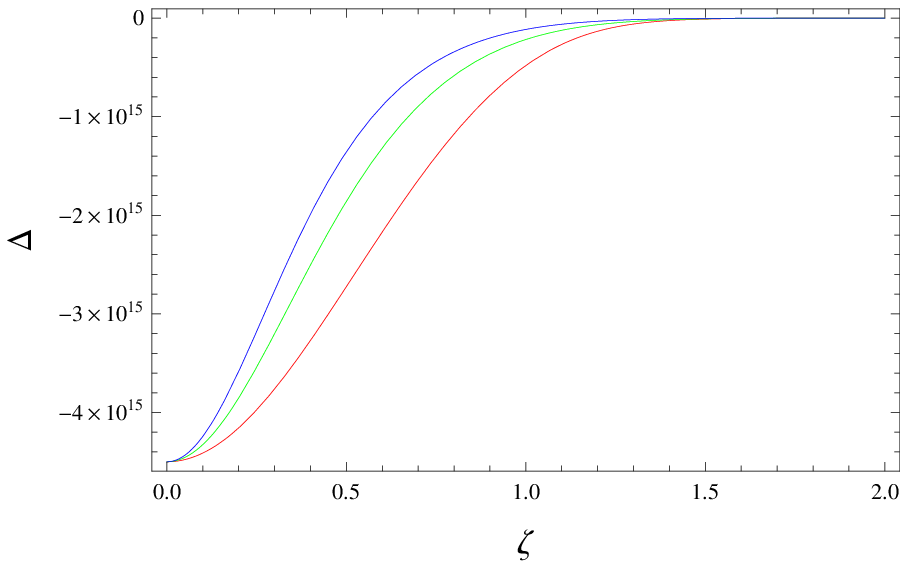,width=0.5\linewidth}\epsfig{file=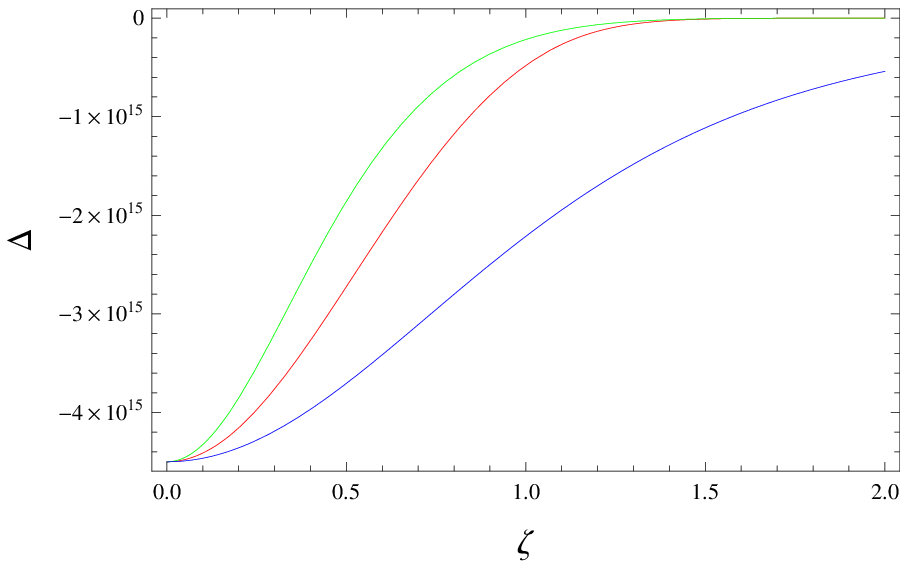,width=0.5\linewidth}
\caption{Plots of $\rho$, $P_r$, $P_\perp$ and $\Delta$
corresponding to anisotropic case \textbf{II} for $\mu=1.5$, $c_1=2$
and $h=0.5$ (red), 1 (green), 1.5 (blue) with $\alpha=1$ (left),
$n=1$ (left) and $\alpha=0.1$ (right), $n=2$ (right).}
\end{figure}

In this subsection, we explore different attributes of the
anisotropic setups for different values of the parameters. The
matter variables of anisotropic polytropes monotonically decrease
away from the center in both scenarios as depicted in Figures
\textbf{11} and \textbf{12}. An increase in the state parameters is
noted as $h$ increases. However, in case \textbf{I}, the density as
well as radial/tangential pressure initially increase and then
decrease corresponding to higher values of $h$ when $\alpha=0.1$ and
$n=1$. Moreover, the anisotropy is negative which indicates the
presence of an attractive force within the spherical structures. The
energy bounds related to anisotropic systems are
\begin{eqnarray*}
&&\text{null energy condition:}\quad\rho+P_r\geq0,\quad\rho+P_\perp\geq0,\\
&&\text{weak energy condition:}\quad\rho\geq0,\quad\rho+P_r\geq0,\quad\rho+P_\perp\geq0,\\
&&\text{strong energy condition:}\quad\rho+P_r+2P_\perp\geq0,\\
&&\text{dominant energy condition:}\quad\rho-P_r\geq0,\quad
\rho-P_\perp\geq0,\\
&&\text{trace energy condition}\quad\rho-P_r-2P_t\geq0.
\end{eqnarray*}
As anisotropic models \textbf{I} and \textbf{II} have positive
energy density and pressures, therefore the null, weak and strong
energy conditions are satisfied. Figures \textbf{13} and \textbf{14}
show that the anisotropic models are viable for the considered
values of the parameters.
\begin{figure}\center
\epsfig{file=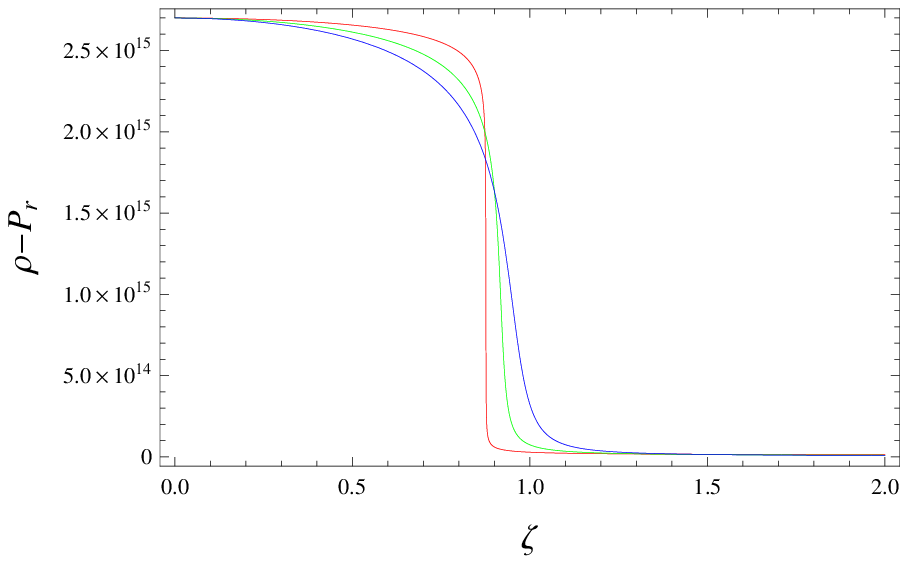,width=0.5\linewidth}\epsfig{file=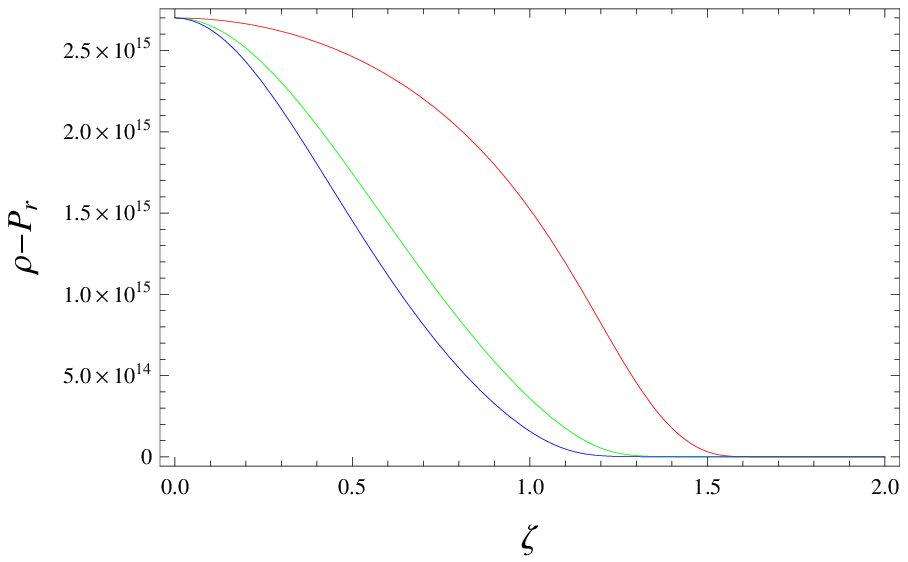,width=0.5\linewidth}
\epsfig{file=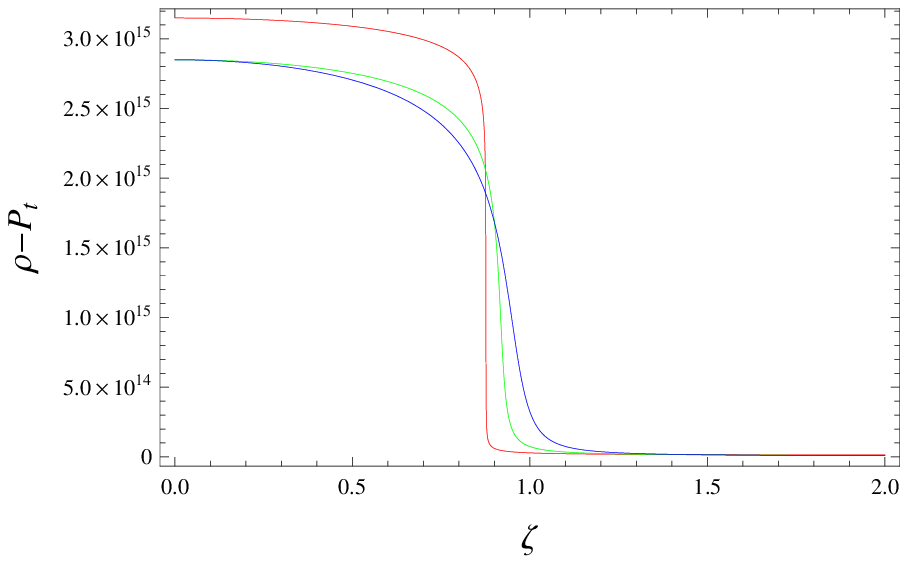,width=0.5\linewidth}\epsfig{file=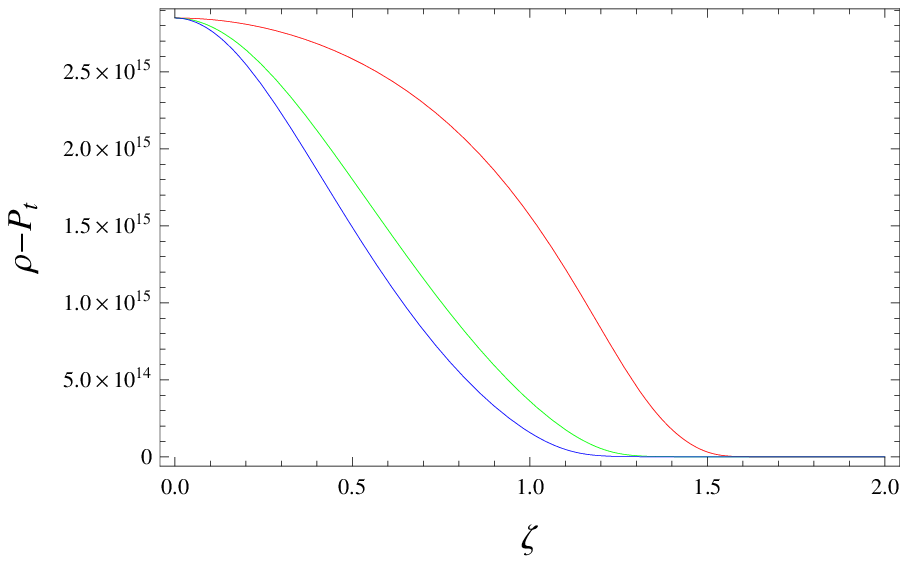,width=0.5\linewidth}
\epsfig{file=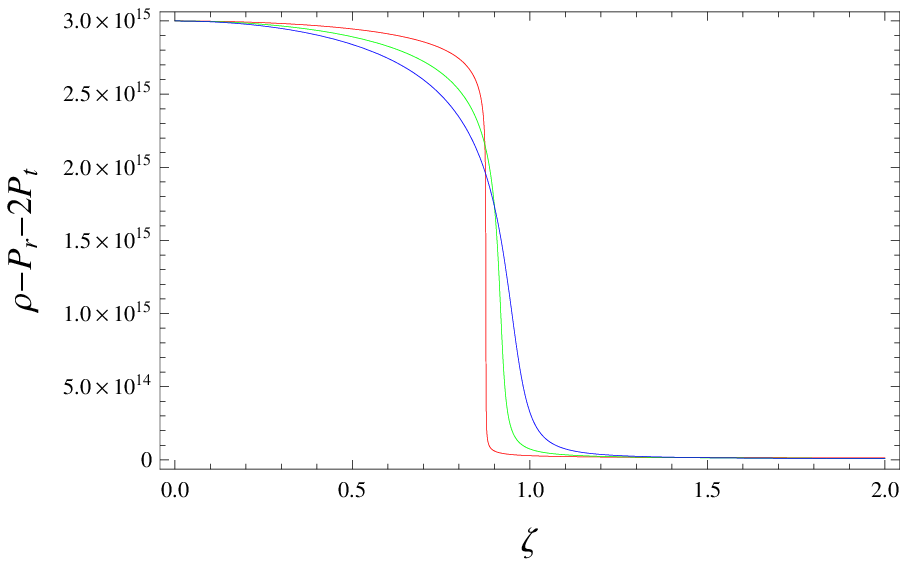,width=0.5\linewidth}\epsfig{file=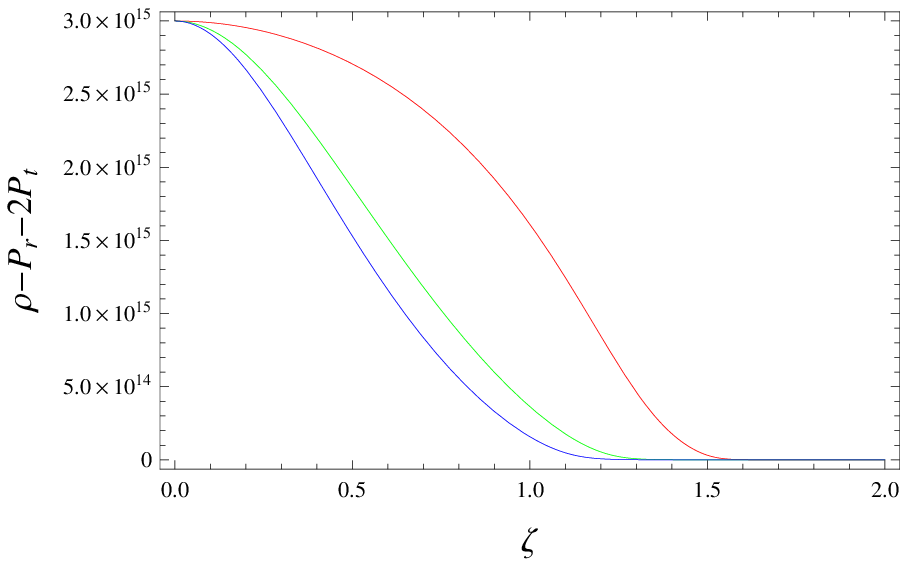,width=0.5\linewidth}\caption{Plots
of $\rho-P_r$, $\rho-P_t$ and $\rho-P_r-2P_t$ corresponding to
anisotropic case \textbf{I} for $\mu=1.5$, $c_1=2$, $h=0.5$ (red), 1
(green), 1.5 (blue) with $\alpha=0.1$ (left), $n=1$ (left) and
$\alpha=1$ (right), $n=0.5$ (right).}
\end{figure}
\begin{figure}\center
\epsfig{file=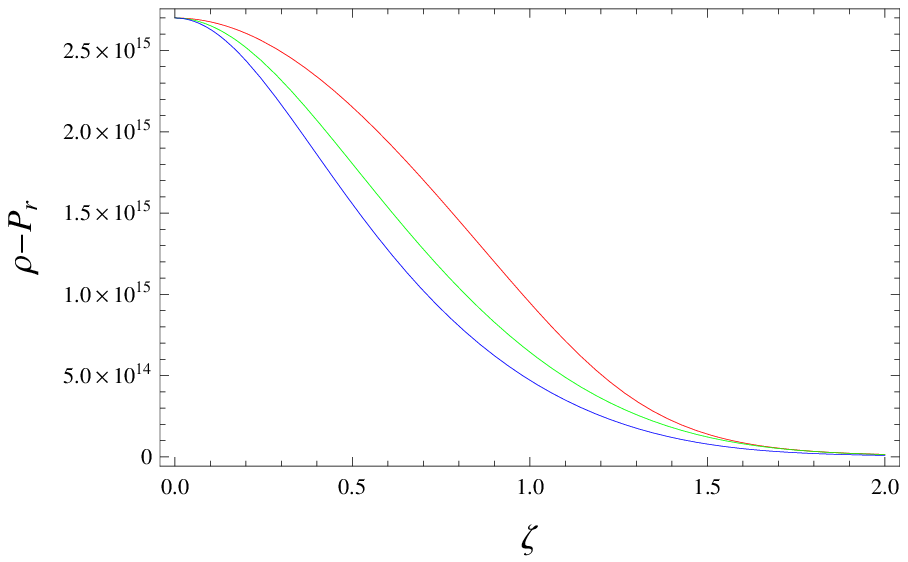,width=0.5\linewidth}\epsfig{file=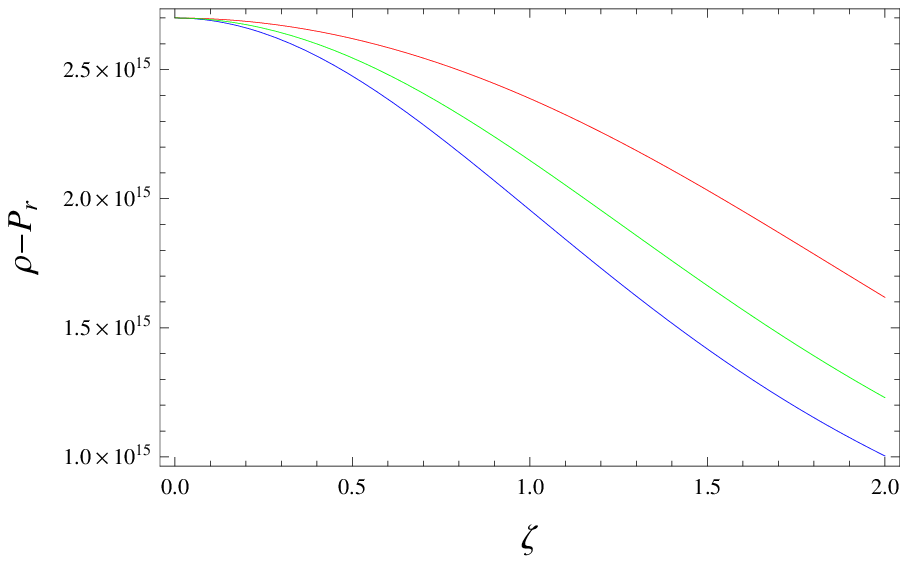,width=0.5\linewidth}
\epsfig{file=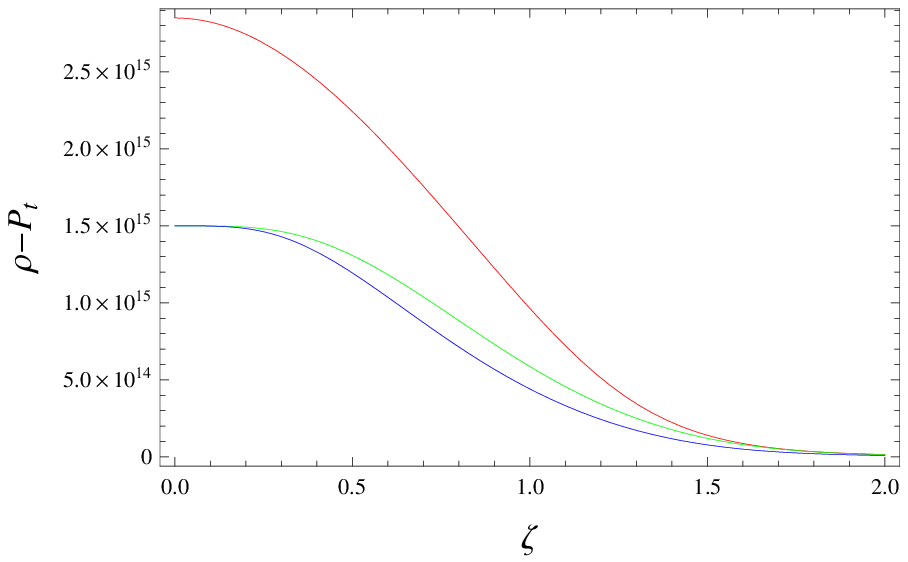,width=0.5\linewidth}\epsfig{file=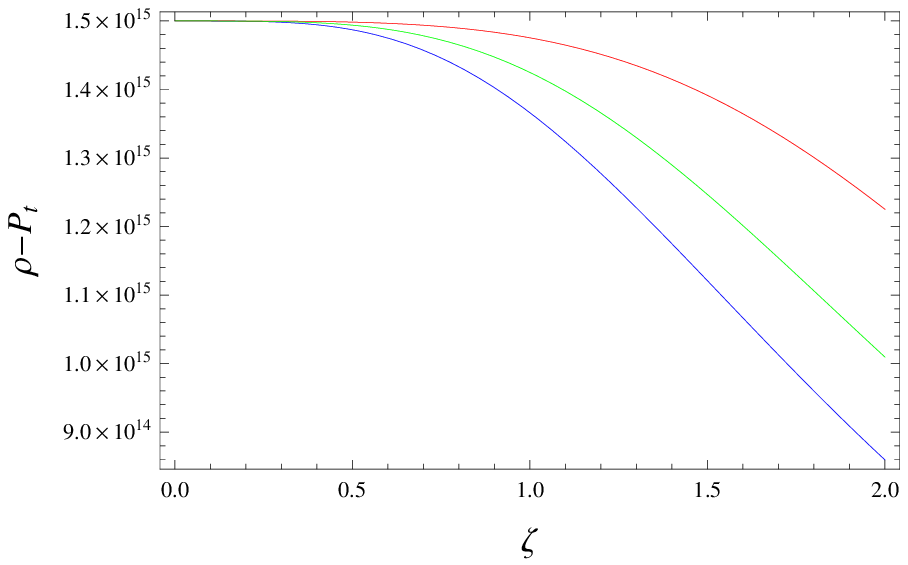,width=0.5\linewidth}
\epsfig{file=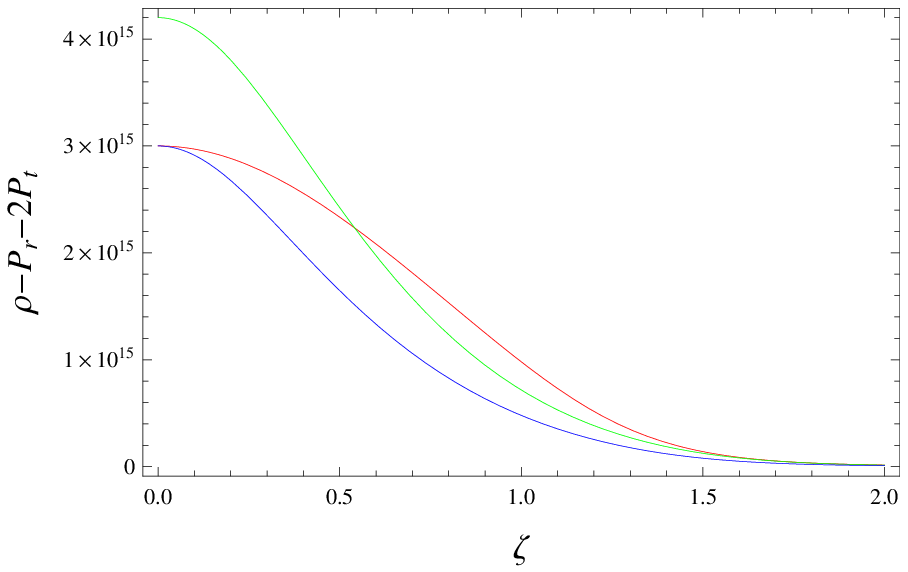,width=0.5\linewidth}\epsfig{file=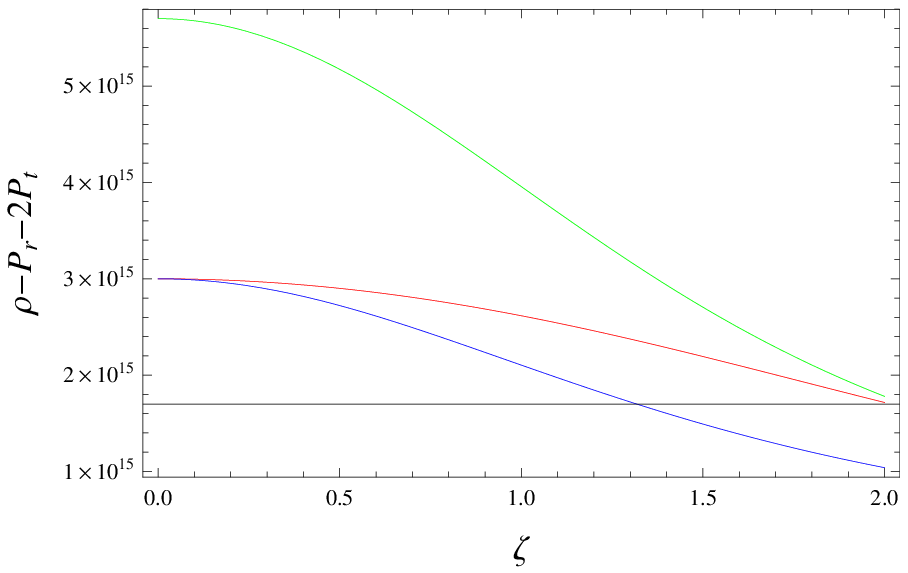,width=0.5\linewidth}
\caption{ Plots of $\rho-P_{r}$, $\rho-P_t$ and $\rho-P_r-2P_t$
corresponding to anisotropic case \textbf{II} for $\mu=1.5$, $c_1=2$
and $h=0.5$ (red), 1 (green), 1.5 (blue) with $\alpha=1$ (left),
$n=1$ (left) and $\alpha=0.1$ (right), $n=2$ (right).}
\end{figure}
\begin{figure}\center
\epsfig{file=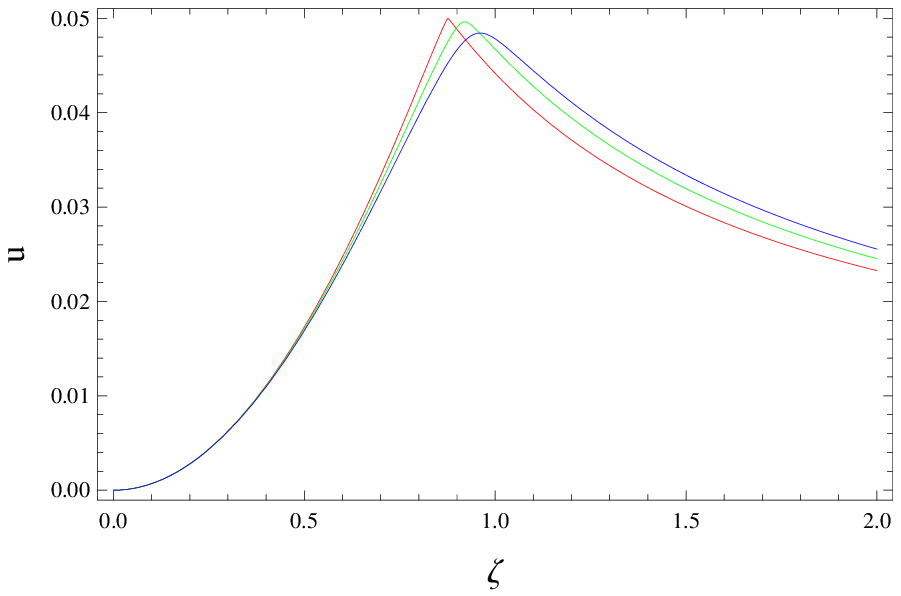,width=0.4\linewidth}\epsfig{file=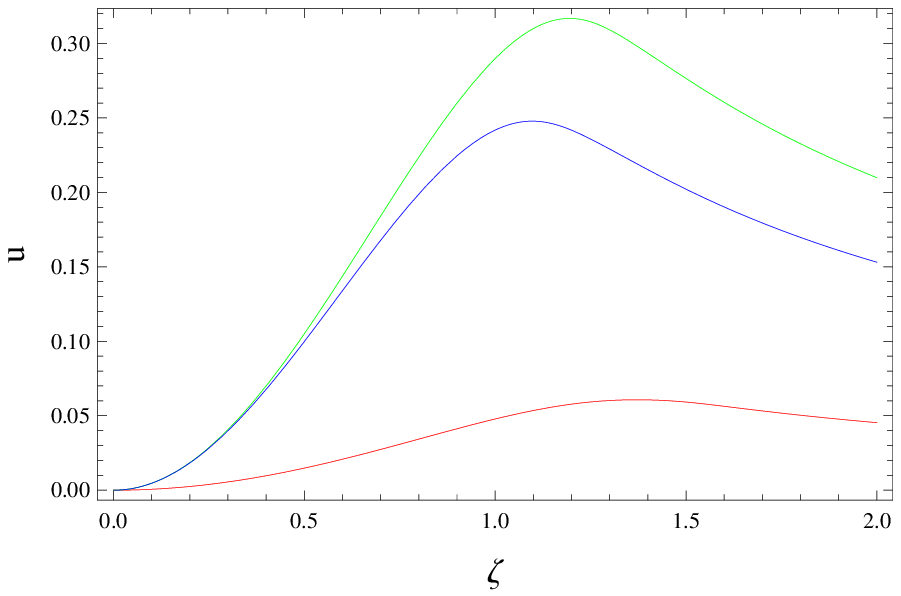,width=0.4\linewidth}
\epsfig{file=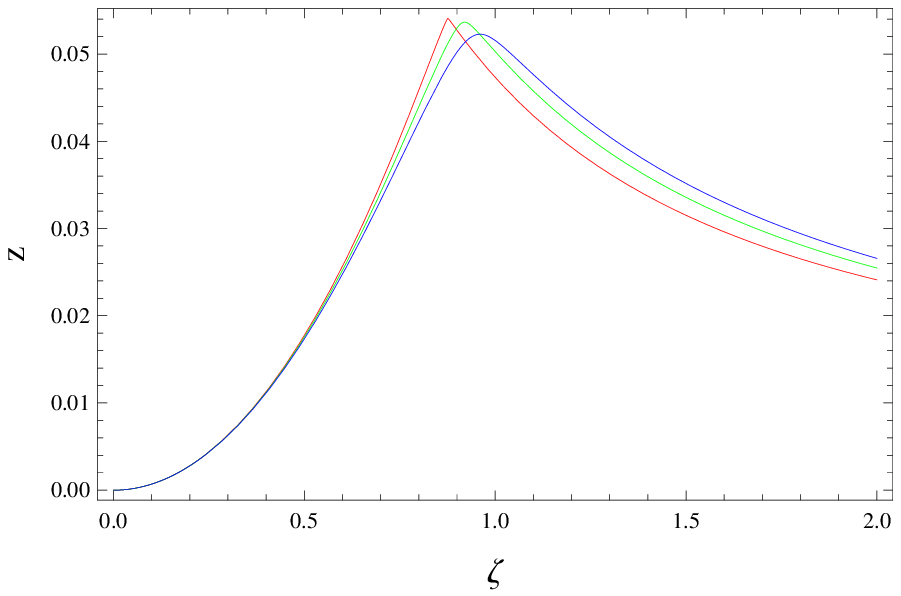,width=0.4\linewidth}\epsfig{file=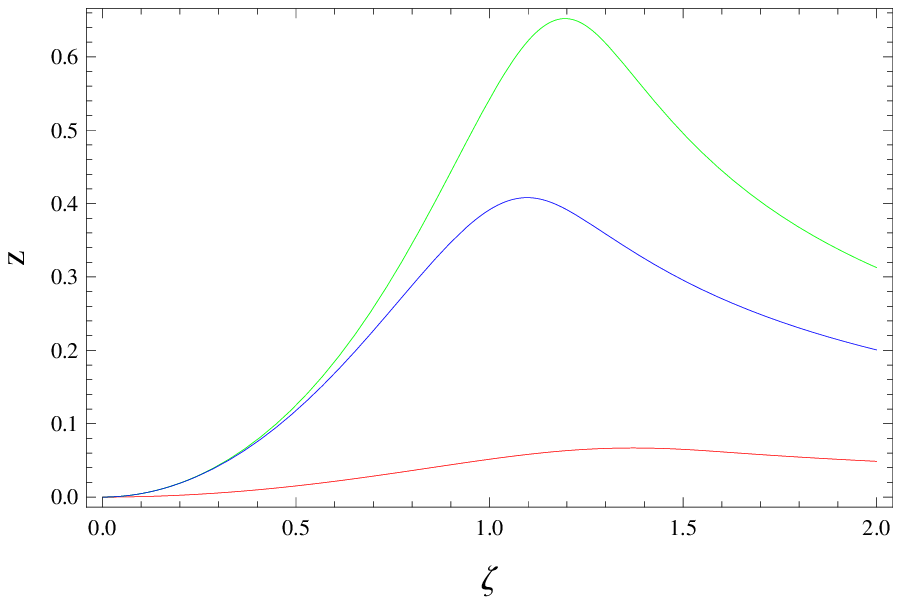,width=0.4\linewidth}
\caption{Plots of $u$ and $Z$ corresponding to anisotropic case
\textbf{I} for $\mu=1.5$, $c_1=2$, $h=0.5$ (red), 1 (green), 1.5
(blue) with $\alpha=0.1$ (left), $n=1$(left) and $\alpha=1$ (right),
$n=0.5$ (right).}
\end{figure}

The upper limit of compactness factor remains unchanged for
anisotropic configurations whereas the upper bound of redshift
increases to 5.211 \cite{31a}. It is observed from Figures
\textbf{15} and \textbf{16} that these parameters comply with their
required limits for cases \textbf{I} and \textbf{II}. We employ
causality condition ($\nu_r^2=\frac{dP_r}{d\rho}<1$ and
$\nu_t^2=\frac{dP_t}{d\rho}<1$ where $\nu_r^2$ and $\nu_t^2$ radial
and tangential speeds of sound, respectively) to determine the
stability of the anisotropic configurations. Figure \textbf{17}
displays that the first anisotropic model is consistent with
causality criterion whereas the second model is unstable for
$\alpha=n=1$ (refer to Figure \textbf{18}). Moreover, the model
constructed in case \textbf{I} is stiff as the radial adiabatic
index $\Gamma_r=\frac{P_r+\rho}{P_r}\frac{dP_r}{d\rho}$ is greater
than $4/3$ throughout the internal configuration. On the other hand,
$\Gamma_r<\frac{4}{3}$ in the second scenario for $\alpha=0.1$ and
$n=2$. In relativistic scenario, Moustakidis \cite{31b} imposed an
additional condition on adiabatic index. He proposed that $\Gamma_r$
must be greater than the critical value
$\Gamma_{r\text{(critical)}}=\frac{4}{3}+\frac{19}{21}u$. Recently,
the critical value of adiabatic index was also used to investigate
the behavior of decoupled solutions \cite{31c}. We have plotted the
critical values of adiabatic index in Figures \textbf{17} and
\textbf{18}. It is noted that $\Gamma_r$ is greater than the
critical value for both case I and II.
\begin{figure}\center
\epsfig{file=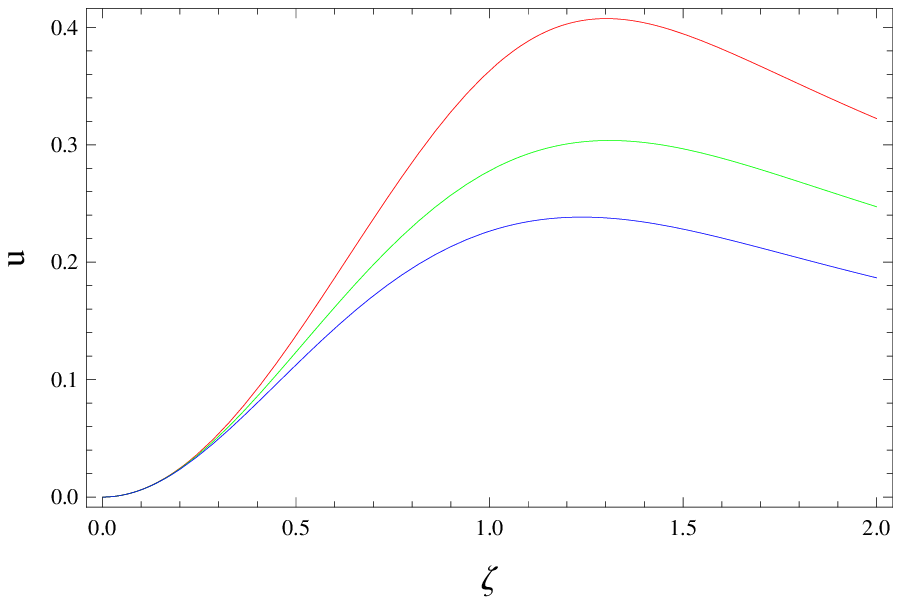,width=0.4\linewidth}\epsfig{file=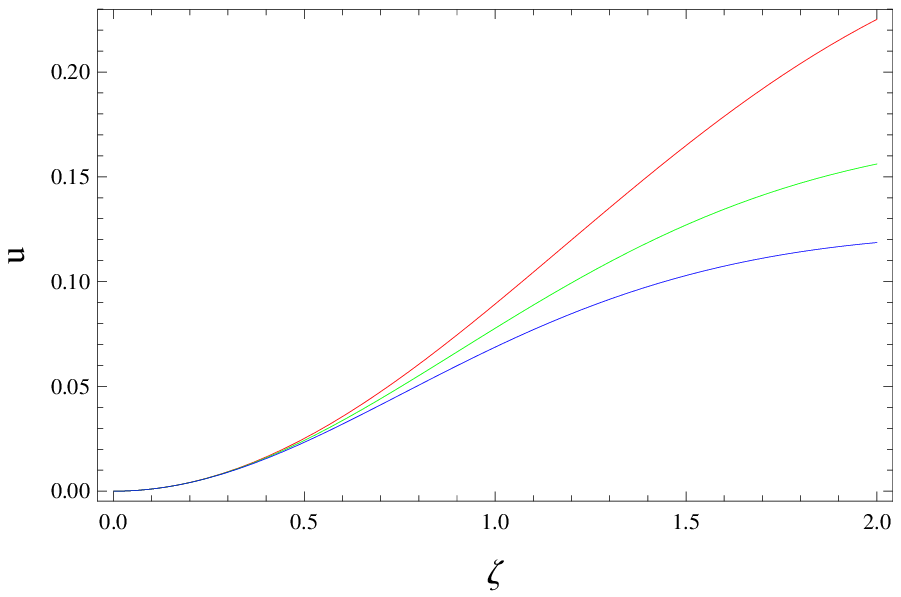,width=0.4\linewidth}
\epsfig{file=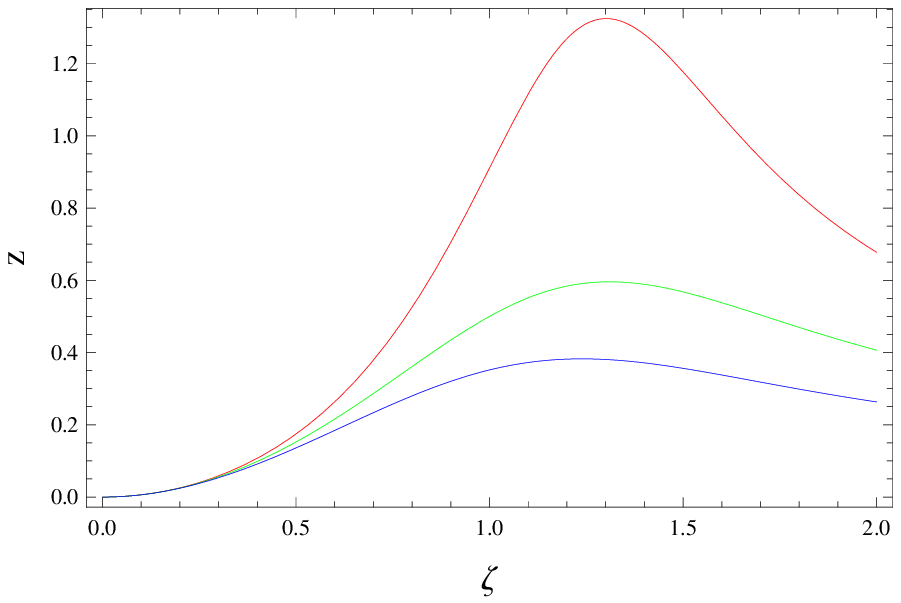,width=0.4\linewidth}\epsfig{file=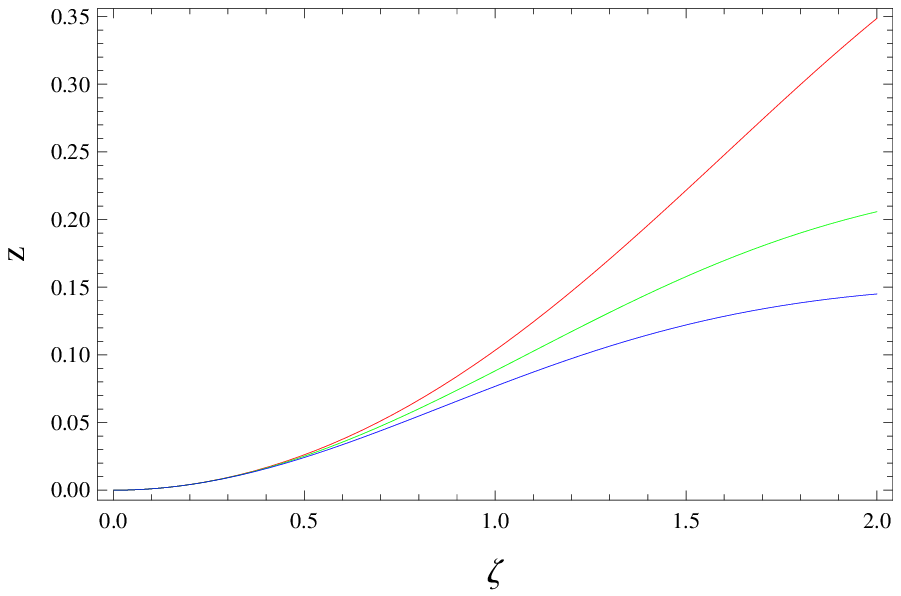,width=0.4\linewidth}
\caption{Plots of $u$ and $Z$ corresponding to anisotropic case
\textbf{II} for $\mu=1.5$, $c_1=2$ and $h=0.5$ (red), 1 (green), 1.5
(blue) with $\alpha=1$ (left), $n=1$ (left) and $\alpha=0.1$
(right), $n=2$ (right).}
\end{figure}
\begin{figure}\center
\epsfig{file=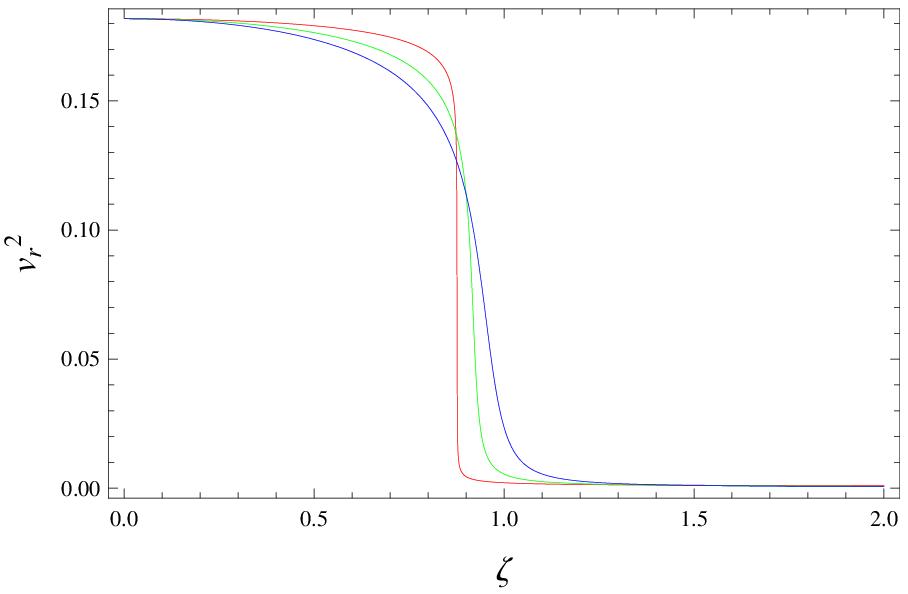,width=0.4\linewidth}\epsfig{file=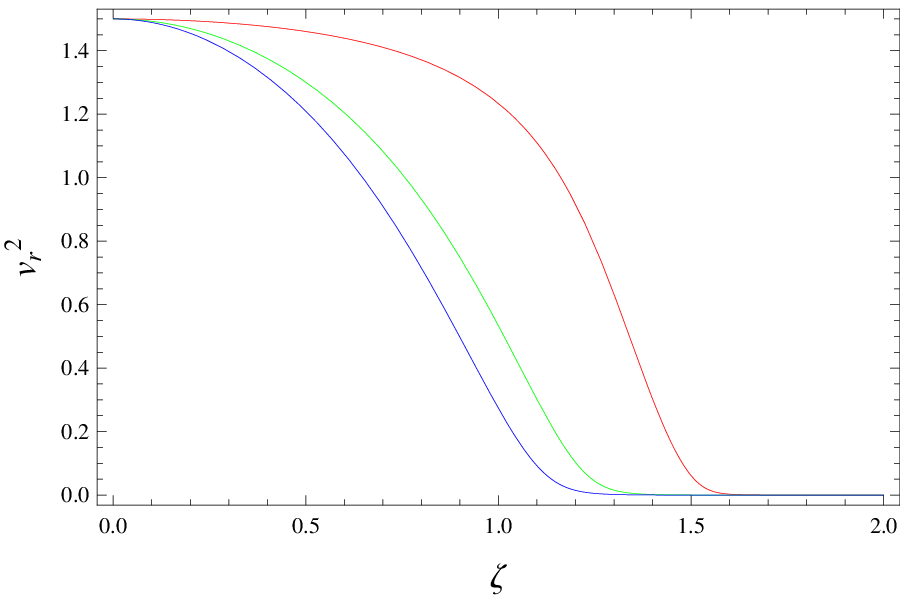,width=0.4\linewidth}
\epsfig{file=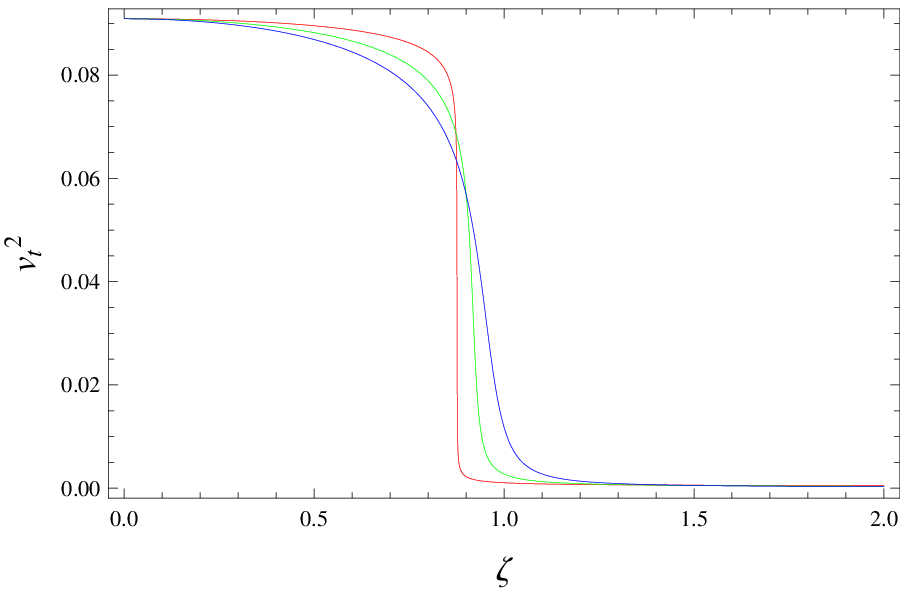,width=0.4\linewidth}\epsfig{file=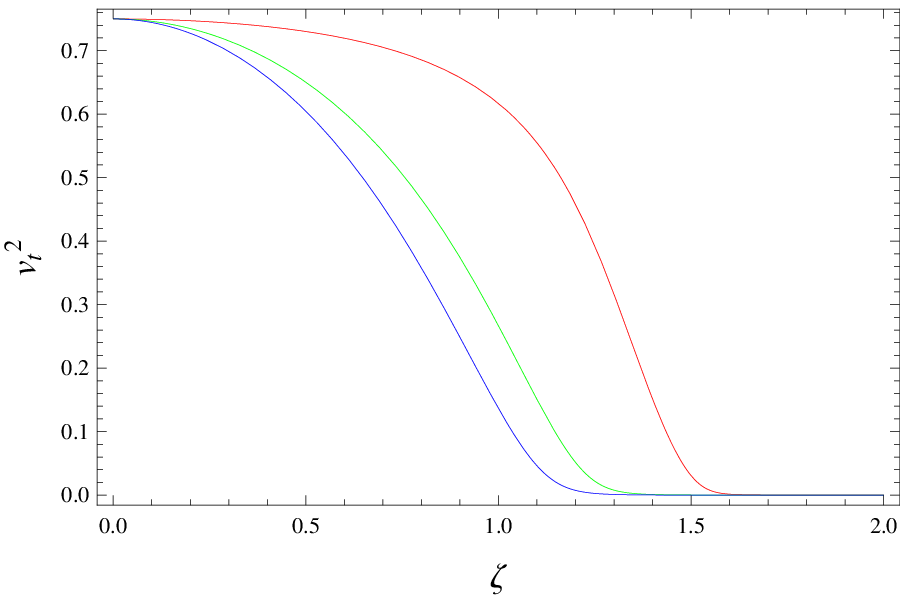,width=0.4\linewidth}
\epsfig{file=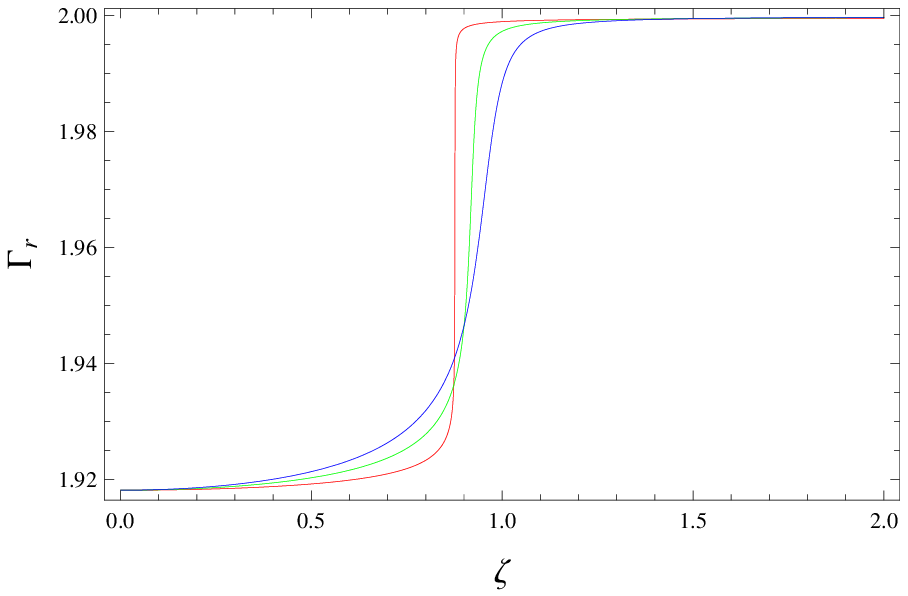,width=0.4\linewidth}\epsfig{file=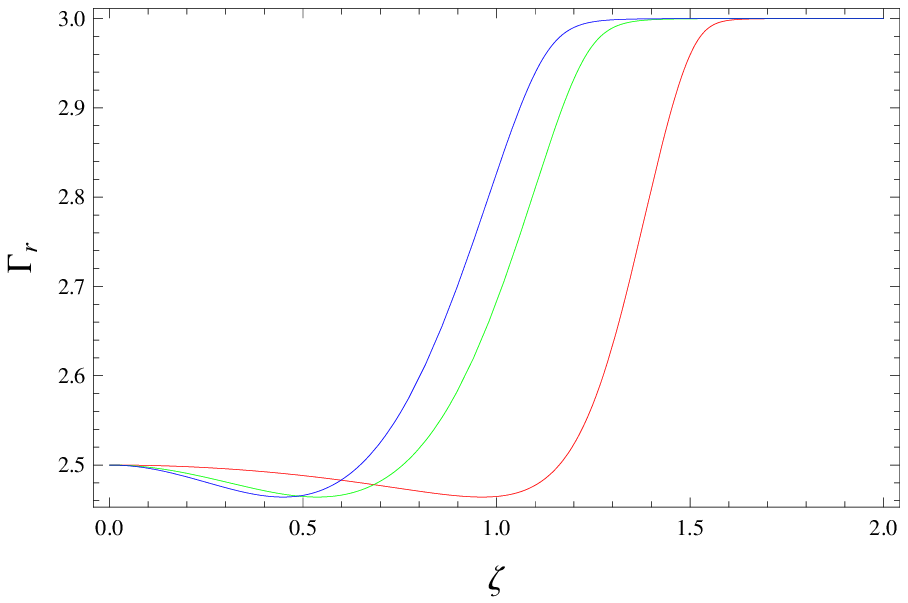,width=0.4\linewidth}
\epsfig{file=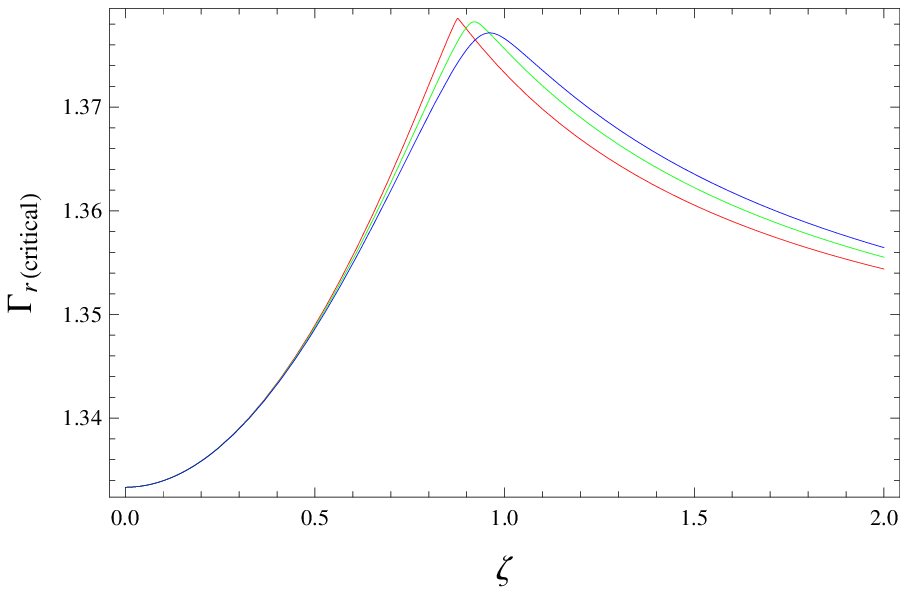,width=0.4\linewidth}\epsfig{file=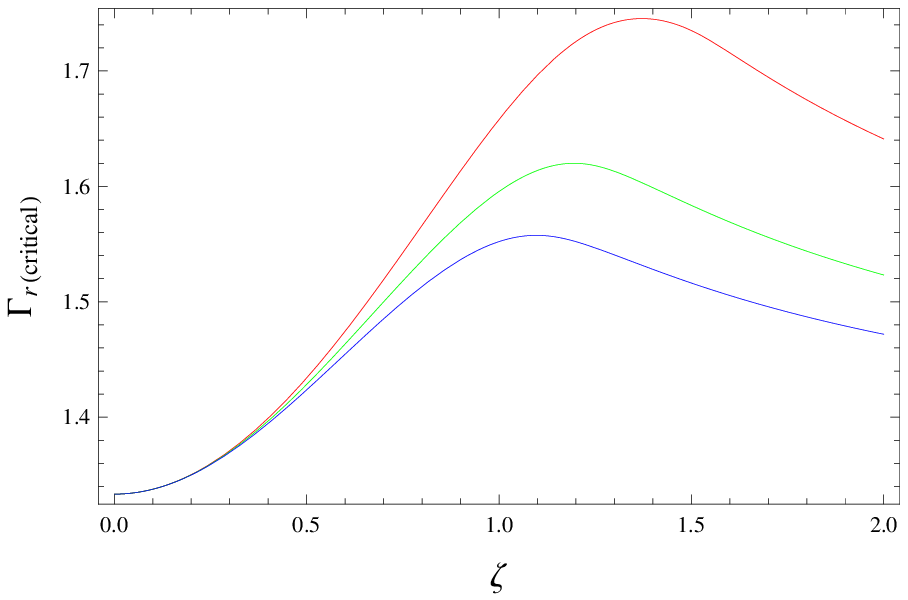,width=0.4\linewidth}
\caption{Plots of $\nu_{r}^{2}$, $\nu_{t}^{2}$, $\Gamma_r$ and
$\Gamma_{r\text{(critical)}}$ corresponding to anisotropic case
\textbf{I} for $\mu=1.5$, $c_1=2$, $h=0.5$ (red), 1 (green), 1.5
(blue) with $\alpha=0.1$ (left), $n=1$ (left) and $\alpha=1$
(right), $n=0.5$ (right).}
\end{figure}
\begin{figure}\center
\epsfig{file=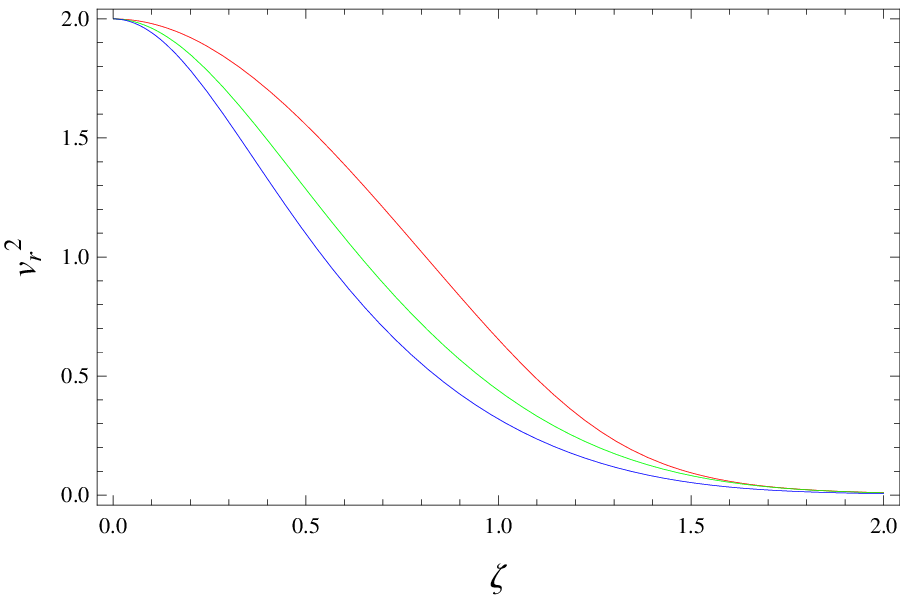,width=0.4\linewidth}\epsfig{file=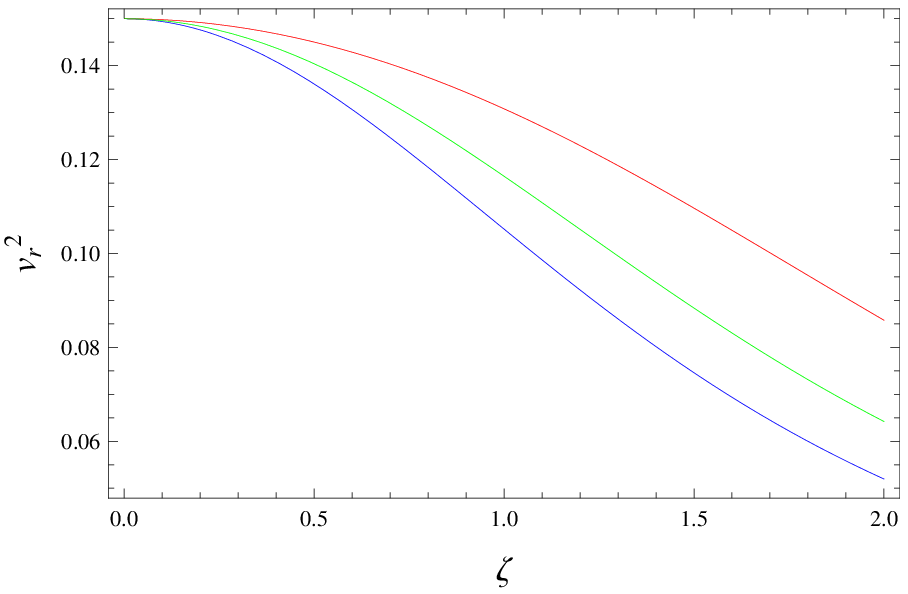,width=0.4\linewidth}
\epsfig{file=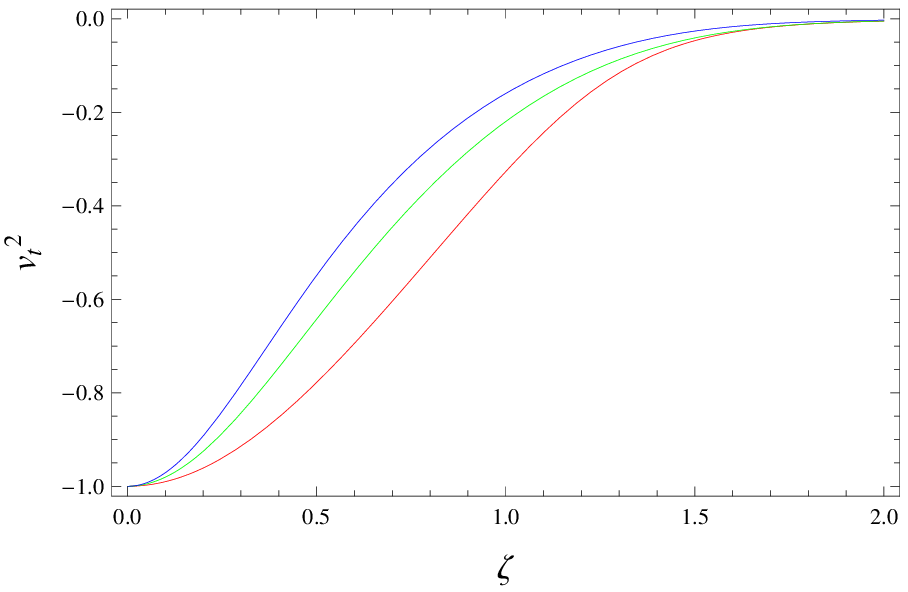,width=0.4\linewidth}\epsfig{file=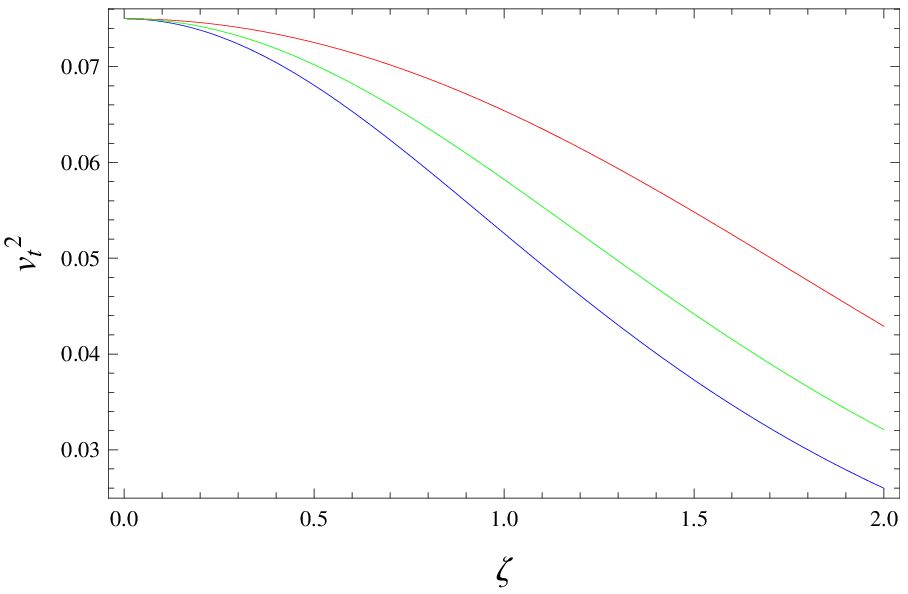,width=0.4\linewidth}
\epsfig{file=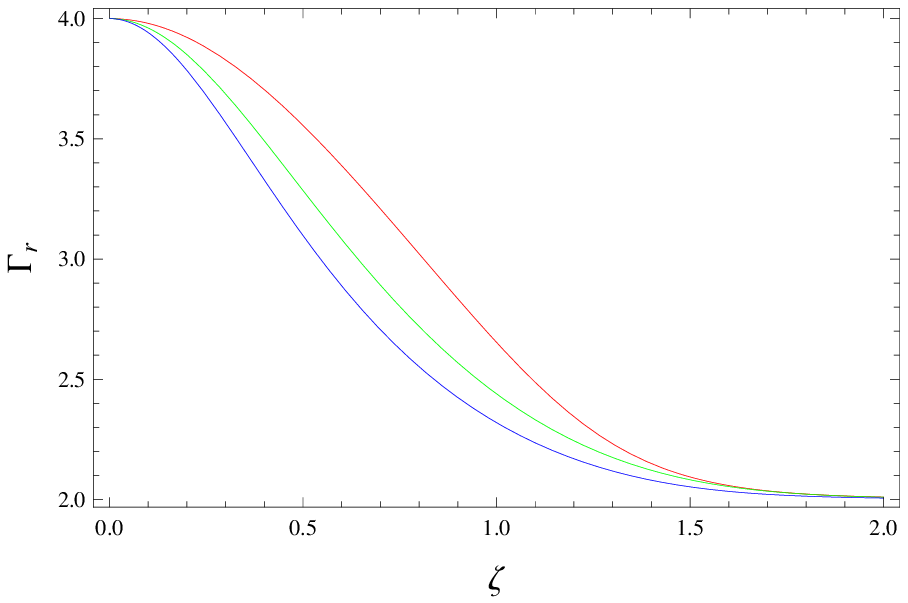,width=0.4\linewidth}\epsfig{file=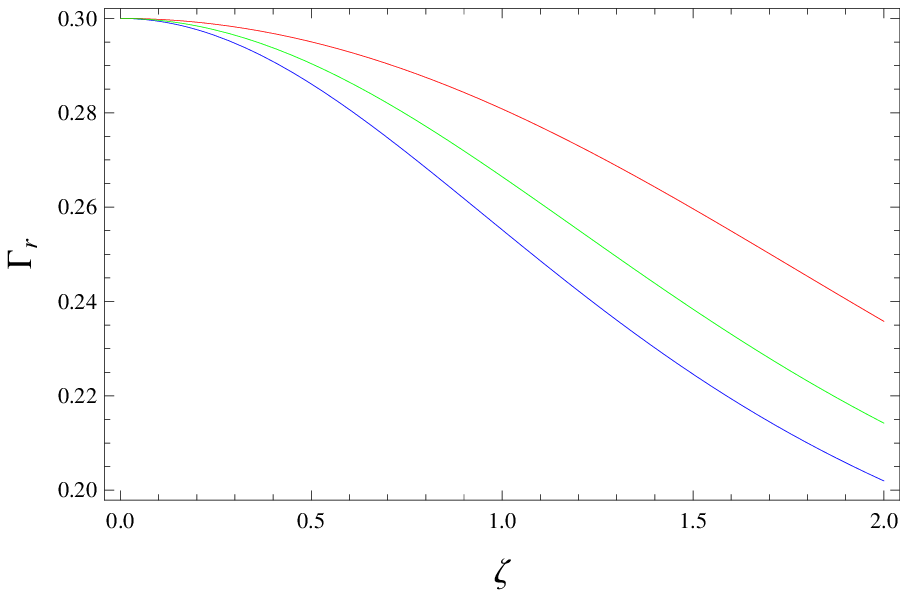,width=0.4\linewidth}
\epsfig{file=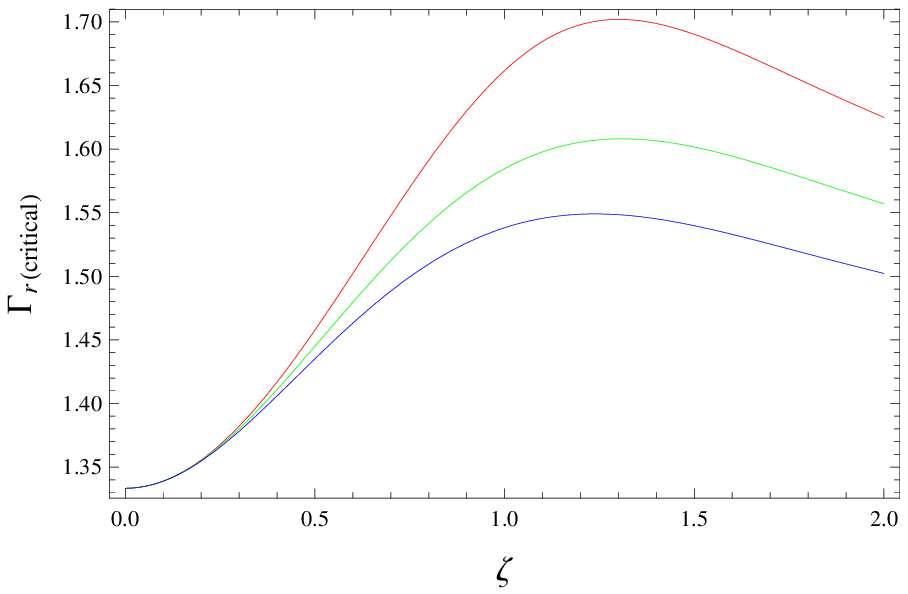,width=0.4\linewidth}\epsfig{file=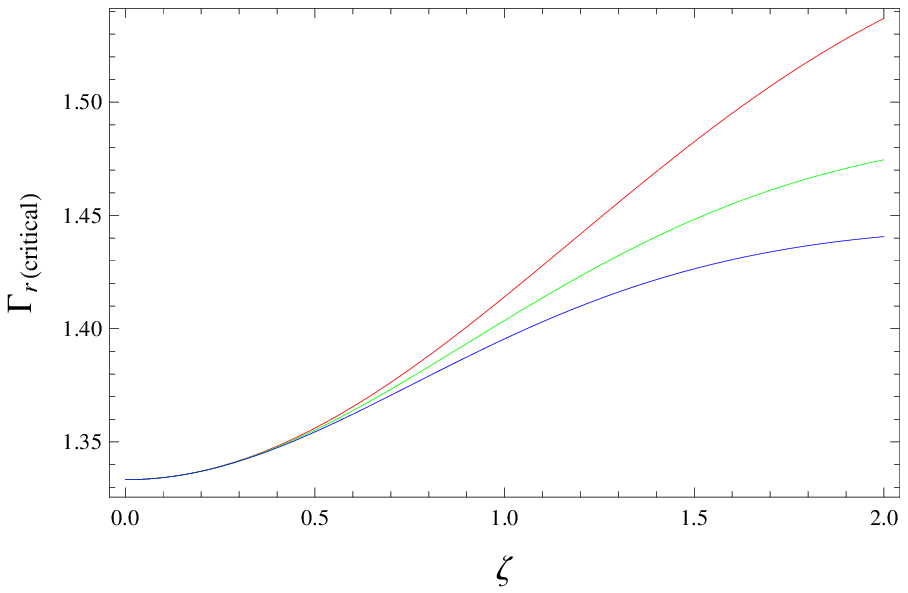,width=0.4\linewidth}
\caption{Plots of $v_{r}^{2}$, $v_{t}^{2}$, $\Gamma_r$ and
$\Gamma_{r\text{(critical)}}$ corresponding to anisotropic case
\textbf{II} for $\mu=1.5$, $c_1=2$ and $h=0.5$ (red), 1 (green), 1.5
(blue) with $\alpha=1$ (left), $n=1$(left) and $\alpha=0.1$ (right),
$n=2$ (right).}
\end{figure}

Now, we compute the Tolman mass, which measures the gravitational
mass, defined as \cite{32}
\begin{equation}\label{52}
m_{\mathrm{T}}=\frac{1}{2}r^{2}e^\frac{{\vartheta-\lambda}}{2}
\vartheta'.
\end{equation}
Substituting Eq.(\ref{34}) with mass function in Eq.(\ref{52}), we
have
\begin{equation}\label{53}
m_{\mathrm{T}}=e^\frac{{\vartheta+\lambda}}{2} \bigg[m+r^{3}(4\pi
P_{r}+\frac{\mu}{2}(\rho+c_2 P_{r}))\bigg].
\end{equation}
For the case \textbf{I}, we use Eq.(\ref{18}) in Eq.(\ref{36}) and
obtain
\begin{eqnarray}\label{54}
P_{r}=K \rho_{rc}^{1+\frac{1}{n}} \psi_{0}^{n+1}, \quad
\rho=\rho_{0c}\psi_{0}^{n}(1+nk\rho_{0c}^{1/n}\psi_{0}).
\end{eqnarray}
The resulting TOV equation yields
\begin{eqnarray}\nonumber
&&2(n+1)\beta d\psi_{0}+h d\vartheta[1+\beta
\psi_{0}(n+1)]+L(n(1+n\beta\psi_{0})\\\label{55} &&
+\beta\psi_{0}(n+c_1(1+n)))d\psi_{0}=0,
\end{eqnarray}
where $\beta=\frac{\alpha}{1-n\alpha}$. Integrating this equation,
it follows that
\begin{equation}\label{56}
e^\vartheta=\frac{H}{[\psi_{0}^{nL}
(1+(n+1)\beta\psi_{0})^{1+c_1L}]^{\frac{2}{h}}},
\end{equation}
where $H$ is an integration constant. Initial condition at the
center ($r=0,~\psi_0=1$) yields $H$ and hence
\begin{equation}\label{57}
e^\vartheta=e^{\vartheta_{c}}\frac{[
(1+(n+1)\beta)^{1+c_1L}]^{\frac{2}{h}}}{[\psi_{0}^{nL}
(1+(n+1)\beta\psi_{0})^{1+c_1L}]^{\frac{2}{h}}}.
\end{equation}
Using the matching conditions defined in Eq.(\ref{7}), we obtain
\begin{equation}\label{58}
e^{\vartheta}=\left(1-\frac{2 M}{r_{\Sigma}}\right)\frac{[(I^{nL}
((1-n\alpha)+(n+1)\alpha I)^{1+c_1L}]^\frac{2}{h}}{[\psi_{0}^{nL}
((1-n\alpha)+(n+1)\alpha \psi_{0})^{1+c_1L}]^\frac{2}{h}}.
\end{equation}
Substituting this value in Eq.(\ref{53}), we have
\begin{eqnarray}\nonumber
&&\upsilon_{\mathrm{T}}=(\upsilon+\zeta^{3}\psi_{0}^{n}(\alpha
\psi_{0}+\frac{\mu}{8
\pi}((1-n\alpha)+(n+c_2)\alpha\psi_{0})))\\\label{59}&&
\times\frac{(a_{\sum})^{1/2}}{(a)^{1/2}}\frac{[(I^{nL}
((1-n\alpha)+(n+1)\alpha I)^{1+c_1L}]^\frac{1}{h}}{[\psi_{0}^{nL}
((1-n\alpha)+(n+1)\alpha \psi_{0})^{1+c_1L}]^\frac{1}{h}},
\end{eqnarray}
where $\upsilon_{\mathrm{T}}=\frac{m_{\mathrm{T}}A^{3}}{4\pi
\rho_{c}}$.

Similarly, for the case \textbf{II}, the TOV equation reduces to
\begin{equation}\label{60}
2\alpha \psi(1+n)d\psi+h\psi(1+\alpha\psi)d\vartheta+L(n+\alpha
c_1(1+n)\psi)d\psi=0.
\end{equation}
Proceeding in the same way as for the case \textbf{I}, we have
\begin{equation}\label{61}
e^{\vartheta}=\left(1-\frac{2 M}{r_{\Sigma}}\right)\frac{[J^{nL}
(1+\alpha J)^{(1+c_1L)(1+n)-nL}]^\frac{2}{h}}{[\psi^{nL} (1+\alpha
\psi)^{(1+c_1L)(1+n)-nL}]^\frac{2}{h}}.
\end{equation}
Substituting this value in Eq.(\ref{53}), it follows that
\begin{eqnarray}\nonumber
&&\upsilon_{\mathrm{T}}=(\upsilon+\zeta^{3}\psi^{n}(\alpha
\psi+\frac{\mu}{8 \pi}((1+c_2\alpha
\psi))))\frac{(a_{\Sigma})^{1/2}}{(a)^{1/2}}\\\label{62}&&\times\frac{[J^{nL}
(1+\alpha J)^{(1+c_1L)(1+n)-nL}]^\frac{1}{h}}{[\psi^{nL} ((1+\alpha
\psi)^{(1+c_1L)(1+n)-nL}]^\frac{1}{h}},
\end{eqnarray}
where $I=\frac{\mu(n \alpha-1)}{\alpha(8\pi+(n+c_2))}$ and
$J=\frac{-\mu}{8\pi +\mu c_2}$.

In order to explore the distribution of Tolman mass through the
sphere during the slow and adiabatic process, the following
dimensionless variables are introduced
\begin{equation}\label{63}
x=\frac{r}{r_{\Sigma}}=\frac{\zeta}{\tilde{A}}, \indent
y=\frac{M}{r_{\Sigma}}, \indent\tilde{m}=\frac{m}{M},
\indent\tilde{A}=r_{\Sigma}A.
\end{equation}
The Tolman mass for both cases (\textbf{I} and \textbf{II}) in terms
of above variables takes the form
\begin{eqnarray}\nonumber
\upsilon_{\mathrm{T}}&=&(\upsilon+x^{3}\tilde{A}^{3}\psi_{0}^{n}(\alpha
\psi_{0}+\frac{\mu}{8 \pi}((1-n\alpha)+(n+c_2)\alpha
\psi_{0})))\\\nonumber&\times&\frac{\left[(I^{nL}
((1-n\alpha)+(n+1)\alpha
I)^{1+c_1L}\right]^\frac{1}{h}}{\left[\psi_{0}^{nL}
((1-n\alpha)+(n+1)\alpha \psi_{0})^{1+c_1L}\right]^\frac{1}{h}}
\\\label{64}&\times&
\left(\frac{x(1-2y)}{x-2\alpha(n+1)
\upsilon/\tilde{A}}\right)^{1/2},
\\\nonumber
\upsilon_{\mathrm{T}}&=&(\upsilon+x^{3}\tilde{A}^{3}\psi^{n}(\alpha
\psi+\frac{\mu}{8 \pi}((1+c_2\alpha
\psi))))\\\label{65}&\times&\left[\frac{x(1-2y)}{x-2\alpha(n+1)
\upsilon/\tilde{A}}\right]^{1/2} \frac{\left[J^{nL} (1+\alpha
J)^{(1+c_1L)(1+n)-nL}\right]^\frac{1}{h}}{\left[\psi^{nL} (1+\alpha
\psi)^{(1+c_1L)(1+n)-nL}\right]^\frac{1}{h}},
\end{eqnarray}
respectively. We note that $\tilde{A}=\zeta_{\Sigma}$ and
Eqs.(\ref{18}) and (\ref{63}), we obtain
\begin{equation}\label{66}
y=\alpha(n+1)\frac{\upsilon_{\Sigma}}{\zeta_{\Sigma}}.
\end{equation}
This shows that $y$ depends on the anisotropy parameter $h$. This
means that $y$ is constant for a pair $(n,\alpha)$ when
$\upsilon_{\Sigma}/\zeta_{\Sigma}=constant$ which is possible only
for every value of $h$. It is noted that the potential at the
surface is uniquely related to each anisotropic and relativistic
polytropes.

The behavior of the surface parameter $y$ corresponding to
anisotropic parameter plays an important role in analyzing the
compactness of the polytropic stars. For case \textbf{I}, the plot
between $y$ and anisotropy parameter is shown in Figure \textbf{19}.
This describes that the value of $y$ decreases as the anisotropy
parameter increases, which means that the degree of compactness of
the model decreases as anisotropy increases and vice-versa. Similar
behavior is obtained for the case \textbf{II} shown in Figure
\textbf{20}.
\begin{figure}\center
\epsfig{file=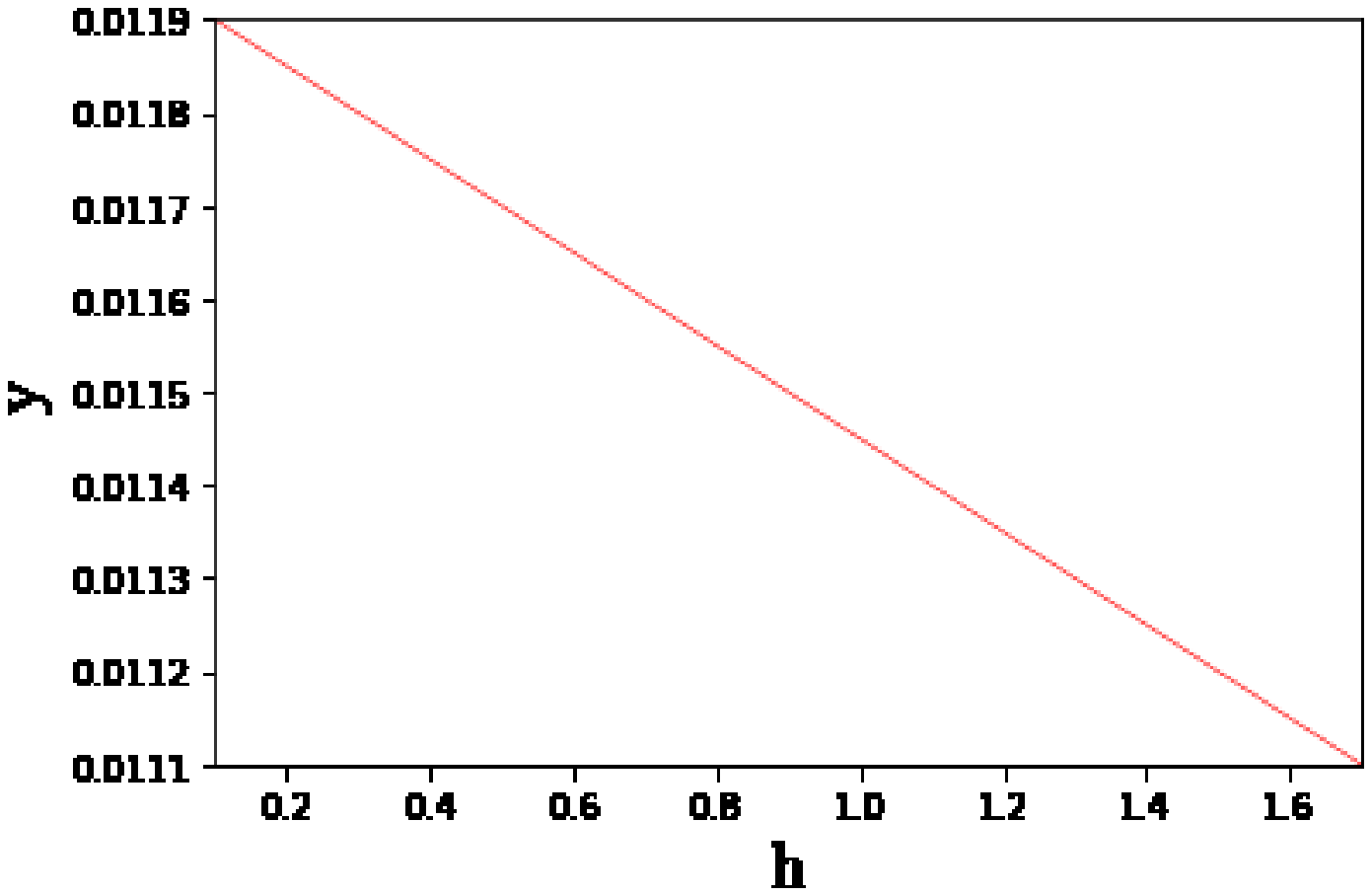,width=0.45\linewidth}\epsfig{file=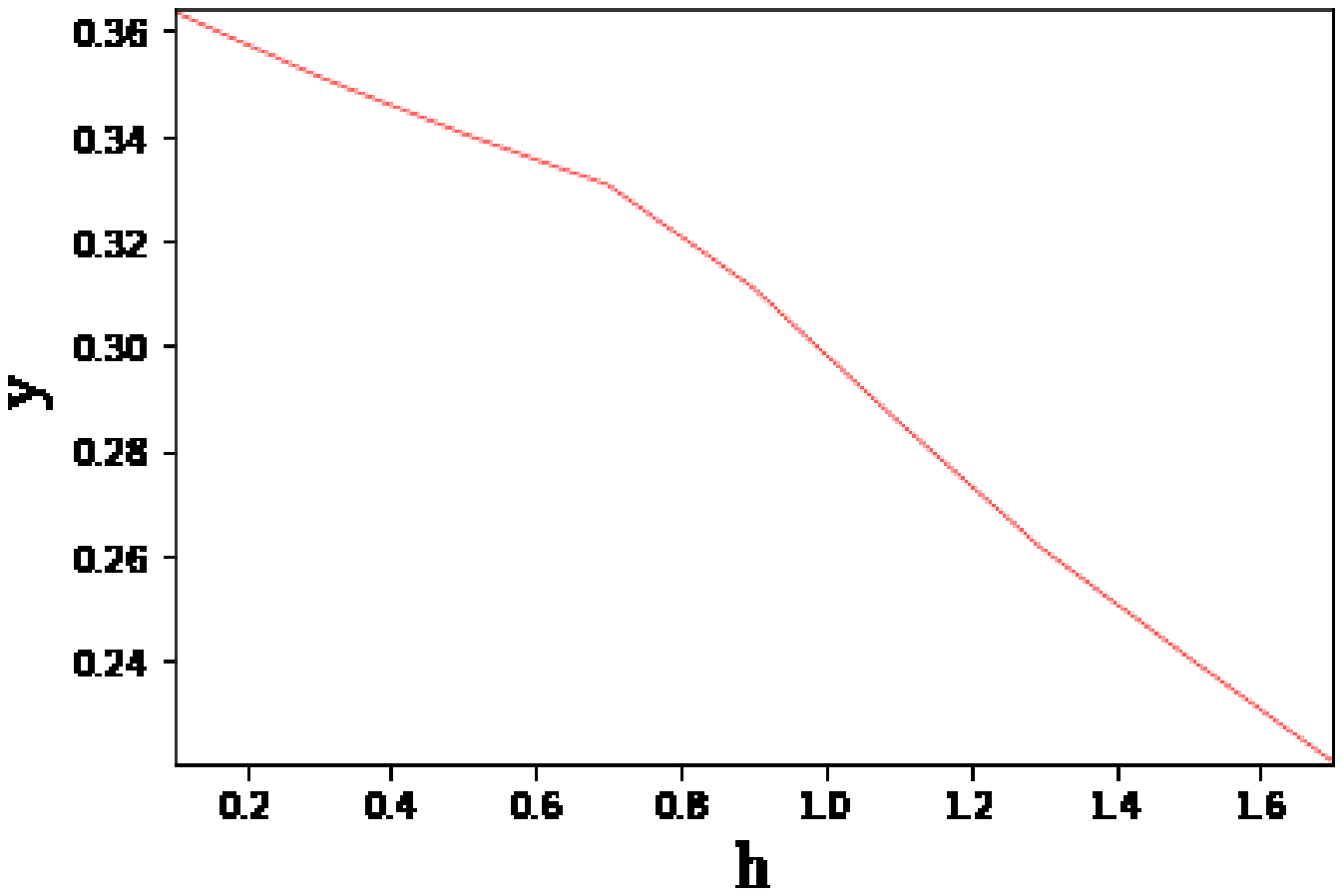,width=0.45\linewidth}\caption{Plots
of $y$ versus $h$ for $\mu=1.5$, $c_1=2$, $\alpha=0.1$ (left), $n=2$
(left), $\alpha=1$ (right) and $n=1$ (right).}
\end{figure}
\begin{figure}\center
\epsfig{file=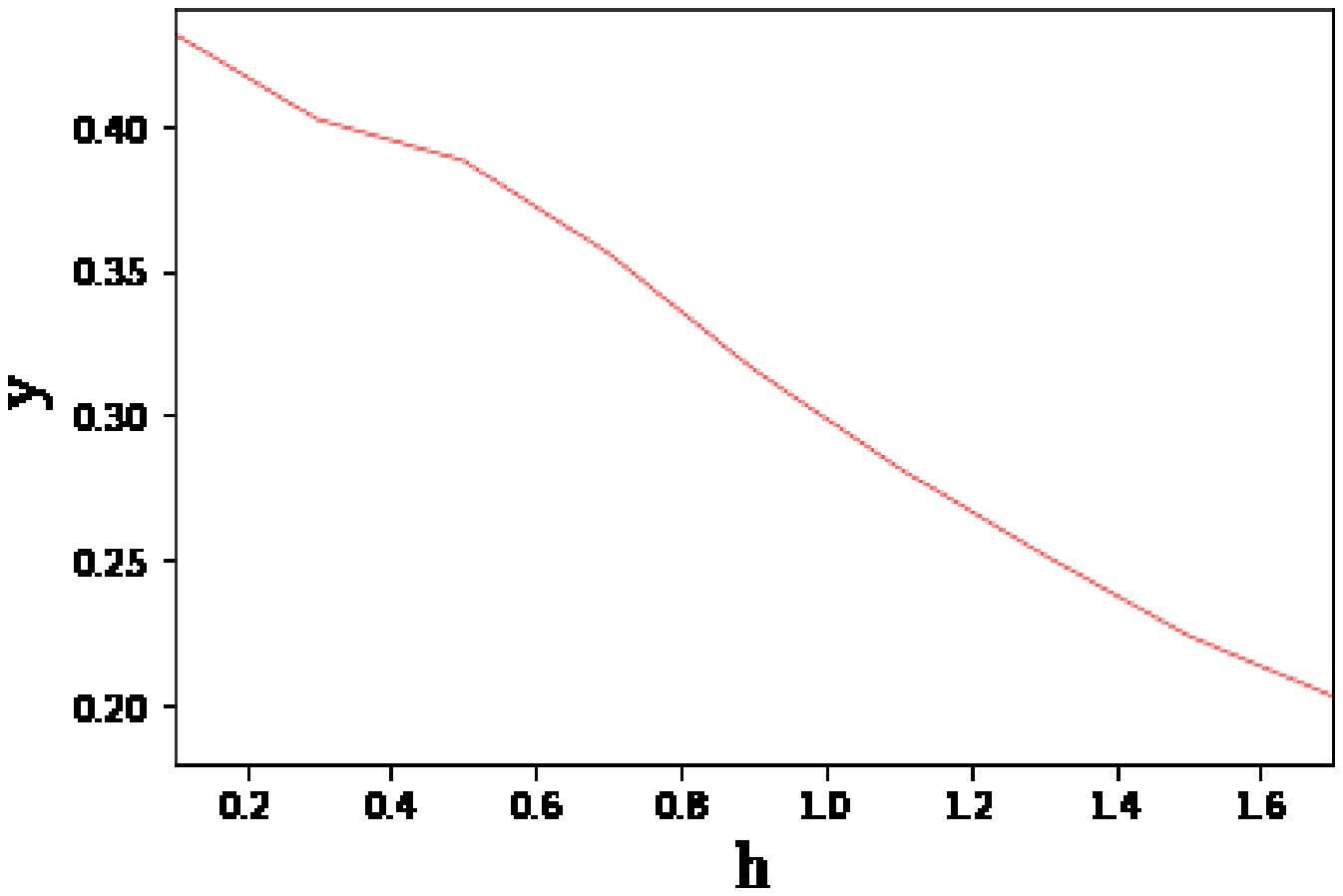,width=0.45\linewidth}\epsfig{file=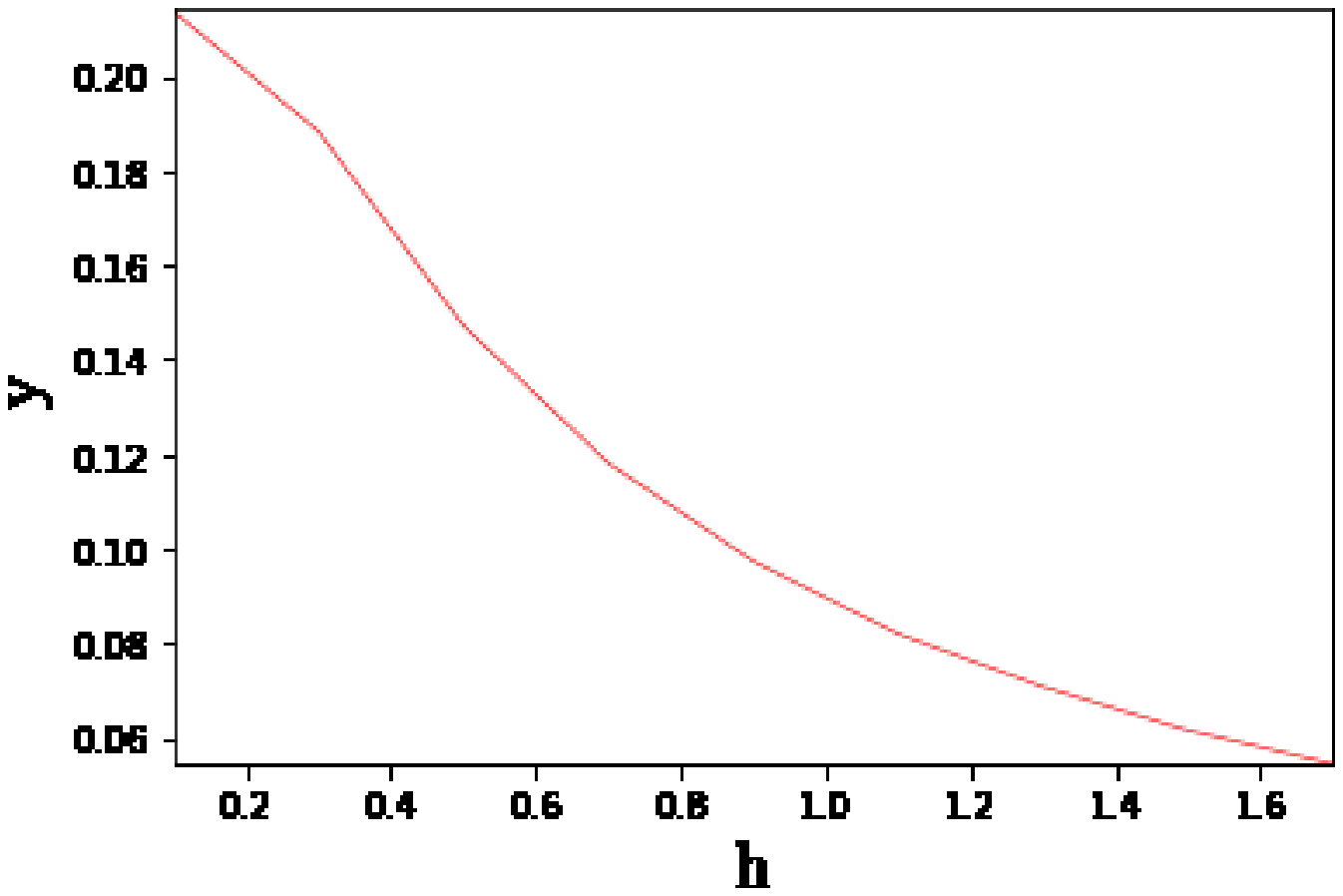,width=0.45\linewidth}
\caption{Plots of $y$ versus $h$ for $\mu=1.5$ and $c_1=2$, $n=1$,
$\alpha=0.1$ (left) and $\alpha=1$ (right).}
\end{figure}
\begin{figure}\center
\epsfig{file=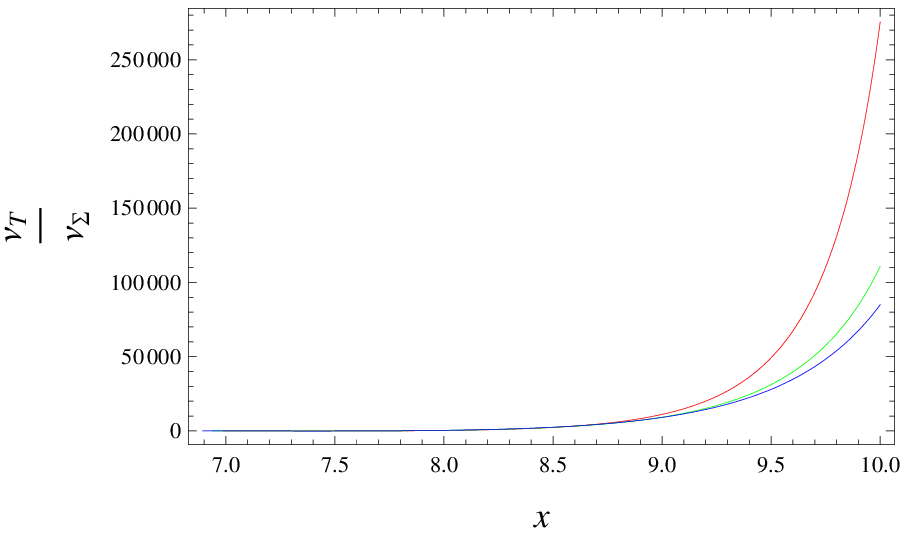,width=0.45\linewidth}\epsfig{file=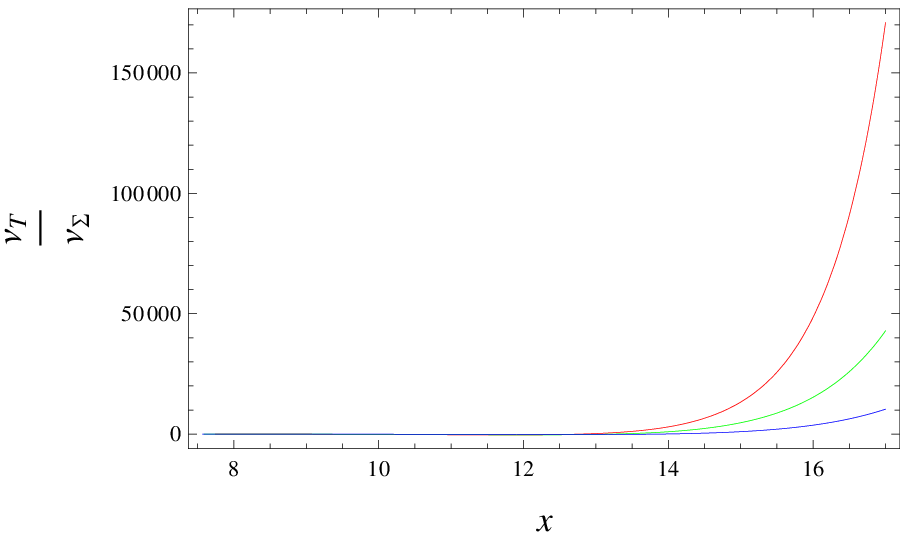,width=0.45\linewidth}
\caption{Plots of $\frac{\upsilon_{\mathrm{T}}}{\upsilon_{\Sigma}}$
versus $x$ for $\mu=1.5$, $c_1=2$, $c_2=4$  and $h=0.5$ (red),
1(green), 1.5(blue), $\alpha=0.1$ (left), $n=2$ (left), $\alpha=1$
(right) and $n=1$ (right).}
\end{figure}

The correspondence between $y$ and $h$ leads to stability of the
models except where the cases for such a correspondence cannot be
obtained. For this purpose, we explore the behavior of Tolman mass
(normalized by the total mass) within the sphere to see the
interesting features of the proposed model. Figure \textbf{21}
(left) implies that when we switch from less compact (large
anisotropy) to the large compact configuration (less anisotropy),
the Tolman mass tends to concentrate at the outer region of the
sphere. The Tolman mass has smaller values in the inner region of a
star as we move from the large anisotropy to the small in the
contraction process. Similar behavior of the Tolman mass is found in
the case \textbf{II} shown in Figure \textbf{21} (right). In this
case, as we move from a small compact to a large compact
configuration, the Tolman mass appears to be condensed at the outer
region of the sphere.

\section{Conclusions}

In this paper, we have studied the isotropic/anisotropic polytropes
for a specific model of $f(\mathcal{R},\mathrm{T})$ theory. We have
considered two polytropic EoS with hydrostatic equilibrium
conditions to construct the LEE. We have used Darmois formalism for
the smooth matching of interior and exterior spacetimes. We have
taken mass (baryonic) density as well as energy density to formulate
the TOV and mass equations in terms of dimensionless variables. The
coupling of these two equations represents a polytrope in
hydrostatic equilibrium. We have also examined different physical
aspects (mass-radius relation, behavior of matter variables,
compactness, redshift, viability and stability) of compact
structures corresponding to isotropic/ anisotropic polytropic EoS
for different values of the parameters. It has been noted that
increase in $\alpha$ decreases the density of the isotropic
solutions whereas more pressure is generated for higher values of
$\alpha$. Moreover, these solutions are viable as well as stable for
$f(\mathcal{R},\mathrm{T})=\mathcal{R}+2\mu \mathrm{T}$ when
$\chi<0$. The compactness, redshift and adiabatic index also comply
with the desired limits.

We have then discussed anisotropic polytropes for two cases. We have
developed the LEE which is then integrated through analytical
approach. The distributions that represent the polytropes, also
determine the compact objects such as neutron stars and
Super-Chandrasekhar white dwarfs. The anisotropic pressure and
relativistic effects cannot be neglected in such configurations. We
have transformed the system into a dimensionless form that reduces
the computational work. The decreasing behavior of the surface
potential for the anisotropic parameter shows compact polytropes.
The graphical analysis of both anisotropic cases reveals that the
related state determinants are positive with a decreasing behavior
towards the surface for $\mu=1.5$. Negative anisotropy implies that
pressure in the radial direction is greater than that in the
transverse direction. Further, the anisotropic polytropes are
composed of normal matter as they are consistent with the energy
bounds. Finally, the first anisotropic solution is stable with
respect to causality criterion whereas the second solution obeys
causality condition for $\alpha=0.1$ and $n=2$. However, the
adiabatic index of the second model is less than $\frac{4}{3}$ for
$\alpha=0.1$ and $n=2$. Thus, the resulting solutions can be used to
construct compact spherical models.

Finally, we have explored the Tolman mass which helps to understand
stability of the models. The efficiency with which the Tolman mass
is reduced in the inner region and concentrated in the outer region
is determined by the anisotropic parameter. The sharp reduction of
the Tolman mass in the interior region of the sphere for smaller
values of anisotropy parameter indicates more compact and more
stable configuration as compared to large values of the anisotropy.
We have found that both isotropic/anisotropic polytropes are more
viable and stable as compared to GR. It is worthwhile to mention
here that all our results reduce to GR \cite{13} when the model
parameter vanishes.

\end{document}